\documentclass[12pt,preprint]{aastex}
\usepackage{amsmath}
\usepackage{amssymb}

\newcommand{\etal}{et~al.~}

\newcommand{\kms}{\ifmmode\,{\rm km}\,{\rm s}^{-1}\else km$\,$s$^{-1}$\fi} 
\newcommand{\be}{\begin{equation}}
\newcommand{\ee}{\end{equation}}
\newcommand{\bea}{\begin{eqnarray}}
\newcommand{\eea}{\end{eqnarray}}

\def \spose#1{\hbox to 0pt{#1\hss}}                                   
\def \ltsim{\mathrel{\spose{\lower 3pt\hbox{$\sim$}}                  
     \raise 2.0pt\hbox{$<$}}}                                                 
\def \gtsim{\mathrel{\spose{\lower 3pt\hbox{$\sim$}}                   
     \raise 2.0pt\hbox{$>$}}}

\def\se#1{\S\ref{Sec:#1}}
 
\def\Fig#1{Figure~\ref{fig:#1}} 
\def\Table#1{Table~\ref{tbl:#1}}

\def\ifm#1{\relax\ifmmode#1\else$\mathsurround=0pt #1$\fi}  
\def\kms{\ifmmode\,{\rm km}\,{\rm s}^{-1}\else km$\,$s$^{-1}$\fi}

\def\ltsima{$\; \buildrel < \over \sim \;$}                    
\def\lsim{\lower.5ex\hbox{\ltsima}}  
\def\gtsima{$\; \buildrel > \over \sim \;$}                    
\def\gsim{\lower.5ex\hbox{\gtsima}}

\def\C28{\rm C_{28}} 

\def\pmb#1{\setbox0=\hbox{#1}% 
\kern-.025em\copy0\kern-\wd0 \kern.05em\copy0\kern-\wd0 
\kern-.025em\raise.0433em\box0} 
 
\def \ion#1#2{#1{\footnotesize{#2}}\relax}  
  
\def \hi {\ion{H}{I} }

\def \littlemm{\ifmmode{\scriptscriptstyle m } 
     \else{\hbox{$\scriptscriptstyle m $ }}\fi}  
\def \topemm{\raise .9ex \hbox{\littlemm}}  
\def \magpoint{\hbox to 2pt{}\rlap{\hskip -.5ex 
     \topemm}.\hbox to 2pt{}}

\usepackage{lscape}
\usepackage{apjfonts}

\slugcomment{For submission to MNRAS. Not for distribution} 

\shorttitle{Formation and Evolution of Virgo Cluster Galaxies - II.} 
\shortauthors{Roediger \etal 2011}

\begin{document}

%%%%%%%%%%%%%%%%%%%%%%%%%%%%%%%%%%%%%%%%%%%%%%%%%%%%%%%%%%%%%%%%%%%%%%%%%%%%%%%%

\title{The Formation and Evolution of Virgo Cluster Galaxies -- \\II. 
 Stellar Populations}

\author{Joel C. Roediger and St\'ephane Courteau}    
\affil{Department of Physics, Engineering Physics \& Astronomy, Queen's 
University, Kingston, Ontario, Canada}

\author{Lauren A. MacArthur}
\affil{Herzberg Institute of Astrophysics, National Research Council of Canada, 
5071 West Saanich Road, Victoria, BC, Canada}

\and

\author{Michael McDonald}  
\affil{Dept. of Astronomy, University of Maryland, College Park, MD}

\email{jroediger,courteau@astro.queensu.ca,Lauren.MacArthur@nrc-cnrc.gc.ca,
mcdonald@astro.umd.edu}

%%%%%%%%%%%%%%%%%%%%%%%%%%%%%%%%%%%%%%%%%%%%%%%%%%%%%%%%%%%%%%%%%%%%%%%%%%%%%%%%

\begin{abstract}
We use a combination of deep optical and near-infrared light profiles for a 
morphologically diverse sample of Virgo cluster galaxies to study the 
radially-resolved stellar populations of cluster galaxies over a wide range of 
galaxy structure. We find that, in the median, the age gradients of Virgo 
galaxies are either flat (lenticulars and Sa--Sb spirals) or positive 
(ellipticals, Sbc+Sc spirals, gas-rich dwarfs, and irregulars), while all 
galaxy types have a negative median metallicity gradient. Comparison of the
galaxy stellar population diagnostics (age, metallicity, and gradients thereof)
against structural and environmental parameters also reveals that the ages of
gas-rich systems depend mainly on their atomic gas deficiencies. Conversely, 
the metallicities of Virgo gas-poor galaxies depend on their concentrations, 
luminosities, and surface brightnesses. The stellar population gradients of all 
Virgo galaxies exhibit no dependence on either their structure or environment.
% Interpreting our results in the context of star formation histories and 
% chemical evolution has allowed us to 
% We have obtained simple constraints on the formation and evolution of Virgo
% galaxies. 
The stellar populations of gas-poor giants (E/S0) are consistent with a 
hierarchical formation, wherein the stars in more massive systems were largely 
formed out of dissipative starbursts associated with gas-rich merging. 
Differences in the stellar population properties of gas-poor dwarfs and giants 
suggests different origins for them. The stellar populations in Virgo dS0's and 
dE's also lack uniformity, suggesting that the formation of Virgo gas-poor 
dwarfs proceeded through at least two different channels (environmental 
transformation and merging). Lastly, the present stellar content of Virgo 
spirals seems to have been largely regulated by environmental effects. Spirals 
with positive age gradients (largely gas-poor types) are likely evolutionary 
remnants of progenitors which were stripped of their gas disks due to prolonged 
exposure to the intercluster medium wind. Spirals with negative age gradients 
are consistent with a traditional inside-out disk growth scenario and have 
likely not been affected by their environment yet. The paucity of flat stellar 
population gradients in Virgo spirals suggests that secular evolution is likely 
not responsible for the formation of their bulges.
\end{abstract}


%%%%%%%%%%%%%%%%%%%%%%%%%%%%%%%%%%%%%%%%%%%%%%%%%%%%%%%%%%%%%%%%%%%%%%%%%%%%%%%%
% SECTION 1: INTRODUCTION
%%%%%%%%%%%%%%%%%%%%%%%%%%%%%%%%%%%%%%%%%%%%%%%%%%%%%%%%%%%%%%%%%%%%%%%%%%%%%%%%

\section{INTRODUCTION}\label{Sec:Intro}

While successes of the $\Lambda$CDM cosmological model \citep{Sp05} suggest 
that galaxies formed through the merging of smaller units, this model still
cannot explain many basic galaxy observables \citep[e.g.,][]{Kl99,No07,Du09},
all of which underscore the importance of the thermal and dynamical interplay
between baryons, feedback, and dark matter during galaxy formation 
\citep{Ba06}. Since this interplay involves the processes of star formation 
(SF) and interstellar medium (ISM) enrichment, study of galaxies' stellar 
populations yields fundamental constraints for galaxy formation models. With 
this in mind, we present a homogeneous analysis of the stellar populations of 
Virgo cluster galaxies, based on an extensive multi-band photometric database 
\citep[Roediger et al. 2011, hereafter Paper I]{Mc11}. The use of a cluster 
sample is advantageous as it encompasses all galaxy types within a relatively 
small angular area and at a common distance. To further motivate our work, we 
highlight below the specific unknowns regarding the formation and evolution of 
basic galaxy types (gas-poor giants and dwarfs, and spirals), and how stellar 
population data may help. The initiated reader may skip to \se{I-Out}.

\subsection{Giant gas-poor galaxies}\label{Sec:I-gGP}
Since the first simulations of binary disk mergers showed that such events 
produce remnants with spheroidal morphologies \citep{TT72}, giant gas-poor 
galaxies (E/S0's) have come to be regarded as the prototypical byproduct of 
hierarchical merging in $\Lambda$CDM cosmology. Supporting evidence for the 
hierarchical origin of E/S0's now includes the time evolution of the galaxy 
merger rate \citep[e.g.,][]{Co03b}, the existence of (ultra-)luminous infrared 
and sub-millimeter galaxies \citep{SM96,Sm97}, and the photometric, structural, 
and kinematical properties of local gas-poor giants on both local and global 
scales \citep[e.g.,][]{Fa97,vD05,Co06,Em07}. Whether the mergers which may have 
created E/S0's were gas-rich, gas-poor, or some combination thereof remains 
unknown, but high-redshift observations of galaxy mergers and sizes, as well as 
input from simulations, suggest that some amount of gas-poor merging is vital 
to the growth, shapes, and kinematics of massive E/S0's \citep{Be06,Na06,vdW08}.

Despite much support for a hierarchical origin of giant gas-poor galaxies, the 
discovery that stellar ages and $\alpha$-enhancements in local E/S0's decrease 
to low masses \citep[i.e., ``downsizing'';][]{Cal03,Ne05,Th05,Cl06,Sm07} has 
been difficult to reconcile within this scenario (but see \citealt{Bow06}, 
\citealt{DeL06} and \citealt{Ne06}). This stems from the fact that downsizing 
implies that more massive E/S0's formed their stars earlier and faster than 
less massive ones, trends which more readily agree with a monolithic collapse 
scenario \citep{Eg62,La74}. The downsizing phenomenon has thus created an 
impetus to measure other stellar population diagnostics for gas-poor giants, 
such as metallicity gradients\footnote{\footnotesize By "gradient," we refer to 
the radial behaviour of a given quantity in a galaxy.} 
\citep[and references therein]{Fo09}, to compare with merger \citep{Ho09b} and 
collapse \citep{Pi08} simulations.

In monolithic collapse, a giant gas cloud undergoes a galaxy-wide starburst as 
it rapidly collapses to its center, leaving remnants with shallow, positive age 
gradients, while their metallicity gradients would be steep and negative due to 
the cumulative enrichment of their central regions by SNe \citep{Ko04,Pi08}. On 
the other hand, the stellar population gradients of merger remnants should span 
a large range, determined primarily by the progenitors' gas fractions. That is, 
the central starburst in a gas-rich merger \citep{HM95} will create a positive 
age and negative metallicity gradient within the remnant's innermost kiloparsec 
\citep{Ho09b}, while violent relaxation will tend to flatten the stellar 
population gradients within a gas-poor merger remnant \citep{Wh80,Ho09c}, 
unless the progenitors' gradients are steep \citep{Di09}. Comparisons of the 
above predictions with stellar population gradients measured in E/S0's has 
however led to conflicting results 
\citep{TO03,Mi05,Wu05,Fo09,Kol09,Sp09,To10a}. Given the difficulty of obtaining 
deep, high signal-to-noise spectra for such galaxies, statistical studies of 
this sort, based on a photometric approach, are highly desirable 
\citep[e.g.,][]{LB10}.

\subsection{Dwarf gas-poor galaxies}\label{Sec:I-dGP}
The formation scenarios for giant gas-poor galaxies would also be relevant to 
their dwarf counterparts (dE/dS0) if the latter simply represented the 
extension of the former to low luminosities. Although the near-orthogonal 
structural scaling relations of gas-poor dwarfs and giants have been used as 
evidence against their common origin (e.g., \citealt{Kor09}; but see 
\citealt{Fe06a} and references therein), \cite{GG03} have shown that all 
gas-poor galaxies populate a continuous manifold when these relations are 
expanded to include the S{\'e}rsic $n$ index. This manifold also explains why 
light profiles vary between gas-poor dwarfs ($n \sim$ 1) and giants ($n \sim$ 
4). Moreover, all gas-poor galaxies follow a single scaling relation between 
the mass of their central compact objects (supermassive black holes or nuclear 
star clusters) and the velocity dispersion of the galaxy spheroids 
\citep{Fe06b}. From the cosmological perspective, a hierarchical formation of 
dwarf gas-poor galaxies appears plausible as $\Lambda$CDM merger trees may 
truncate at low masses \citep{Va08}. A summary of additional evidences that 
favour a common origin for all gas-poor galaxies is presented in \cite{GG03}.

Alternatively, gas-poor dwarfs may represent the evolutionary remnants of 
gas-rich dwarfs which had their gas supply removed. This scenario is supported 
by several evidences: (i) both dwarf galaxy types exhibit exponential light 
profiles \citep{LF83}, (ii) unsharp masking of cluster gas-poor dwarf galaxy
images reveals a high frequency of weak stellar disks, spiral arms and bars 
\citep{Je00,Ba02,Gra03,Li06a}, (iii) the size-luminosity trends exhibited by 
dwarf and giant gas-poor Virgo cluster galaxies disagree \citep{JL08}, (iv) 
many cluster dE's possess significant rotational support (\citealt{Pe02}, 
\citealt{De03}, \citealt{vZ04a}; but see \citealt{Ge02}), (v) unlike Virgo 
E/S0's, dE's \citep{vZ04b} had extended star formation histories as gauged by 
their low stellar $\alpha$-enhancements, (vi) some Virgo dE's possess gas disks 
\citep{Co03a,De03,Li06b}, and (vii) gas-poor dwarfs are mainly found in 
high-density environments, as opposed to gas-poor giants \citep[e.g.,][]{Ha06}.

An evolutionary origin of gas-poor dwarfs would likely be the result of an 
external (environmental) trigger. Supporting evidence of such a trigger 
includes the segregation of gas-poor dwarfs to the centers of galaxy groups and 
clusters \citep{Bi90} and the susceptibility of both the structure and gas 
content of dwarf galaxies to external influences \citep{Ma01,Ma06}. Internal 
triggers of such an evolution, like gas exhaustion and/or expulsion, may only 
play secondary roles as suggested by simulations \citep{MLF99}, simple physical 
arguments \citep{Gre03} and the extended star formation histories of Local 
Group dwarf galaxies \citep{Tols09}. Environmental triggers come in one of two 
broad flavours, tidal \citep{Mo96} and hydrodynamic \citep{GG72}, but the 
observed stripping of (more massive) cluster spirals \citep{Chu09} and the 
correlation of cluster gas-rich dwarfs' H$\alpha$ emission with their gas 
deficiency but not their cluster-centric location \citep{Gav06}, both suggest 
that gas-rich dwarfs would more likely be victims of the latter. Indeed, 
\cite{Bo08a,Bo08b} have demonstrated that ram-pressure stripping and subsequent 
fading of gas-rich dwarfs explain the integrated, multi-wavelength SEDs and 
structural scaling relations of Virgo dE/dS0's. A study of some of these 
galaxies' stellar populations \citep{vZ04b} yielded a similar conclusion.

Despite the compelling similarities between gas-rich and gas-poor dwarfs, 
\cite{Li06a,Li06b,Li07,Li09}, \cite{Tolo09}, \cite{Ki10}, and \cite{Pa10a} have 
shown that morphological subgroups (i.e., ``normal'', disky, blue center, etc.) 
exist amongst Virgo dE's and that this diversity correlates with their colours, 
ages, locations, and orbits. These studies all suggest that the young, 
disk-like dwarfs which reside in the cluster outskirts and move on radial 
orbits evolved from gas-rich progenitors while the old, nucleated ones which 
are more centrally-concentrated and follow circular orbits may have formed 
similarly to the gas-poor giants. As exampled by \cite{Li08}, \cite{Tolo09} and 
\cite{Pa10a}, a stellar population analysis of Virgo gas-poor dwarfs would help 
elucidate the possible bimodal origins of these galaxies.

\subsection{Spiral galaxies}\label{Sec:I-Sp}
Since spiral galaxies generally have two principal components, understanding 
their formation necessarily involves the consideration of both bulges and 
disks. Large spiral bulges resemble giant gas-poor galaxies in many ways 
\citep[hereafter M09]{KI83,Dr87b,An95,Ja96,Xi99,Fa02,Gu09,Ma09}, even out to 
intermediate redshift \citep{Ma08}, so that the formation of such ``classical'' 
bulges has long been discussed in terms of monolithic collapse or hierarchical 
merging \citep{Be92}. Alternatively, ``pseudo-bulges'' may be formed through a 
radial redistribution of disk material due to torques from a bar and/or spiral 
arms, a process known as secular evolution \citep{KK04}.

Early studies of spiral bulges failed to provide a consistent picture of their 
formation \citep[e.g.,][]{Co96,Pe99}. These discrepant results may reflect 
a morphological trend in bulge formation mechanism \citep{Ba03,KK04}, but this 
clashes with the existence of exponential bulges across the Hubble sequence 
\citep{Ma03}. Based on a spectroscopic analysis of late-type bulges, M09 have 
suggested that no conflict exists between the above studies because the 
formation signature inferred depends on the means by which bulges are observed. 
For instance, since secular evolution induces SF (see below), optical images 
of bulges would likely be biased to younger stellar populations (when present), 
while spectra can more fully sample the entire stellar content, thereby 
revealing any underlying classical nature. Unfortunately, M09's study was 
limited to only eight galaxies.

The origin(s) of spiral bulges should be testable, in principle, through these 
galaxies' stellar populations. In the classical bulge formation scenario, 
a gas disk would accrete around the assembled bulge, thereby making the bulge 
older and more enriched than the disk; that is, one expects to find negative 
age and metallicity gradients within such galaxies. On the other hand, the 
redistribution and scattering of disk stars into a pseudo-bulge would make 
it appear roughly coeval and isometallic with respect to its (parent) disk; 
that is, one would expect flat stellar population gradients within these 
galaxies. Although secular evolution also redistributes the gas of the disk 
and likely turns it into stars, it is unclear what effect (if any) this will 
have on a galaxy's stellar population gradients as the gas should end up 
spatially confined to a nucleus and rings \citep{KK04}. In spite of this, it is 
clear that important constraints on spiral galaxies' formation may be obtained 
by identifying the relative ages and metallicities of their bulges and disks.

Spiral galaxies' stellar population gradients also reflect on their disk 
formation. In the traditional scenario \citep{FE80}, disks form inside-out 
through the collapse of a rotating gas cloud, such that their resulting age and 
metallicity gradients should both be negative. Simulations and observations 
both suggest that an outside-in formation scenario, while physically plausible, 
ought to be rare \citep{SL03,Mu07}. A spiral disk stripped of its gas, however, 
may appear like it formed from outside-in since it should exhibit a positive 
age gradient (its metallicity gradient would still be negative though) 
\citep{Bos06}. The possible evolution of disks due to environment is thus an 
important element in the analysis of clusters spirals's stellar populations. 
This is supported by the relative paucity of star-forming systems in regions of 
high density \citep{Dr80}, the existence of red disks at $z \sim$ 1 
\citep{Ko05}, and the dependence of the star formation histories of Virgo 
spirals on their neutral gas deficiencies \citep{Ga02}.

Although the possible impact of environment on the evolution of cluster spirals 
is \textit{a priori} unknown, most evidence to date seems to favour 
hydrodynamic over tidal effects \citep{BG06}. For instance, observations of 
neutral gas displaced from the disks of many Virgo cluster spirals suggest that 
their gas deficiencies result from ram pressure stripping \citep{Chu09}. 
Distinguishing between hydrodynamic and tidal effects via the radially-resolved 
stellar populations of gas-rich cluster galaxies is however not trivial as both 
are expected to leave a galaxy with a young central stellar population 
\citep{Mo96,Vo01}. Further complicating this issue is the fact that the popular 
scenario of SF being quenched within gas-rich galaxies upon their entry into 
dense environments due to gas removal is not strictly true for cluster galaxies 
\citep{Ga91}. Such deviant behaviour may arise if the gas removal mechanism(s) 
can actually trigger SF via ISM compression. Diagnosing the formation and/or 
evolution of spiral galaxies based on stellar population data alone may thus be 
challenging. 

\subsection{Outline}\label{Sec:I-Out}
The study of stellar populations is central to understanding galaxy formation 
and evolution. Broadband colours remain the most efficient means for studying 
the stellar populations of large galaxy samples. For this, a wide and 
well-sampled wavelength baseline is optimal for the modeling of galaxy colours 
\citep[e.g.,][]{Ga02}; the use of optical and near-infrared (NIR) imaging alone 
already overcomes the age-metallicity degeneracy at optical wavelengths 
\citep{Car03}. In this paper we use a large, morphologically complete database 
of optical and NIR photometry for cluster galaxies (Paper I), down to $M_B 
\gtrsim$ -15 mag, to study the stellar populations of Virgo cluster galaxies. 
This database combines imaging from the SDSS \citep{Ad07,Ad08}, 2MASS 
\citep{Sk06}, GOLDMine database \citep{Ga03} and our own H-band observations 
(Paper I).

The layout of this paper is as follows: In \se{Models} we discuss our approach 
to interpreting the colour gradients of Virgo galaxies (Paper I) with stellar 
population models. In \se{Results} we extract stellar population profiles and 
diagnostics for these galaxies to understand their star formation histories 
(SFHs), chemical evolution, and the parameters that govern them. We interpret 
these results in \se{Disc} as constraints for the formation and evolution 
scenarios of basic galaxy types (gas-poor giants and dwarfs, and spirals). We 
conclude in \se{Concs} with a summary of major results. Throughout this paper, 
we express metallicity, $Z$, on a logarithmic scale relative to the solar value 
($Z_{\odot}$ = 0.02) and assume a distance of 16.5 Mpc to all Virgo cluster 
galaxies \citep{Me07}. We also refer to both S0 and S0/a galaxy types simply as 
``S0''.

%%%%%%%%%%%%%%%%%%%%%%%%%%%%%%%%%%%%%%%%%%%%%%%%%%%%%%%%%%%%%%%%%%%%%%%%%%%%%%%%
% SECTION 2: STELLAR POPULATION MODELING
%%%%%%%%%%%%%%%%%%%%%%%%%%%%%%%%%%%%%%%%%%%%%%%%%%%%%%%%%%%%%%%%%%%%%%%%%%%%%%%%

\section{STELLAR POPULATION MODELING}\label{Sec:Models}

This study uses the technique of stellar population synthesis to infer the 
stellar population properties of Virgo galaxies based upon their colour 
profiles (Paper I). In so doing, we adopt the approach of 
\citep[hereafter M04]{Mac04}, providing references where appropriate for the 
sake of brevity.

\subsection{Choice of model and fitting procedure}\label{Sec:CM&FP}
\begin{figure*}
 \begin{center}
  \begin{tabular}{c c}
   \includegraphics[width=0.45\textwidth]{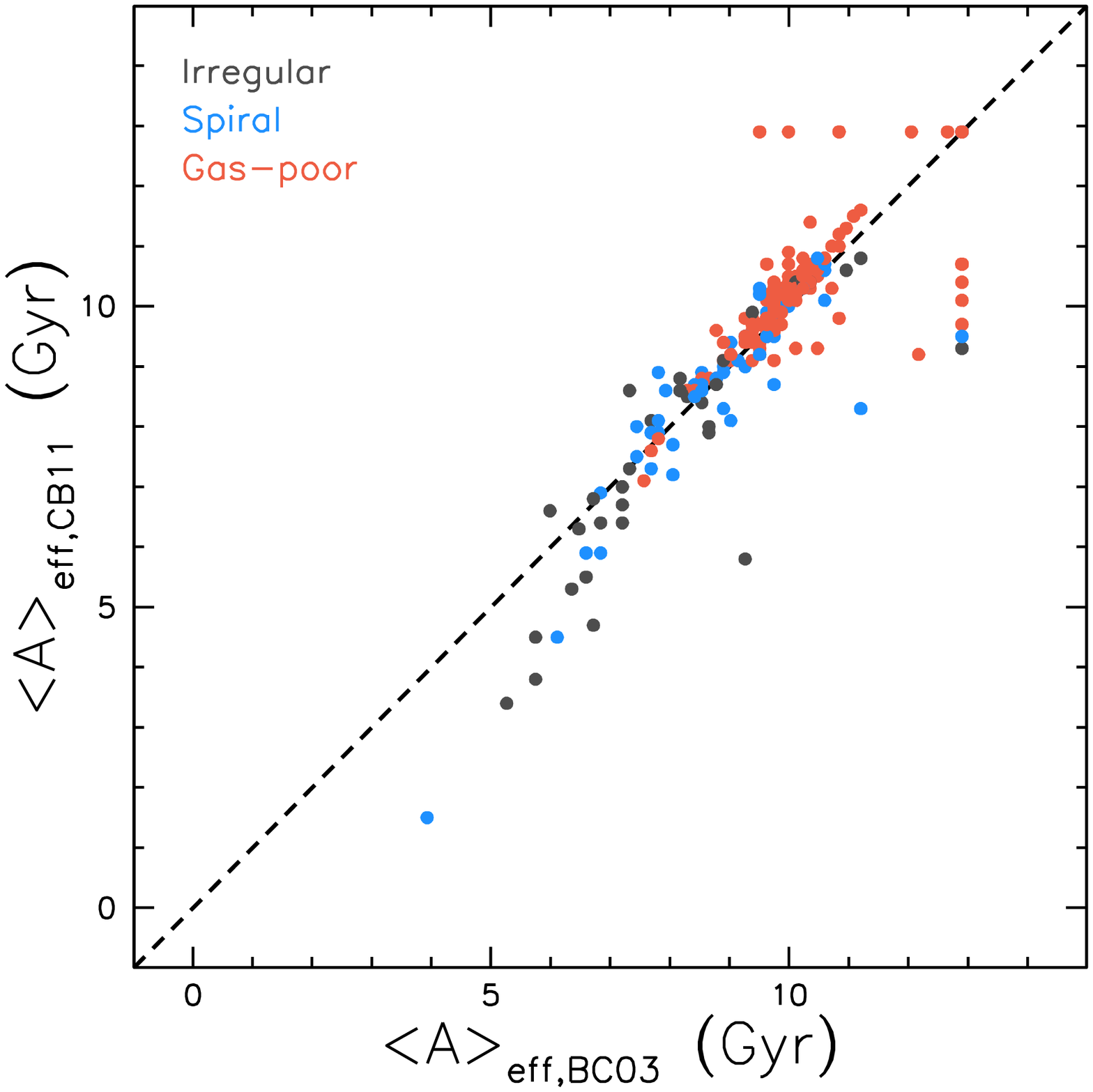} & 
   \includegraphics[width=0.45\textwidth]{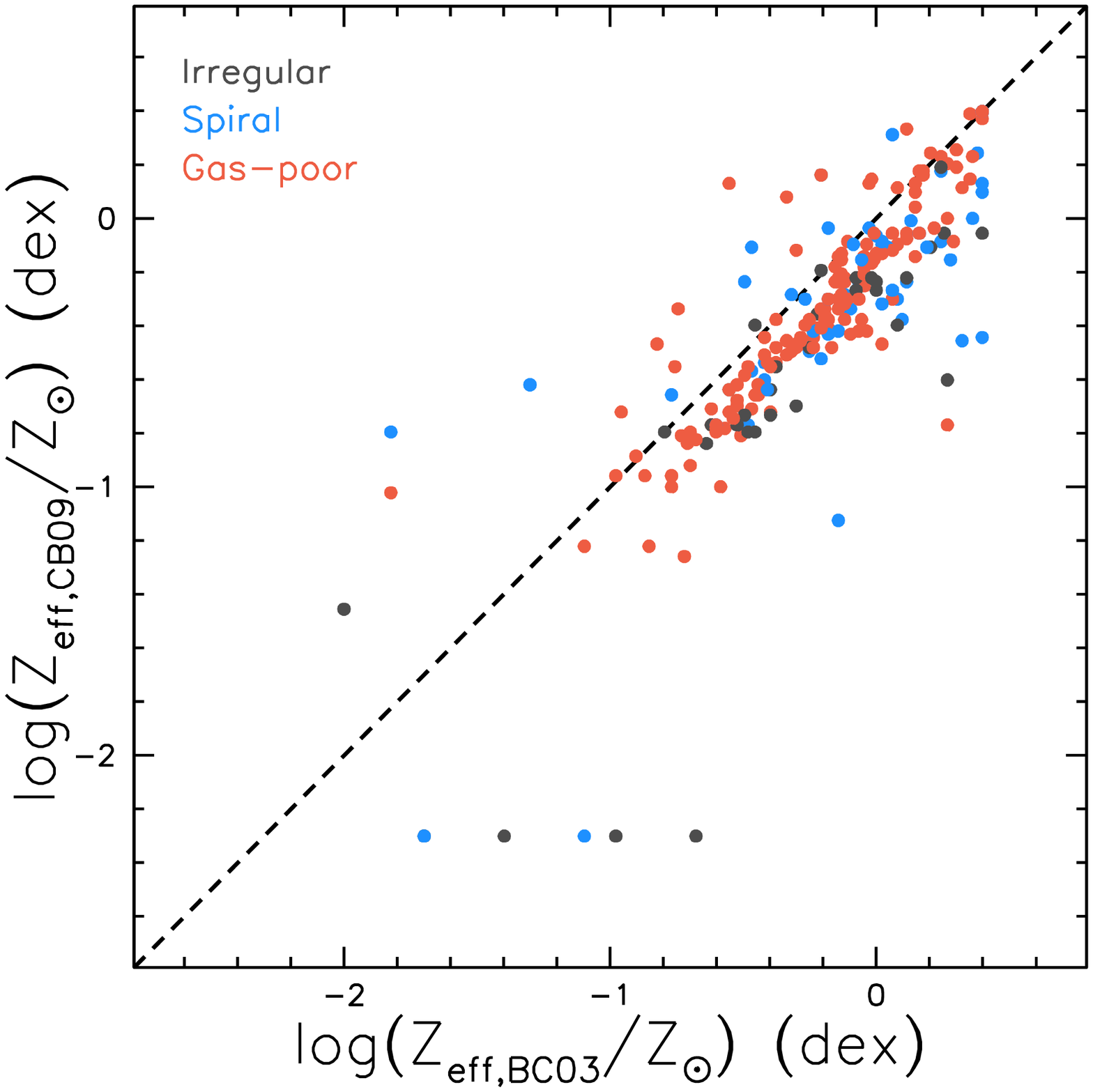} \\
  \end{tabular}
  \caption{(\textit{left}) Comparison of the mean ages measured at the 
effective radii, $r_e$, of all Virgo galaxies based on either the Bruzual \& 
Charlot (2003) or Charlot \& Bruzual (2010) models. The data points have been
coloured according to basic galaxy type (gas-poor, spiral, irregular).
The dashed line shows the locus of equality.
(\textit{right}) Same as (\textit{left}) but for metallicities.}
  \label{fig:Mdl-Comp}
 \end{center}
\end{figure*}

The simplest use of stellar population synthesis involves the comparison of 
galaxy colours to predictions for the passive evolution of a coeval, 
isometallic population of stars (a ``simple stellar population'', SSP) to 
determine light-weighted SSP-equivalent ages and metallicities for the observed 
stellar populations. Of the several public SSP models we prefer those of 
Charlot \& Bruzual (2010, in preparation; hereafter CB10) for their revision to 
the treatment of thermally-pulsating AGB (TP-AGB) stars from the preceding 
version of their models \citep{BC03}\footnote{\footnotesize When present, 
TP-AGB stars dominate a galaxy's infrared emission \citep{Mar05} and thus 
significantly affect its optical-NIR colours \citep{Br07,Ma10}.}. Although 
efforts to improve the TP-AGB treatment in stellar population models, like 
those of CB10, may be inadequate \citep{CG10}, our results are essentially 
invariant under either \cite{BC03} or the CB10 models, as demonstrated in Fig. 
\ref{fig:Mdl-Comp}. Other than the TP-AGB issue, competing SSP models differ 
primarily in details which do not result in significant discrepancies in 
predicted galaxy colours (M04), though careful examination of the systematic 
uncertainties in SSP models have only recently begun in earnest 
\citep[e.g.,][]{Co09}.

Rather than assume that the stars in Virgo galaxies came from single bursts of 
SF (as implicit in the SSP description), we model their colours with synthetic 
populations from an extended SFH. A single-burst SFH not only conflicts with 
the idea of hierarchical galaxy growth but there is also much evidence that an 
SSP is a poor representation of the stellar content of any galaxy, even the 
reddest spheroids (\citealt{ST07}; \citealt{Rog08}; \citealt{Sm09}; but see 
\citealt{Al09}). We opt to model SFHs with simple functional forms, thereby 
neglecting any possible stochasticity in the SF process \citep{Le07}. While the 
combination of simple SFHs, observational errors and systematic model 
uncertainties guarantees that our estimated ages and metallicities are not 
absolutely accurate, our homogeneous approach to modeling Virgo galaxy colours 
ensures that these estimates are meaningful in a relative sense \citep{BdJ00}.

We compute the colours of a composite, isometallic stellar population model via 
the convolution of SSP spectra of many different ages with a chosen SFH (see 
Eq. 5 of M04), assuming that SF began 13 Gyr ago. We do not invoke any chemical 
evolution in our models as this process involves complicated poorly understood 
astrophysics \citep{SP99}. Instead, we interpolate between the colours from the 
metallicity basis set of the CB10 models to create a fine metallicity grid. Our 
models also do not include any treatment of dust attenuation, for reasons 
discussed in M04, Paper I and the appendix.

\begin{figure*}
 \begin{center}
  \includegraphics[width=0.9\textwidth]{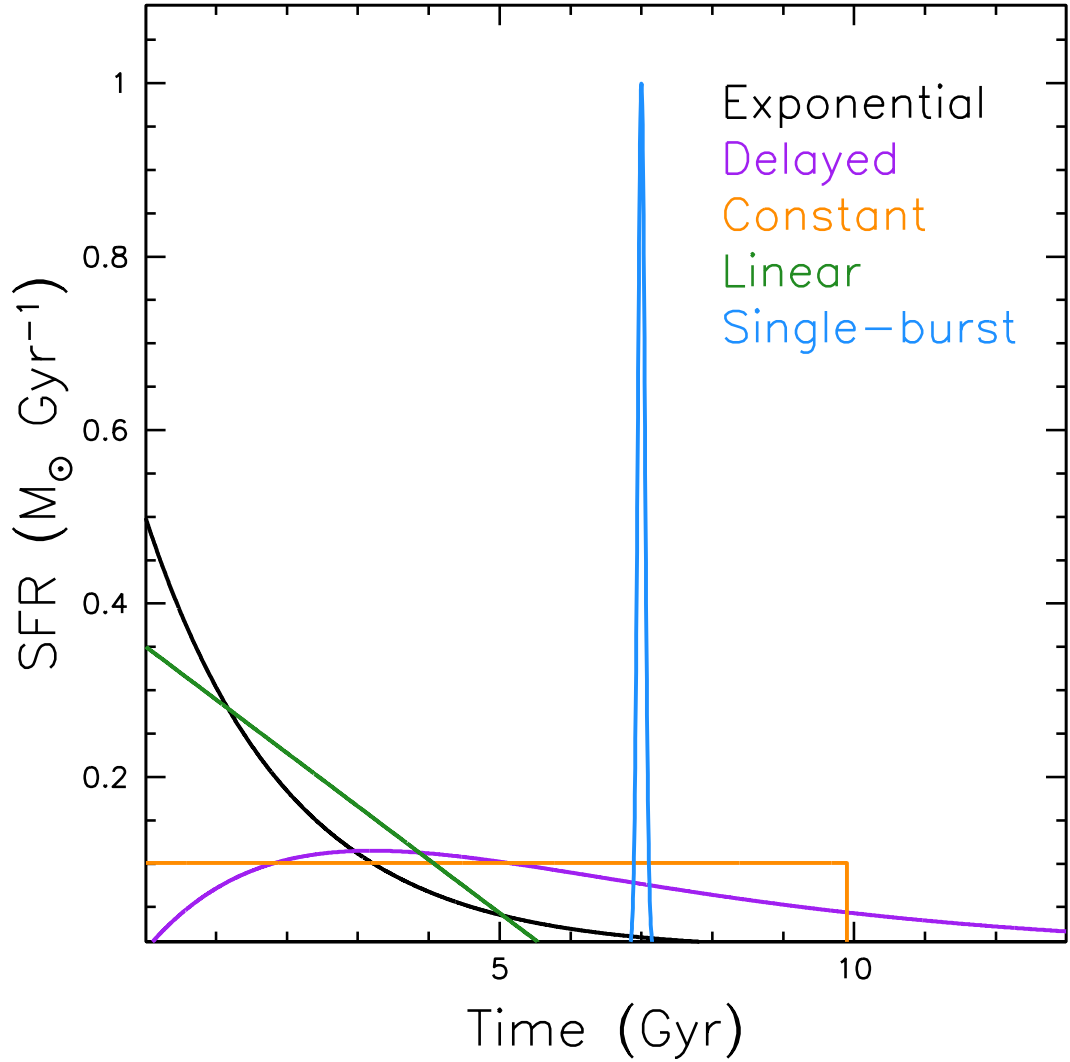}
  \caption{Representative star formation histories (SFHs) explored in our 
stellar population modeling. Each curve is normalized to produce a stellar
mass of unity at 13 Gyr.}
  \label{fig:SFHs}
 \end{center}
\end{figure*}

In the spirit of M04, we have explored several stellar population models 
corresponding to the following SFHs: constant, delayed, exponential, linear, 
and single-burst. The essential information for all models is summarized in 
\Table{SFHs} and \Fig{SFHs}; the formulae representing the star formation rate 
and mean age in each model are quoted in columns 2-3 of the table, while the 
range in model parameters (star formation timescale/mean age) are given in 
columns 4-5. The mean age of a given stellar population model is computed from 
Eq. 10 of M04. Despite assuming an extended SFH for Virgo galaxies, our 
modeling is still light-biased, such that the fits we obtain may not be 
representative of a galaxy's true stellar population if a recent SF burst ($<$1 
Gyr old) has occurred (\citealt{dJD97}; \citealt{ST07}; M09). However, M04 
demonstrated that such a scenario shrinks the colour coverage of the stellar 
population models so much as to be inconsistent with observations, unless the 
burst is small by mass ($\le$10\%) and/or heavily obscured by dust.

\begin{figure*}
 \begin{center}
  \begin{tabular}{c c}
   \includegraphics[width=0.45\textwidth]{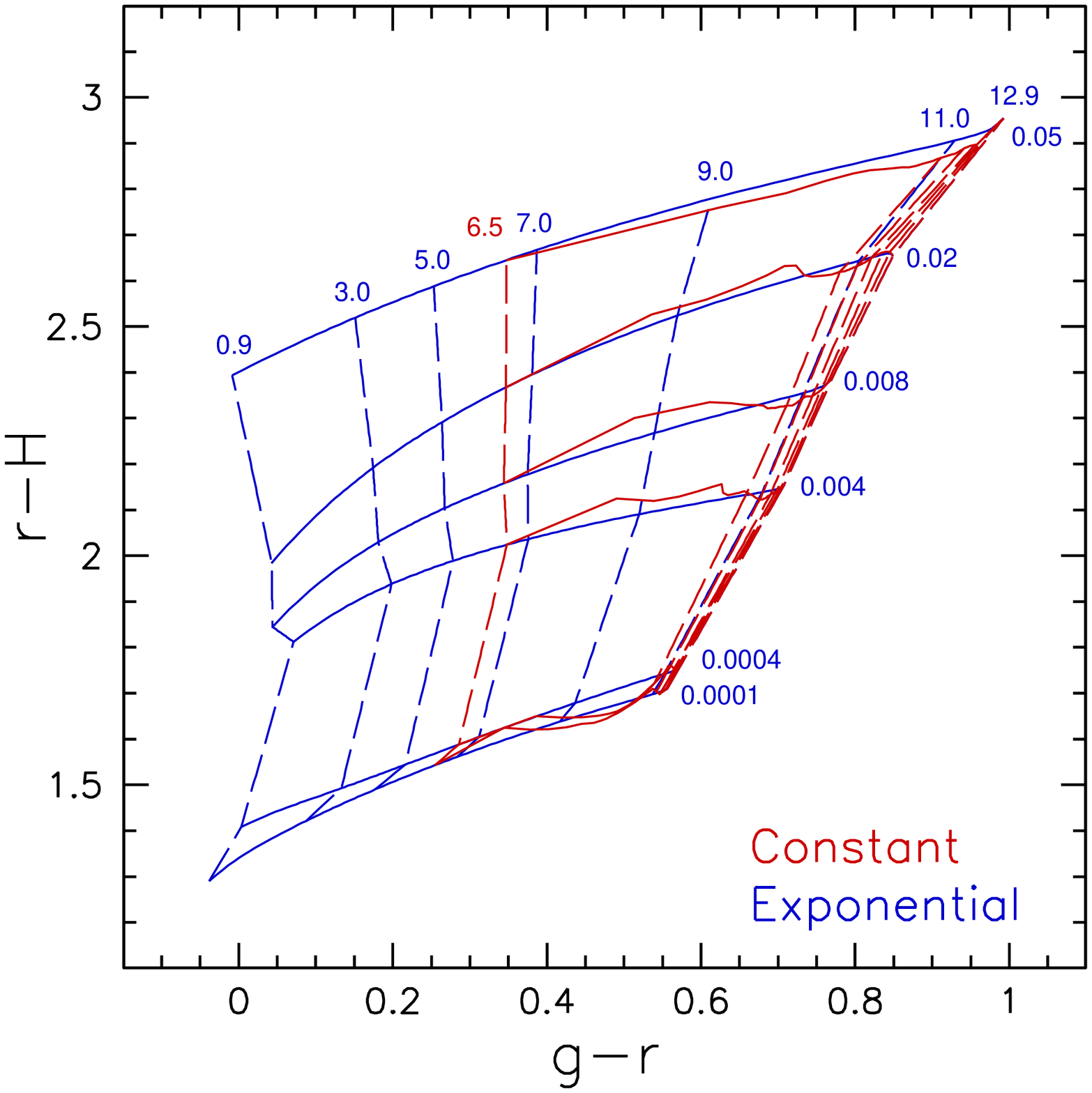} & 
   \includegraphics[width=0.45\textwidth]{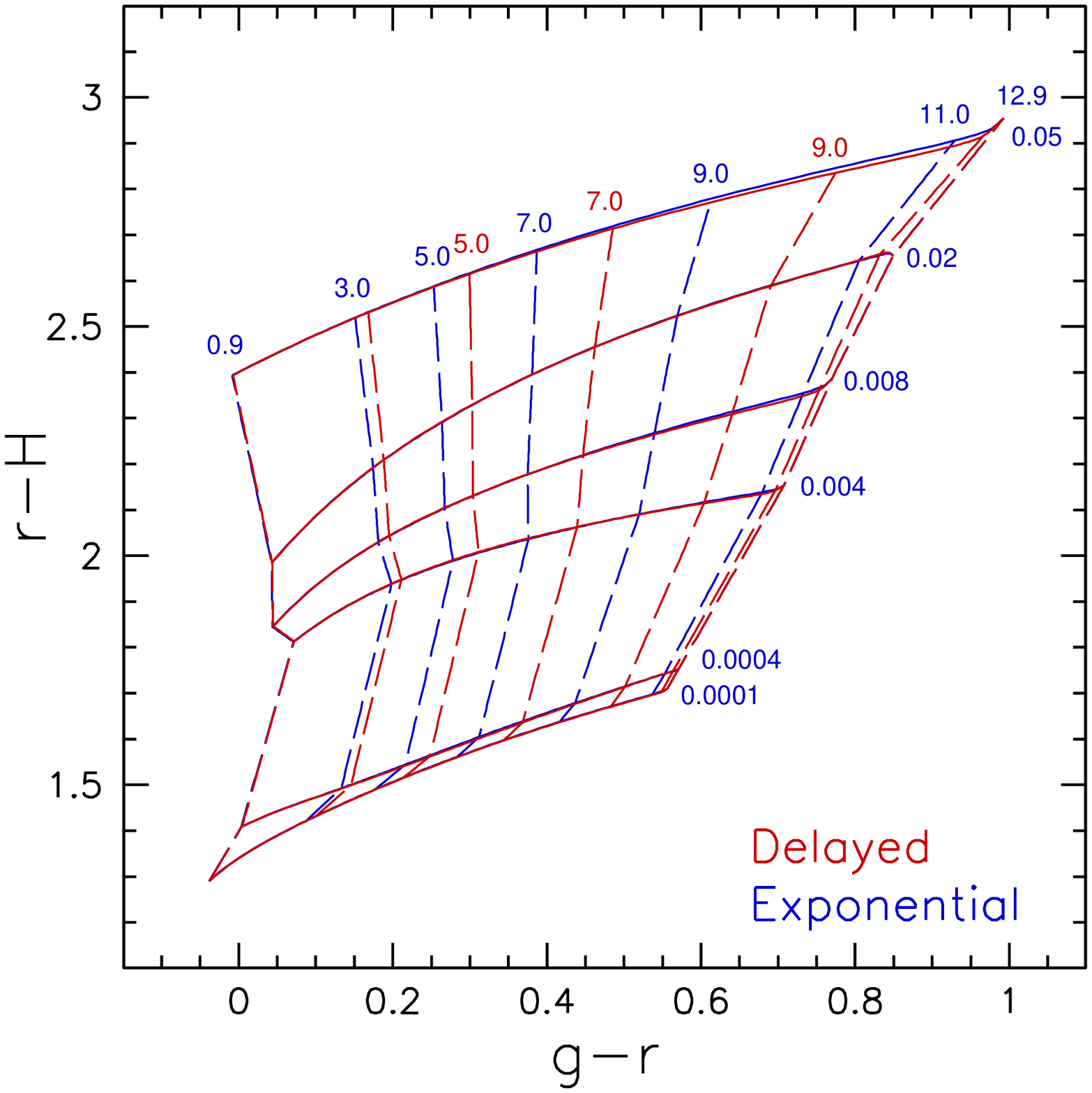} \\
   \includegraphics[width=0.45\textwidth]{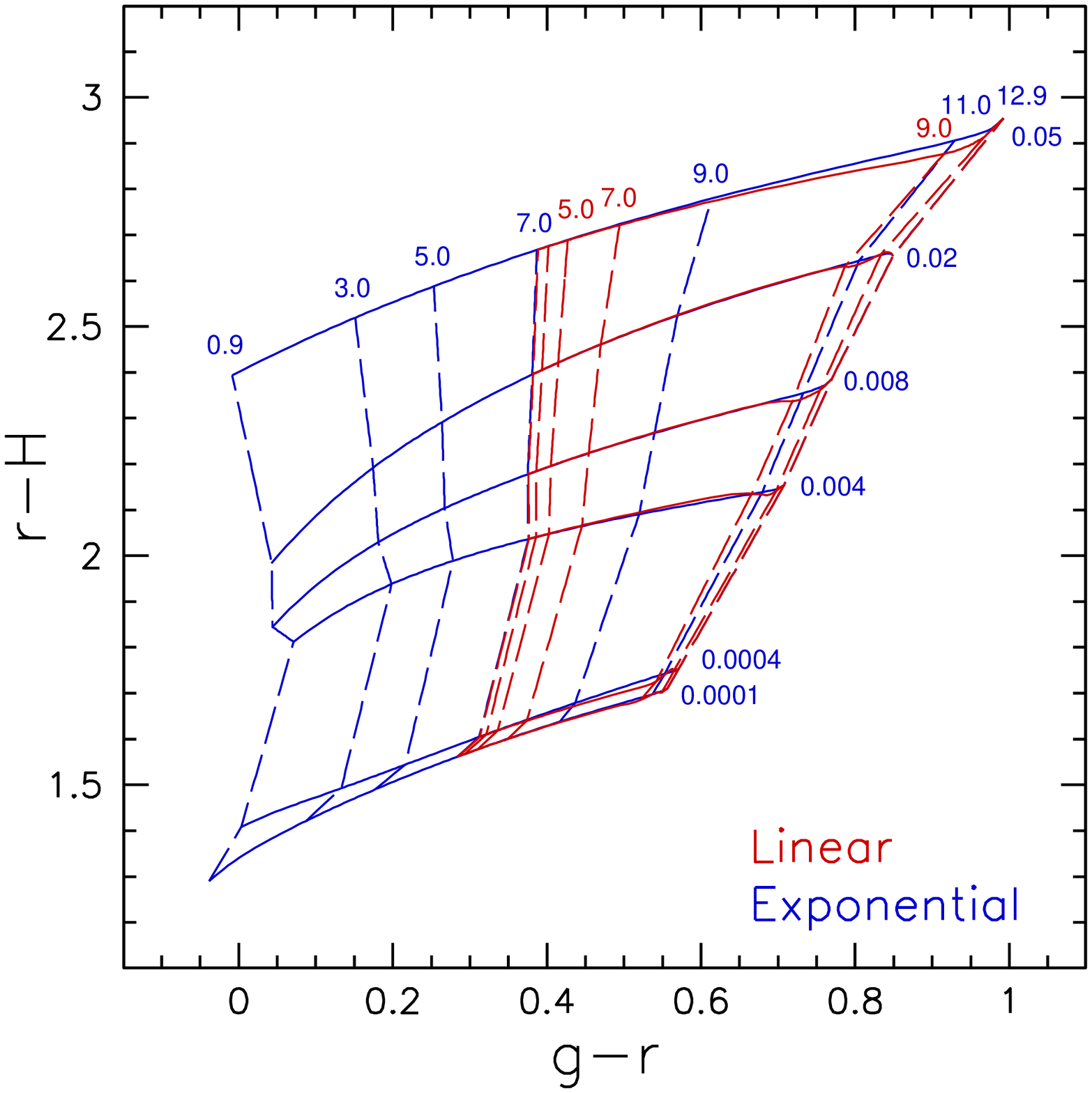} & 
   \includegraphics[width=0.45\textwidth]{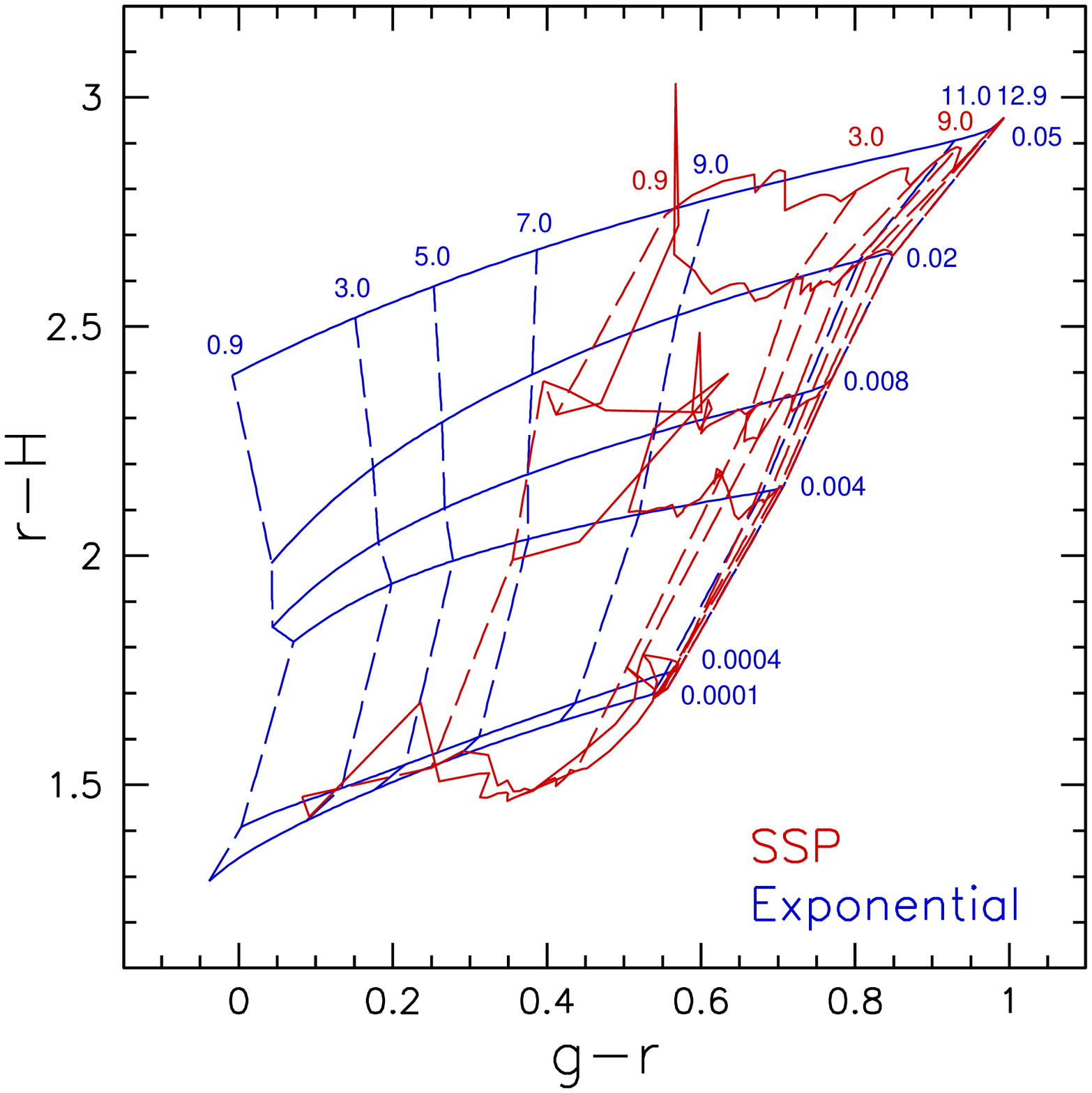}
  \end{tabular}
  \caption{Predictions of the $r$-$H$ versus $g$-$r$ colour grids from 
Charlot \& Bruzual (2010) stellar population models for a (\textit{upper left}) 
constant, (\textit{upper right}) delayed, (\textit{lower left}) linear, and 
(\textit{lower right}) single-burst SFH. The dashed and solid lines represent 
loci of constant mean age and metallicity, respectively.  The predictions
of an exponential SFH model are shown in each panel for comparison.}
  \label{fig:SFH-Comp-1}
 \end{center}
\end{figure*}

\begin{figure*}
 \begin{center}
  \begin{tabular}{c c}
   \includegraphics[width=0.45\textwidth]{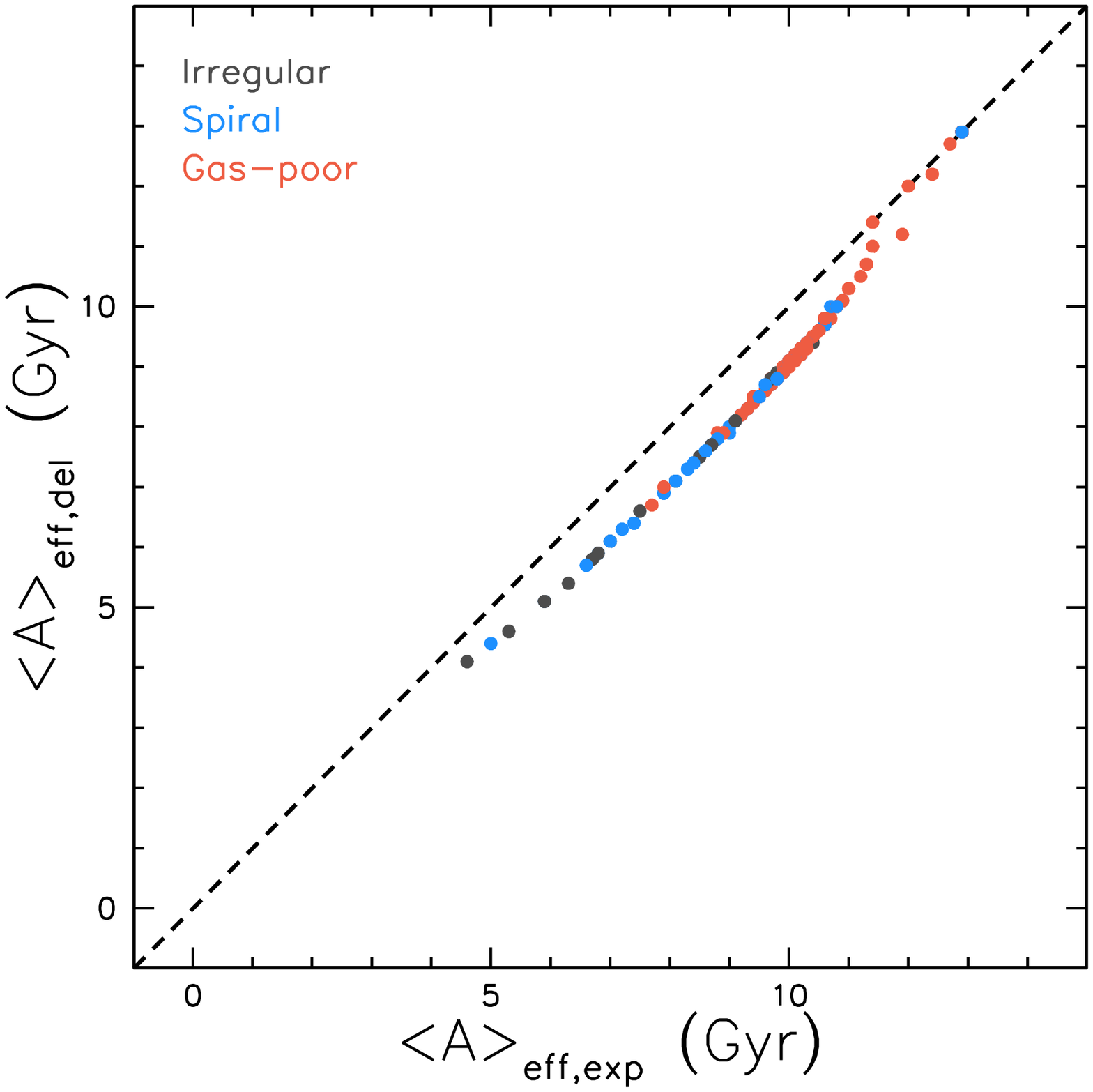} & 
   \includegraphics[width=0.45\textwidth]{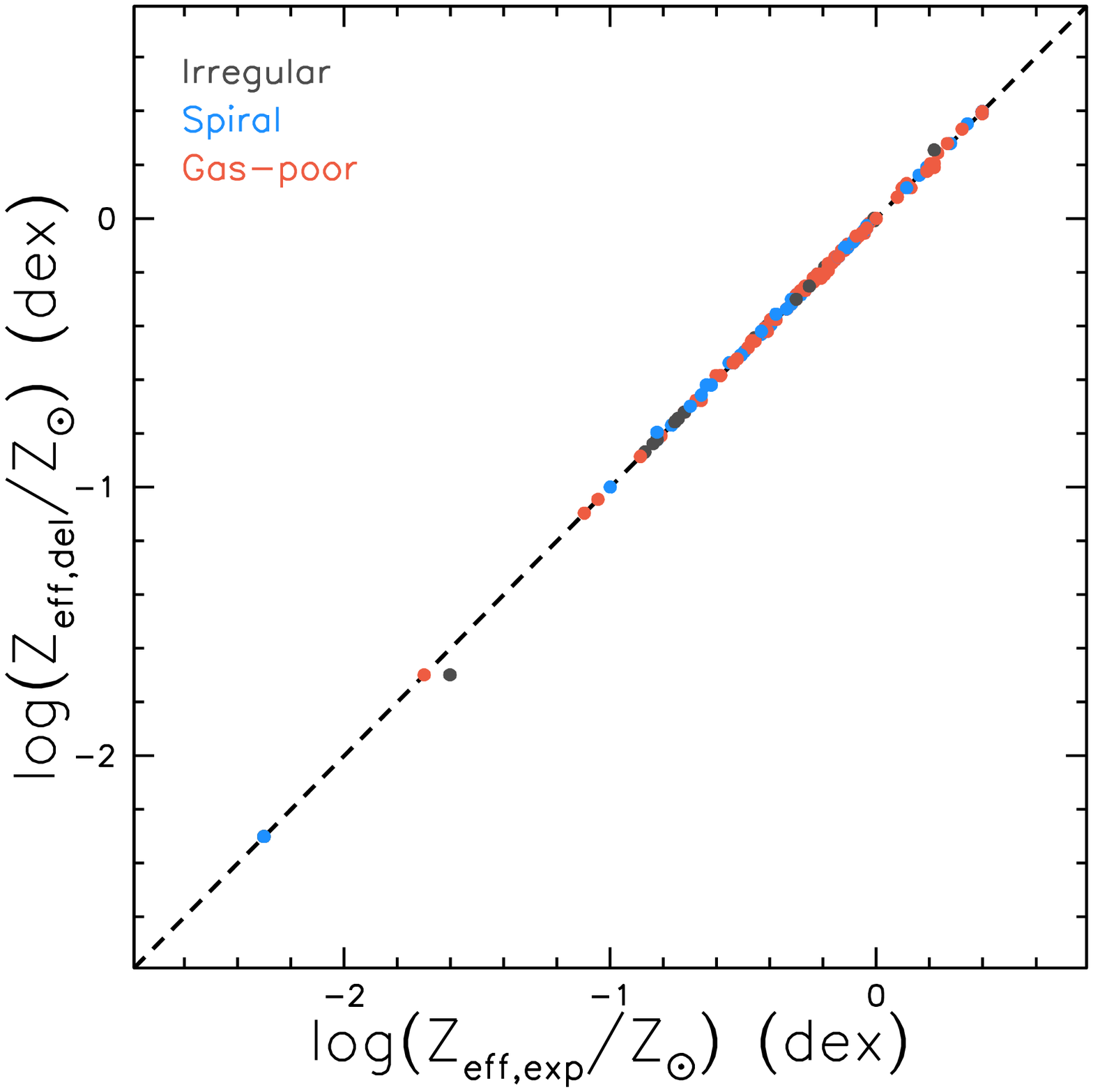} \\
  \end{tabular}
  \caption{(\textit{left}) Comparison of the mean ages measured at $r_e$
for all Virgo galaxies based on either a delayed or exponential SFH. The
data points have been coloured according to basic galaxy type (gas-poor,
spiral, irregular).  The dashed line shows the locus of equality.
(\textit{right}) Same as (\textit{left}) but for metallicities.}
  \label{fig:SFH-Comp-2}
 \end{center}
\end{figure*}

Since we can only hope to quantify relative differences in the stellar 
populations of Virgo galaxies, we opt for a single SFH in our modeling. The 
best choice of SFH is then the one for which the colour coverage is wide enough 
to match that of as many galaxies in our sample as possible. The SFH colour 
coverages shown in \Fig{SFHs} are compared against each other in 
\Fig{SFH-Comp-1}. While the star formation rates in most models decline with 
time, the time of peak SF in the delayed and exponential models is variable, 
such that only they can reproduce the very blue optical colours found in many 
Virgo gas-rich galaxies (Paper I). For consistency with M04, we adopt the 
exponential SFH model but, as shown in \Fig{SFH-Comp-2}, this choice has little 
effect on our inferred mean ages and metallicities.

We determine the age and metallicity for the stars of a given radial bin 
through an explicit grid search for the combination which minimizes the 
model--data colour residuals. The $\chi_r^2$ estimate for the fit includes 
error contributions from calibration, sky, and shot noise uncertainties. For 
each fit, the model-data normalization is determined from a weighted mean of 
their differences in the $griH$ bands (Eq. 16 of M04), while the age and 
metallicity errors are estimated from the envelope about the 68 most reliable 
solutions from fits to 100 Monte Carlo realisations of the measured errors 
(assuming a normal distribution for each). Our error estimates also flag 
outlier bins, as their fits will correspond to the edge(s) of the model grid, 
with zero associated error. We also compute the $\Delta \chi_r^2 \le$ 1 
age-metallicity envelopes about the $\chi_r^2$ minima for all fits and use 
their sizes as weights when determining the age and metallicity gradients of 
Virgo galaxies via linear fits to their stellar population profiles.

%%%%%%%%%%%%%%%%%%%%%%%%%%%%%%%%%%%%%%%%%%%%%%%%%%%%%%%%%%%%%%%%%%%%%%%%%%%%%%%%
% SECTION 3: RESULTS
%%%%%%%%%%%%%%%%%%%%%%%%%%%%%%%%%%%%%%%%%%%%%%%%%%%%%%%%%%%%%%%%%%%%%%%%%%%%%%%%

\section{RESULTS}\label{Sec:Results}

We now consider the age and metallicity profiles of Virgo galaxies, and scaling 
relations between diagnostics of their stellar populations (age, metallicity, 
and their gradients) and their structural and environmental parameters. These 
data are not only sensitive to the relative formation epochs of different 
regions in galaxies (e.g., bulge versus disk) but also to the physical 
mechanisms that control their SFHs and chemical evolution. We interpret these 
data in terms of the formation and evolution of basic galaxy types in \se{Disc}.

\subsection{Age and metallicity profiles}\label{Sec:AZPrfs}
\begin{figure*}
 \begin{center}
  \includegraphics[width=0.9\textwidth]{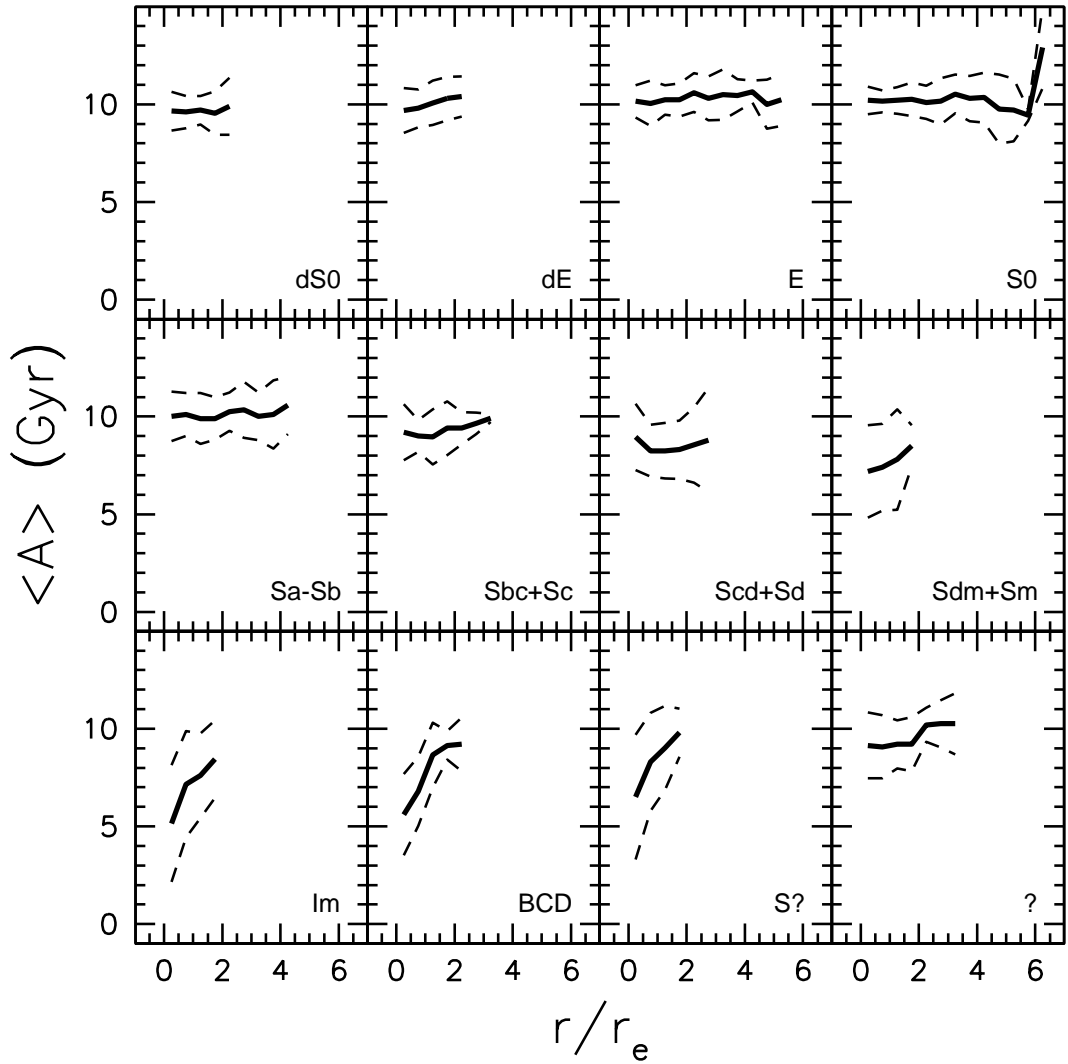}
  \caption{Mean age profiles as a function of scaled radius for Virgo galaxies, 
binned by morphology. The median profiles and their rms dispersions are 
indicated by the thick solid and dashed lines, respectively.}
  \label{fig:Aprf}
 \end{center}
\end{figure*}

We show in \Fig{Aprf} the typical radial age variations for Virgo galaxies, 
binned by morphological type, where the solid and dashed lines represent the 
median stack and rms dispersion, respectively, of the individual age profiles 
in each morphological bin. Having radially binned our galaxies' light profiles 
(Paper I) before applying stellar population models to them, we note that the 
median age profile for each galaxy type extends to at least 2.0 $r_e$ (where 
$r_e$ represents the $H$-band effective radius) at a minimum signal-to-noise 
ratio of ten. \Table{Agrd} lists the median age gradient (columns 2-4) and 
central age (column 5) for each morphological bin (with column 6 providing the 
number of galaxies in each bin), both of which largely reflect our impressions 
from \Fig{Aprf}.

We find from \Fig{Aprf} that the most galaxy types (E$-$Sd, dE/dS0, ?) have old 
central mean ages (9-10 Gyr), while gas-rich dwarfs (Sdm+Sm, Im+BCD) and 
peculiars (S?) are considerably younger (5-7 Gyr). A bimodality seems apparent 
in the age gradient distribution for Virgo galaxies, whereby the different 
galaxy types have either flat (dS0, S0, Sa-Sb) or positive (dE, E, Sbc+Sc, 
Sdm+Sm, Im+BCD, S?+?) median age gradients (a negative median age gradient is 
only found in Scd-Sd spirals); note that the age gradient distributions for the 
dS0, S0, and Sa-Sb galaxy types actually encompass both positive and negative 
values (\Table{Agrd}). The gas-rich dwarfs and peculiars have the steepest 
median age gradients of all galaxy types, which we attribute to their ongoing 
SF, particularly within their central regions. We speculate that the age 
upturns in the outskirts of the median age profiles for the Sbc-Sd galaxy types 
may represent inner disk stars that were scattered outward by spiral arms 
\citep{Deb06,Fo08,Ros08} or gas stripped disks. Finally, despite having similar 
median age gradients (when measured in terms of $r_e^{-1}$), Kolmogorov-Smirnov 
(KS) tests reveal that the age gradient distributions of the gas-poor dwarfs 
and giants are in fact dissimilar.

\begin{figure*}
 \begin{center}
  \includegraphics[width=0.9\textwidth]{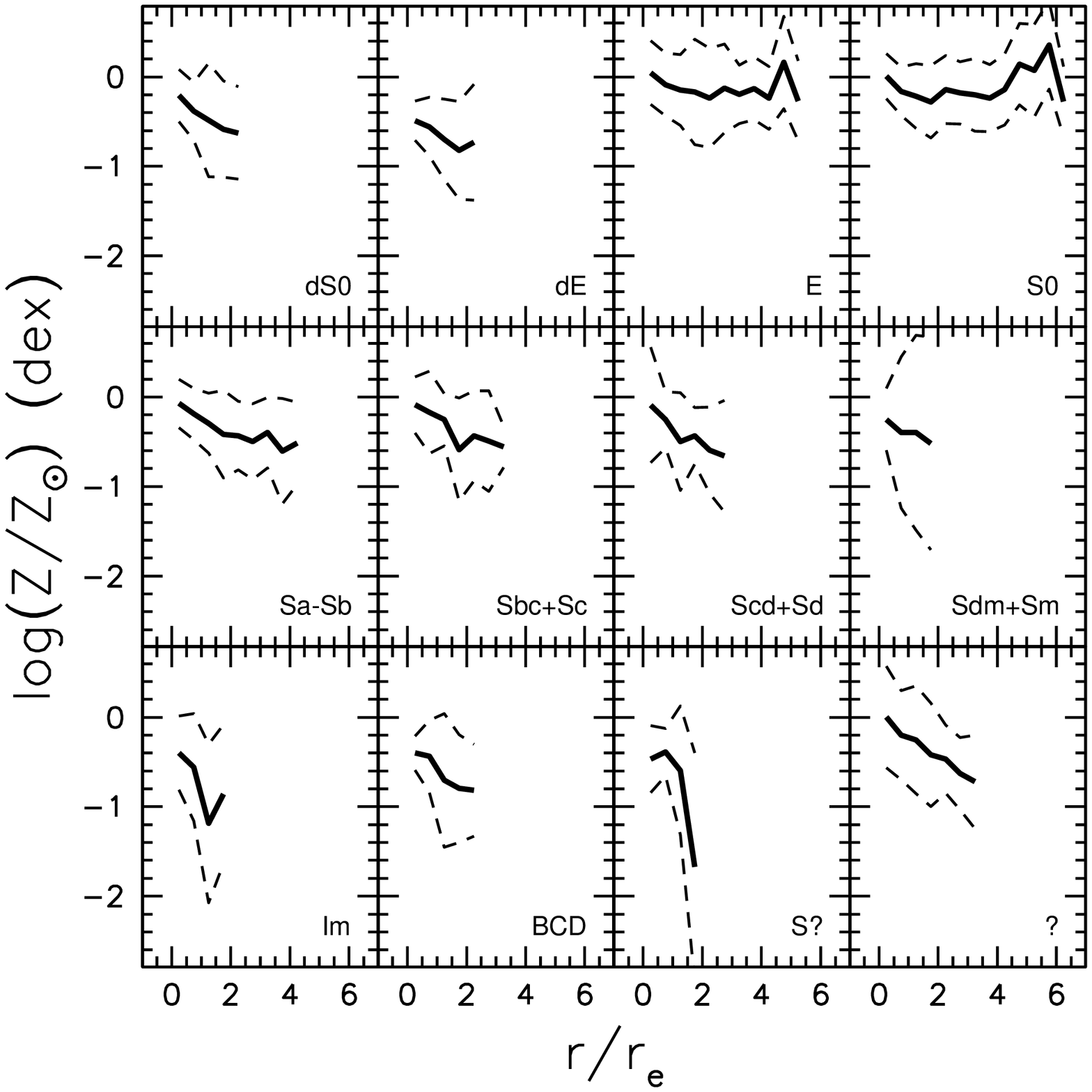}
  \caption{As in \Fig{Aprf} but for metallicity profiles.}
  \label{fig:Zprf}
 \end{center}
\end{figure*}

The median metallicity profiles of Virgo galaxies are shown in \Fig{Zprf}, 
while the median metallicity gradients and central metallicities for all 
morphological bins are listed in \Table{Zgrd} (the structure of which is 
identical to \Table{Agrd}). The median metallicity profiles of Virgo galaxies 
are more uniform than their age profiles in the sense that all galaxy types, in 
the median, exhibit a negative metallicity gradient, albeit spread over a 
considerable range from $\sim$--0.06 dex $r_e^{-1}$ for S0's to $\sim$--0.68 dex 
$r_e^{-1}$ for Im's. Interestingly, the median dlog($Z/Z_{\odot}$)/d$r$ amongst 
S0-Sd galaxies seems to correlate with morphology, such that later types have 
more negative gradients, a fact confirmed through KS tests of their gradient 
distributions. On the other hand, the central metallicities of Virgo galaxies 
appear to segregate the different galaxy types into either metal-rich 
[log($Z_0/Z_{\odot}$) $>$ --0.1 dex; E$-$Scd] or metal-poor [log($Z_0/Z_{\odot}$) 
$<$ --0.2 dex; dS0/dE, Sm, Im+BCD, S?+?] groups, consistent with brighter (more 
massive) galaxies tending to be more enriched than the dimmer (less massive) 
ones \citep{Za94,Tr04}. The latter point is also supported by KS tests which 
yield a probability less than $\sim$10\% that the gas-poor dwarfs and giants 
have similar metallicity gradient distributions. However, the low central 
metallicities of the gas-rich dwarfs, coupled with their steepest of all 
metallicity gradients, suggests that these galaxies are less enriched than the 
gas-poor dwarfs, as expected from studies of the Local Group \citep{Gre03}.

Knowledge of the typical stellar population gradients within galaxies of all 
types enables an interpretation of the source(s) of their colour gradients. 
Comparing the median age, metallicity, and colour (Paper I) profiles of Virgo 
galaxies, we find that the colour gradients of most galaxy types primarily 
reflect the outward decrease in their metallicities, save for a few notable 
exceptions. For one, the negative colour (bluing outward) gradients in Scd+Sd 
spirals are created by both their negative age and metallicity gradients, while 
the outward reddening seen in gas-rich peculiar (S?) galaxies is undoubtedly 
due to their positive age gradients. On the other hand, the positive age and 
negative metallicity gradients found in E, Sdm+Sm, and BCD galaxies couple so 
as to generate their quasi-flat colour profiles. Therefore, the null colour 
gradients within galaxies should not be taken as a reflection of a stellar 
population with constant age and metallicity profiles.

\subsubsection{Comparison with literature}\label{Sec:SPGrd-Comp}
As in Paper I, we compare our results with similar published studies. The most 
studied galaxy types in the context of stellar populations is by far the 
gas-poor giants. Qualitatively speaking, several studies have reported both 
positive \citep{LB10,To10a} and flat \citep{Fr89,KA90,Sa00,TO03,dP05,Wu05} age 
gradients within giant gas-poor galaxies, similar to what we have found in the 
median for Virgo E's and S0's, respectively. In addition, these studies 
unanimously show that these galaxy types possess negative metallicity 
gradients, spread over a range from $\sim$--0.2 to $\sim$--0.4 dex per decade 
radius. Our measured median metallicity gradient for Virgo giant gas-poor 
galaxies, --0.33 dex per decade radius, falls nicely within this range. 
% Negative metallicity and negative age gradients - BG90,TO04

For the remaining galaxy types (dwarfs and spirals), the literature appears 
short of quantitative estimates of their stellar population gradients. Any 
comparison of with our results must be kept strictly qualitative. Despite this 
limitation, there exists studies of spiral galaxies that report age and 
metallicity gradients which span a range of positive to negative values 
(\citealt{dJ96}, M04, \citealt{MH06}, \citealt{Ja07}, M09), as is the case with 
our Virgo spirals. Age and metallicity gradients have been inferred in dwarf 
galaxies as well \citep{Ch07,Chi09,Kol09,Pa11}, in qualitative agreement with 
our results.

\subsection{Stellar population scaling relations}\label{Sec:SPSR}
\begin{figure*}
 \begin{center}
  \begin{tabular}{c c}
   \includegraphics[width=0.45\textwidth]{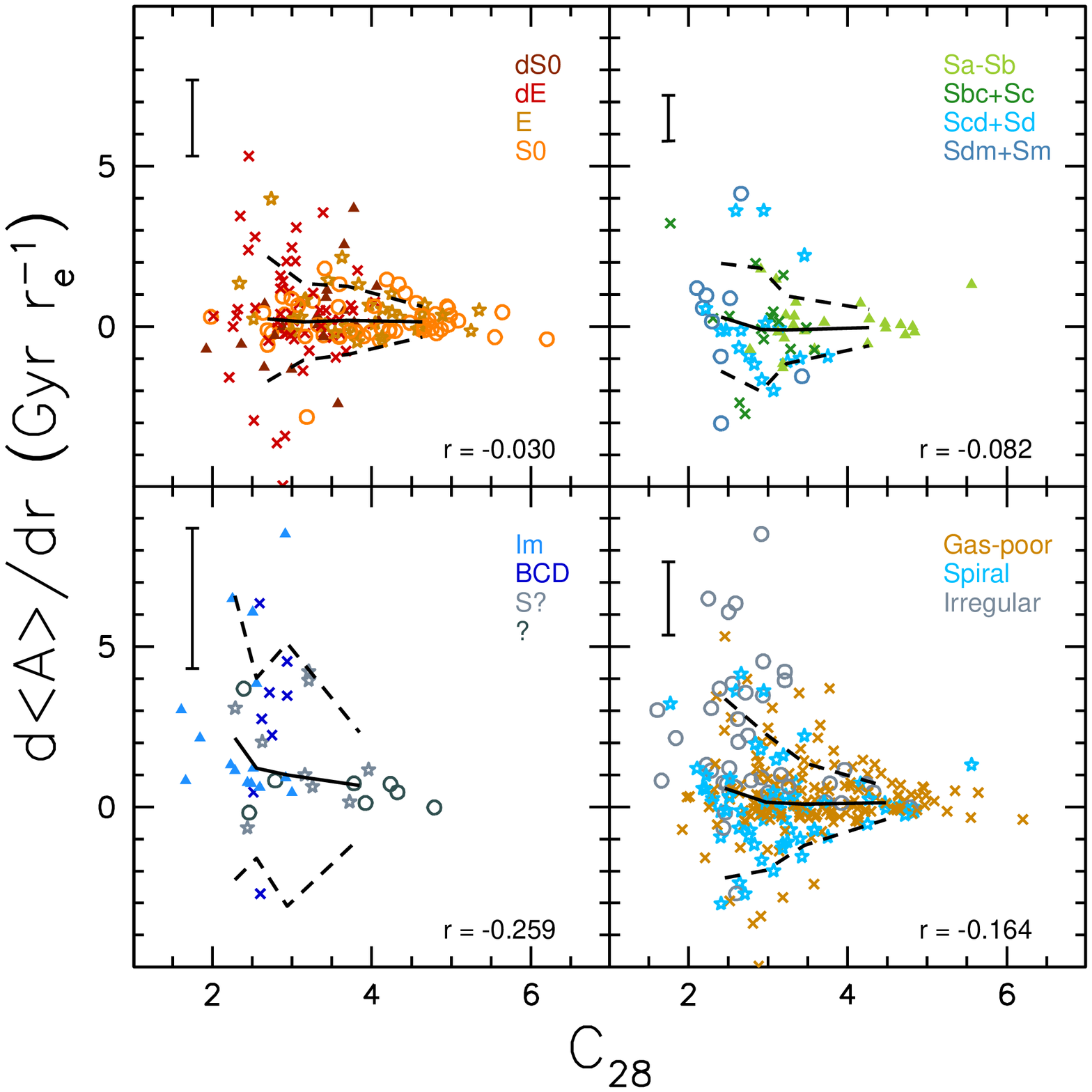} & 
   \includegraphics[width=0.45\textwidth]{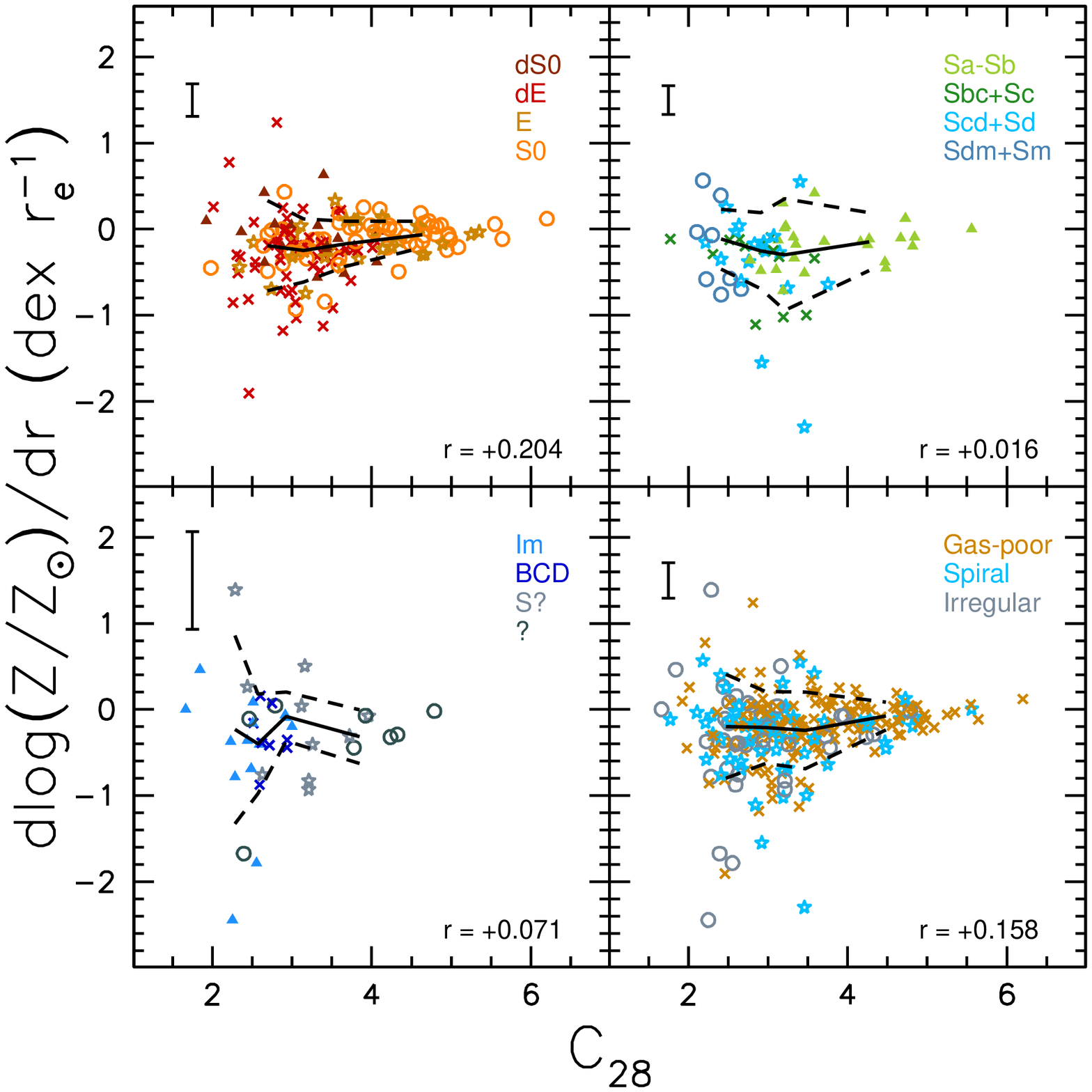} \\
   \includegraphics[width=0.45\textwidth]{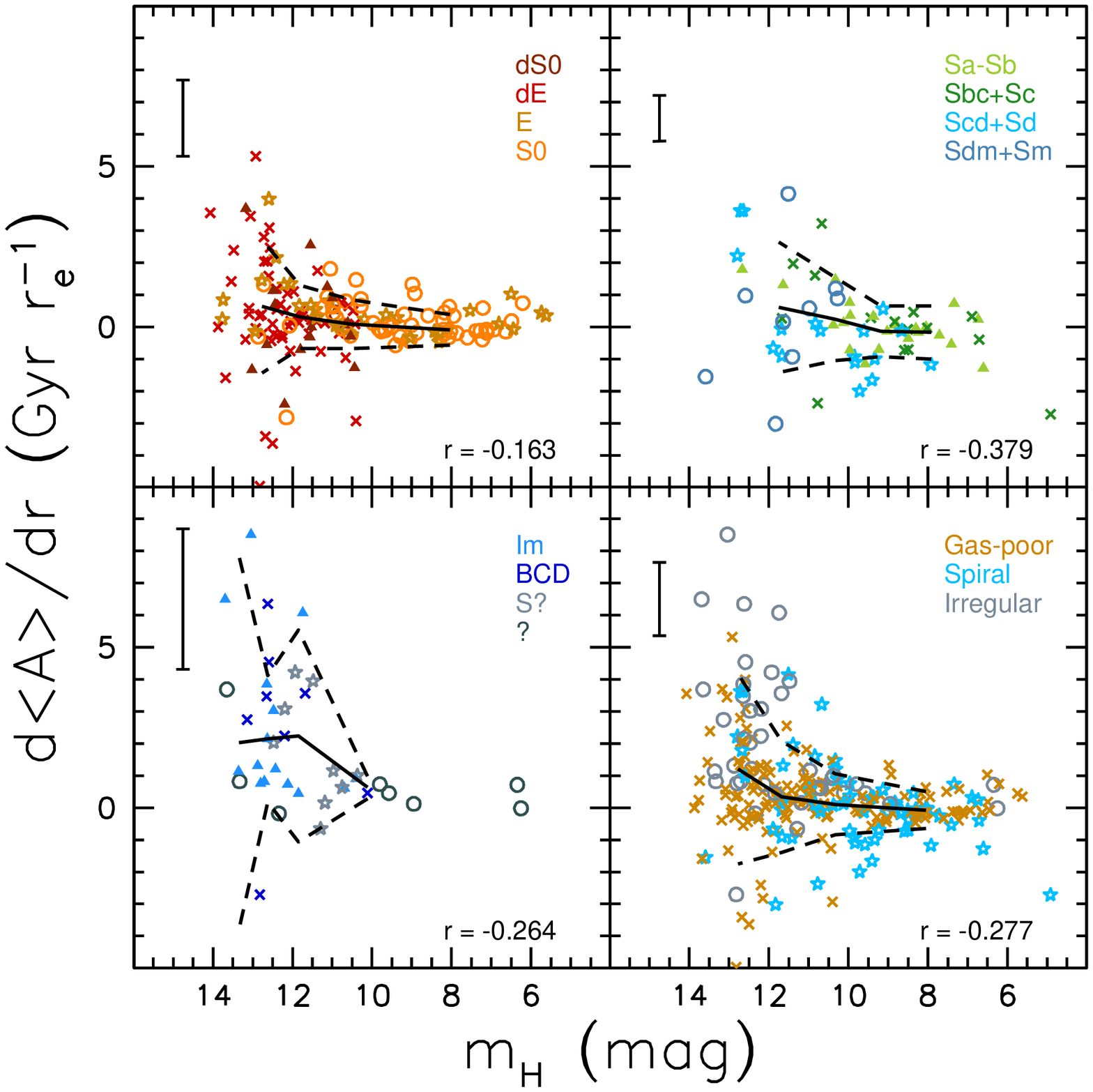} & 
   \includegraphics[width=0.45\textwidth]{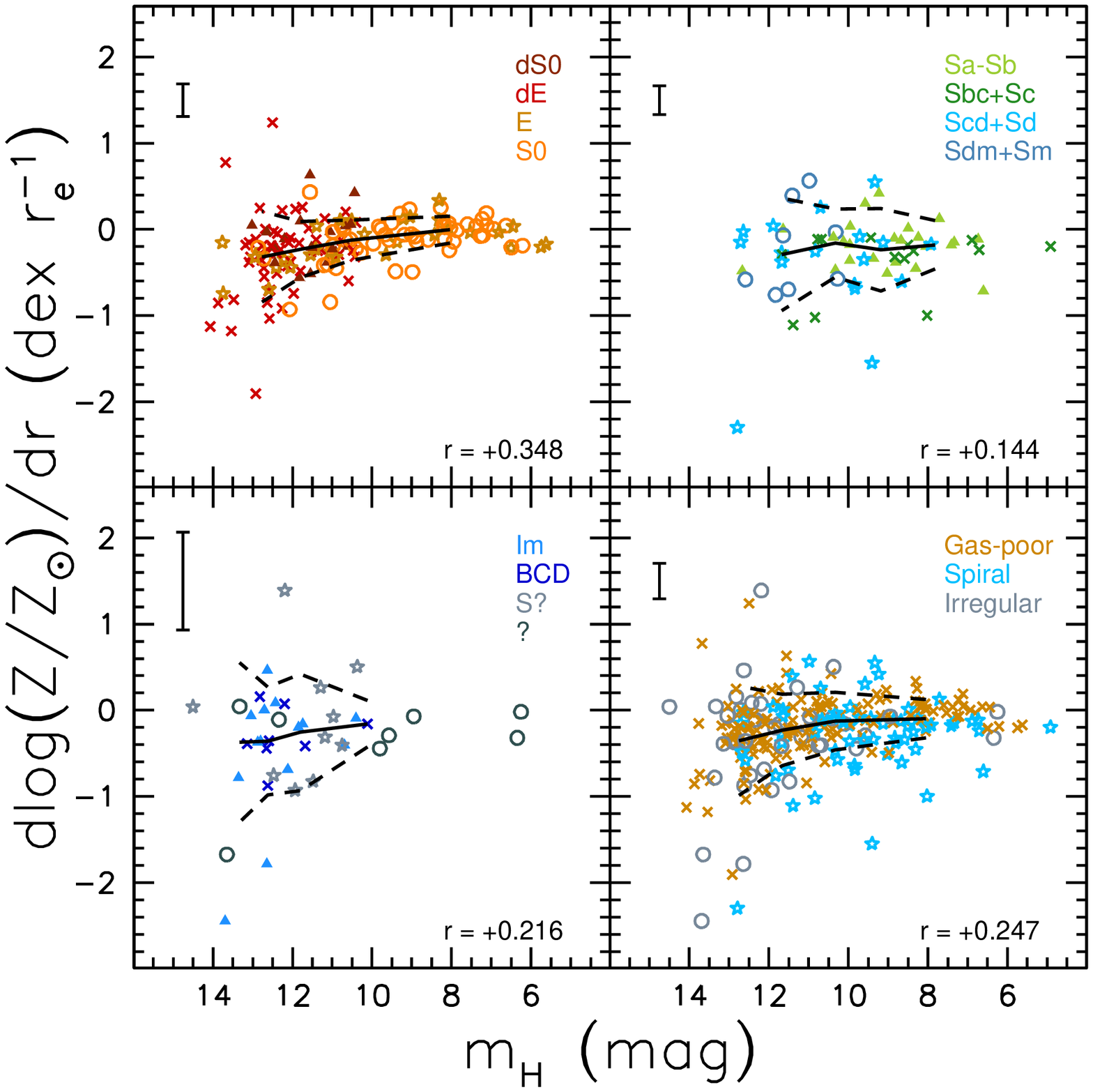}
  \end{tabular}
  \caption{Mean age gradients (\textit{top-left}) and metallicity gradients 
(\textit{top-right}) versus $H$-band concentrations for Virgo galaxies, binned 
by basic galaxy type (gas-poor, spiral and irregular). The gradients are 
expressed in terms of the scaled radius $r/r_e$, while individual galaxies
are separated by specific morphology.  The median trend (solid line) and its
rms disperison (dashed lines), the Pearson correlation coefficient, and the
typical error per point are shown in each window. The bottom two plots are
as at top, but shown versus apparent $H$-band magnitudes.}
  \label{fig:MAZ-C28-H}
 \end{center}
\end{figure*}

\begin{figure*}
 \begin{center}
  \begin{tabular}{c c}
   \includegraphics[width=0.45\textwidth]{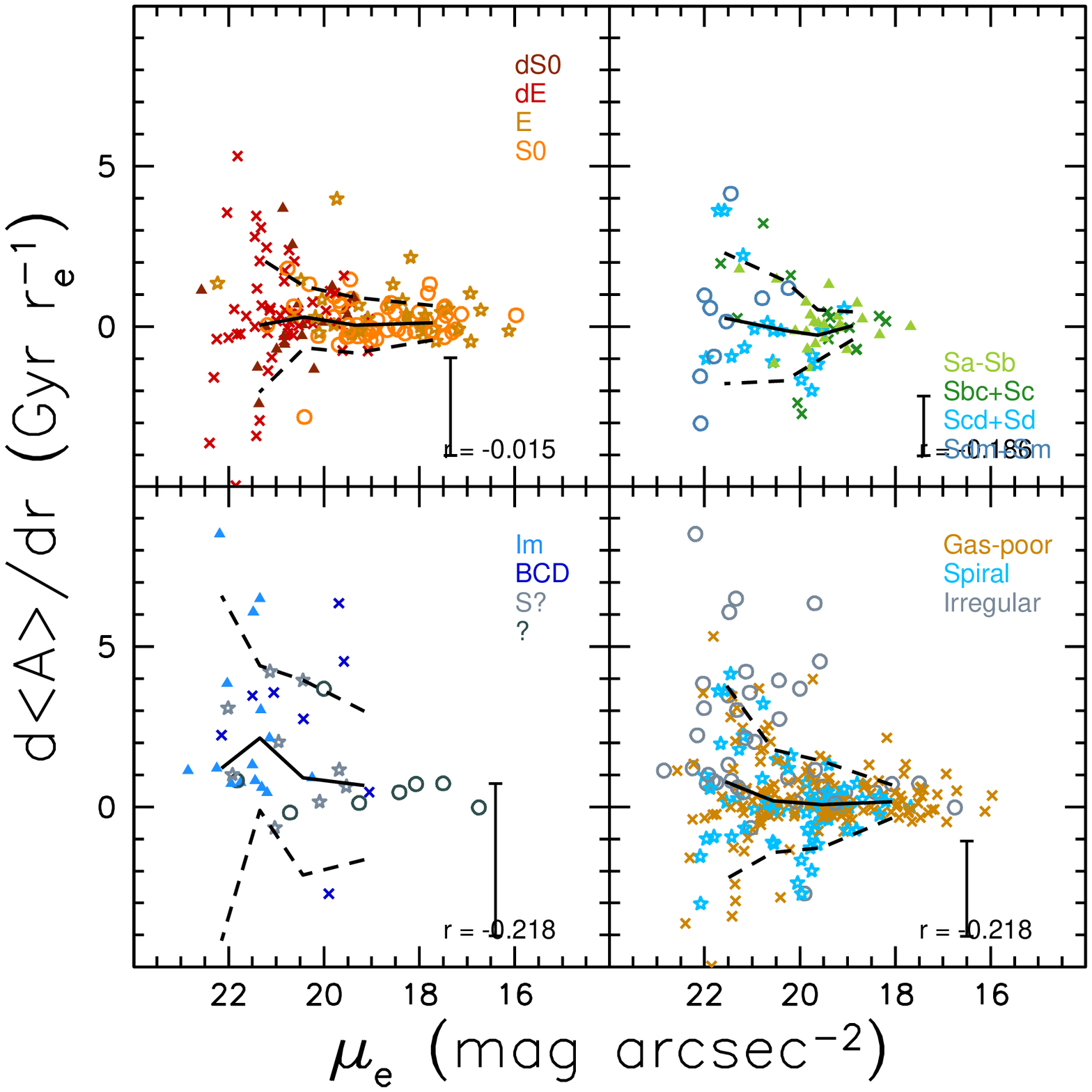} & 
   \includegraphics[width=0.45\textwidth]{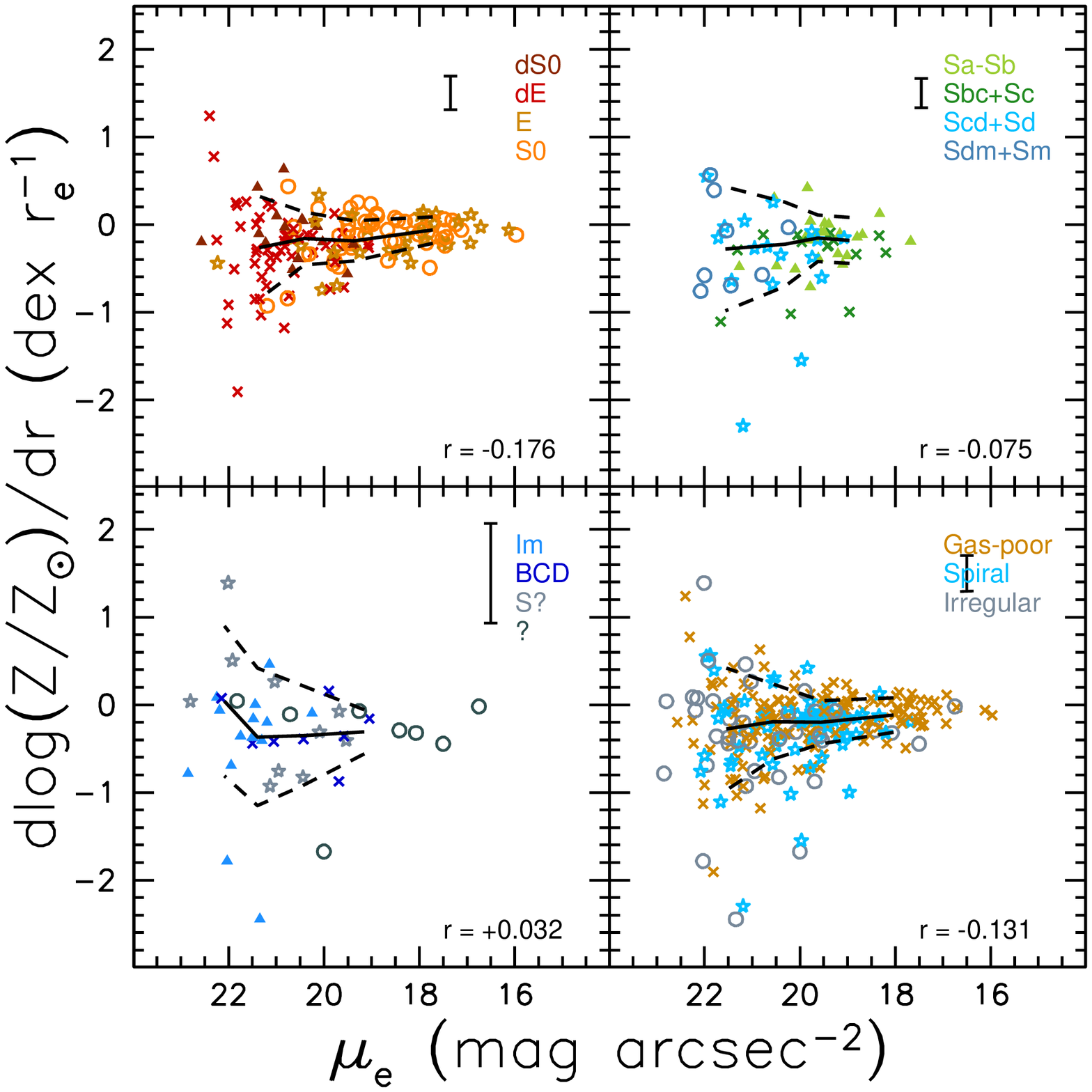} \\
   \includegraphics[width=0.45\textwidth]{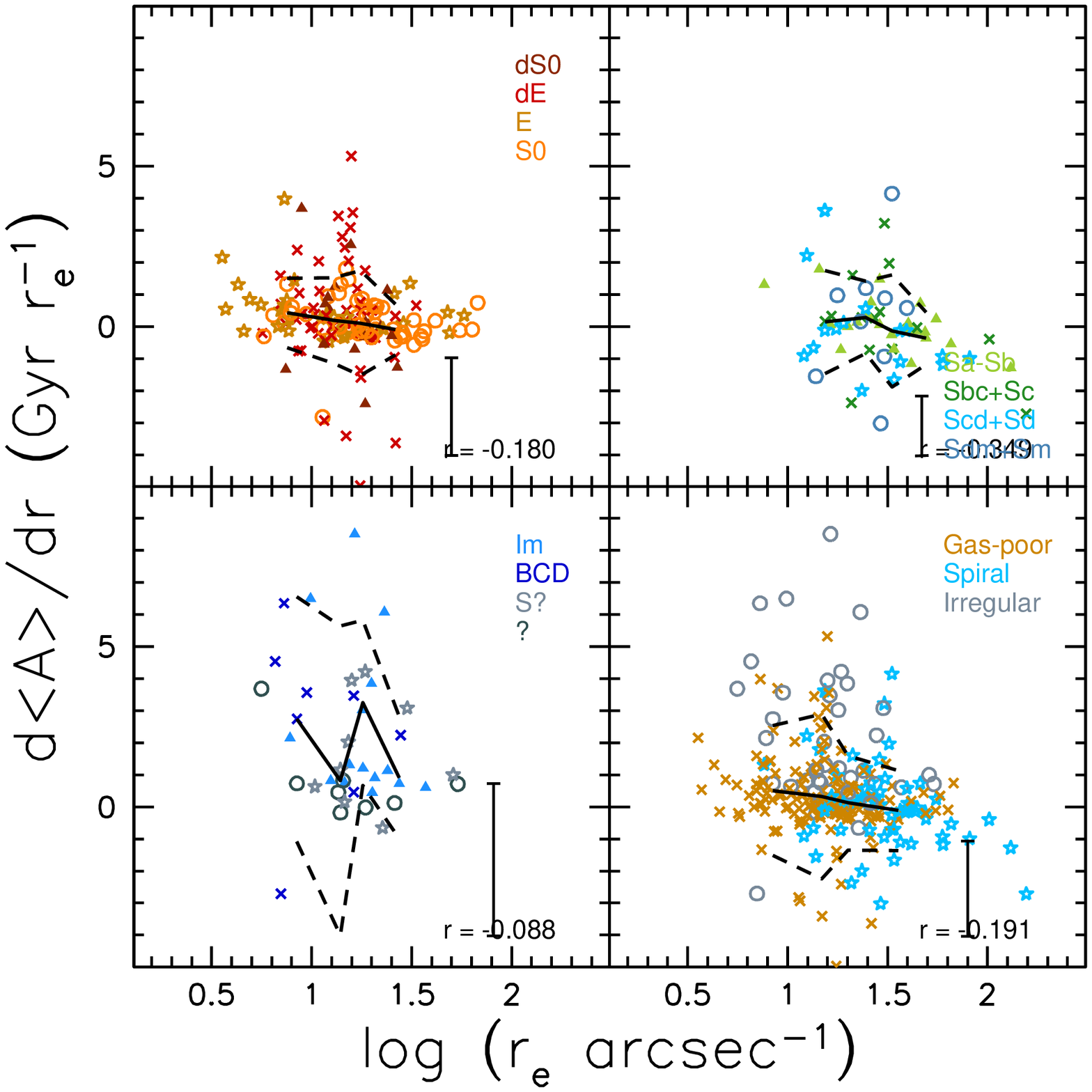} & 
   \includegraphics[width=0.45\textwidth]{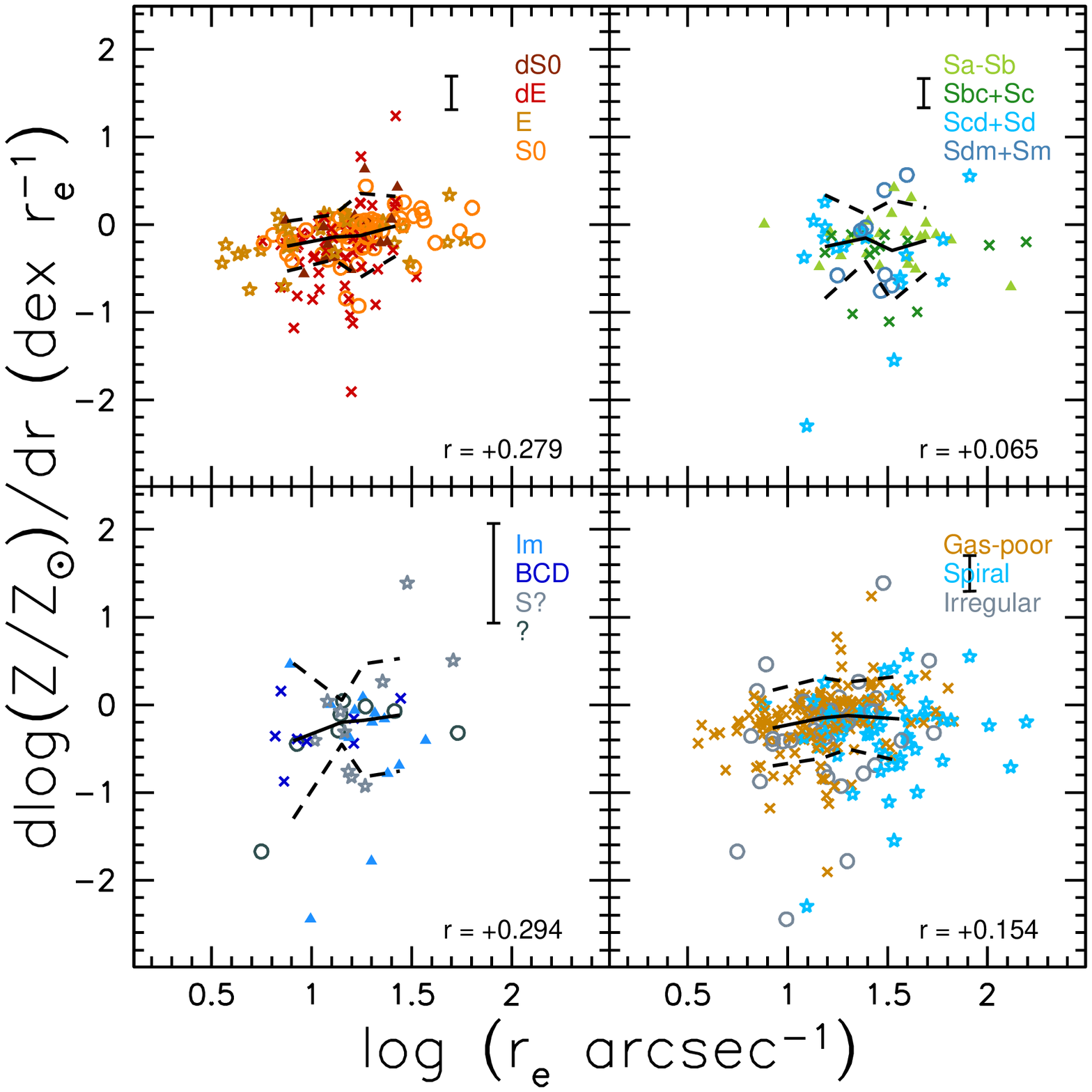}
  \end{tabular}
  \caption{As in \Fig{MAZ-C28-H} but versus $H$-band effective surface 
brightnesses (\textit{top}) and effective radii (\textit{bottom}).}
  \label{fig:MAZ-mue-re}
 \end{center}
\end{figure*}

\begin{figure*}
 \begin{center}
  \begin{tabular}{c c}
   \includegraphics[width=0.45\textwidth]{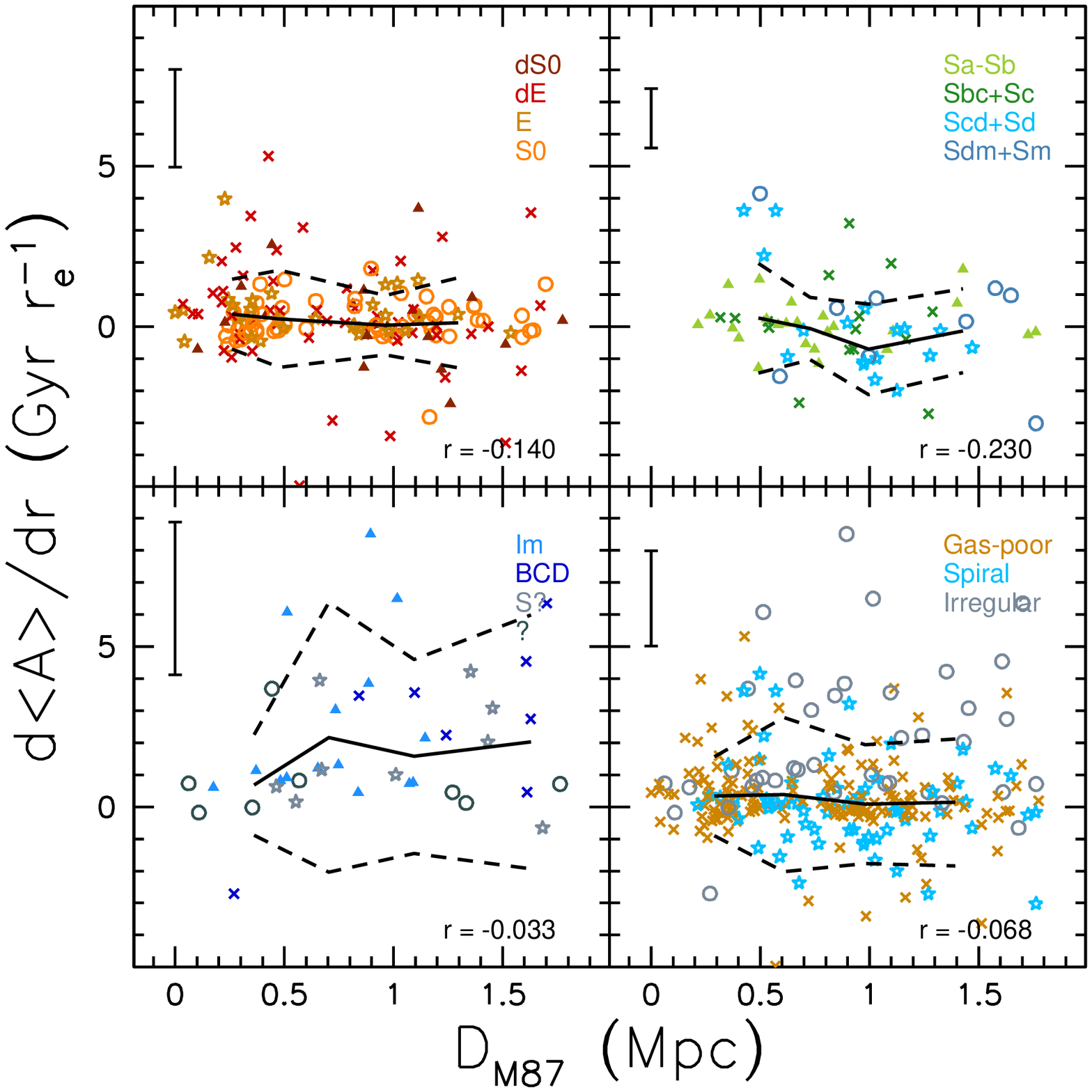} & 
   \includegraphics[width=0.45\textwidth]{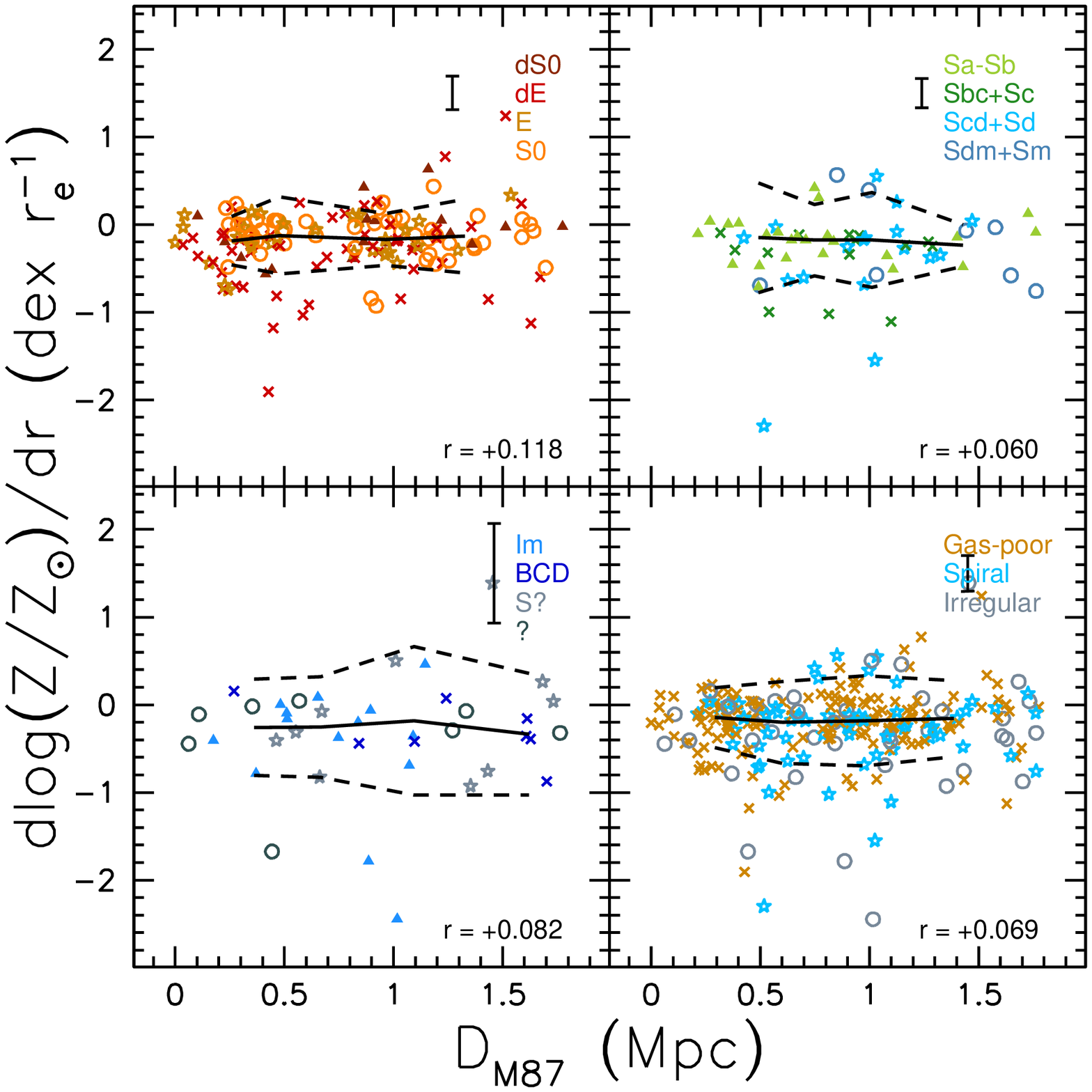} \\
   \includegraphics[width=0.45\textwidth]{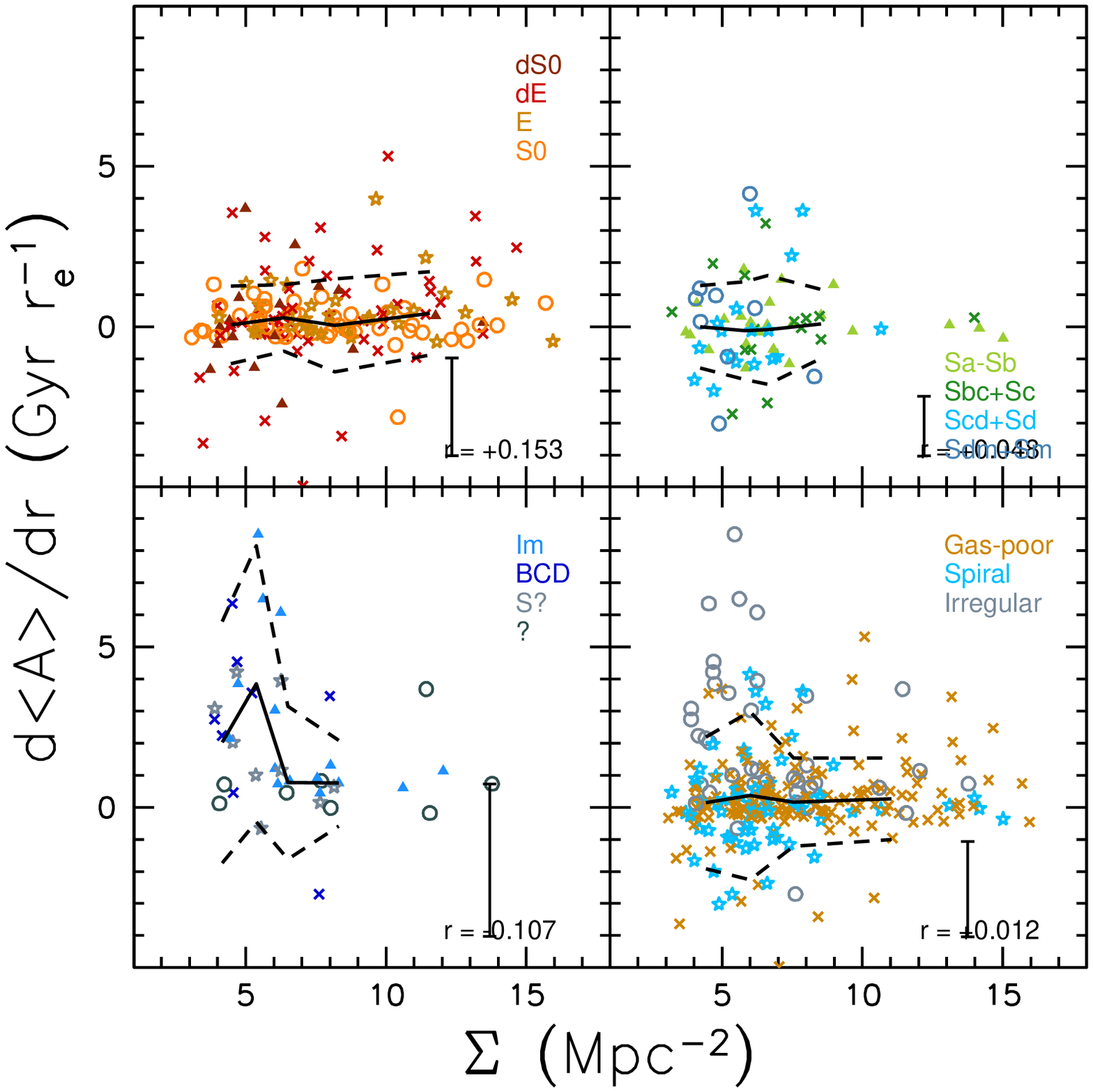} & 
   \includegraphics[width=0.45\textwidth]{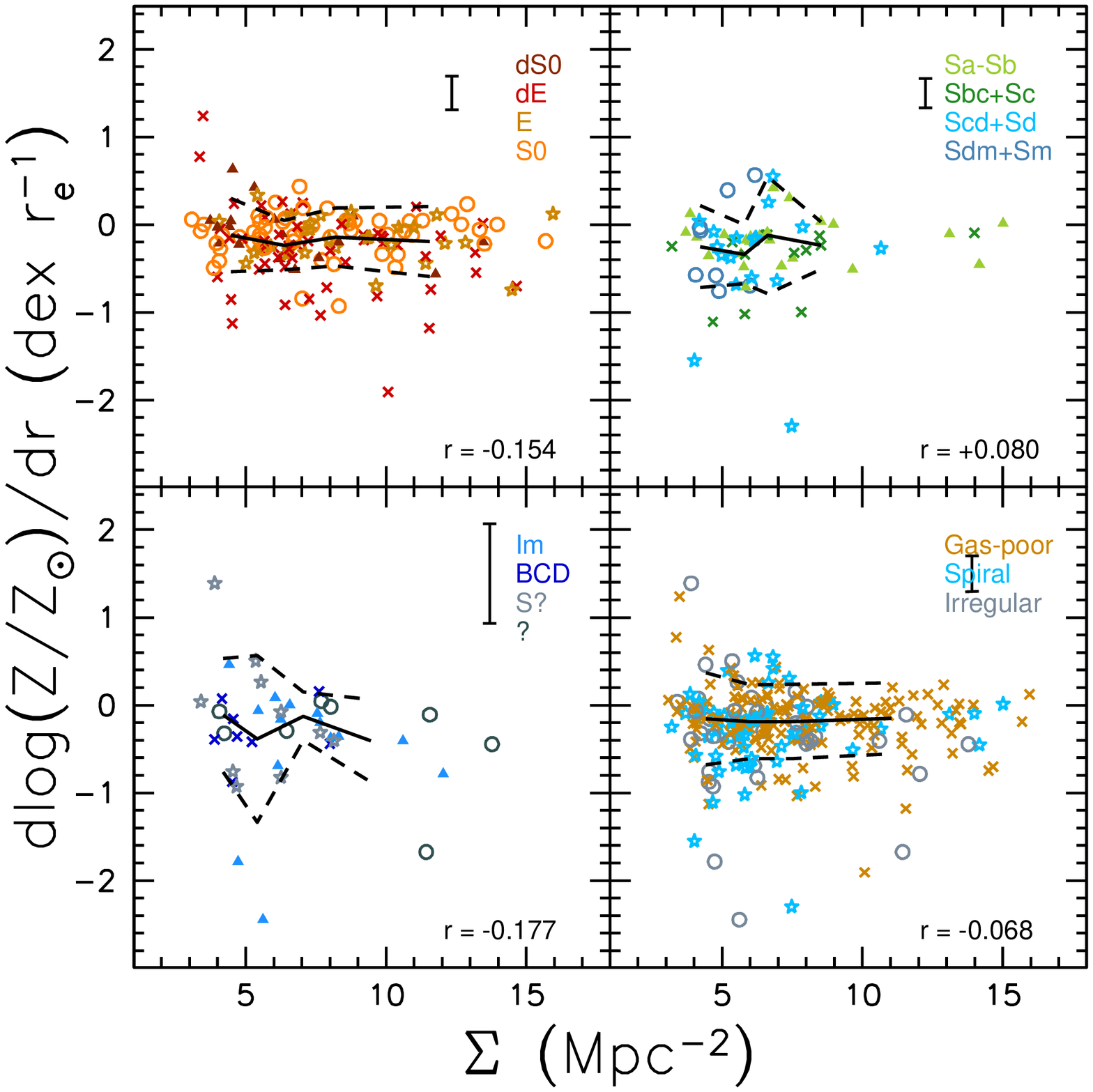} \\
   \includegraphics[width=0.45\textwidth]{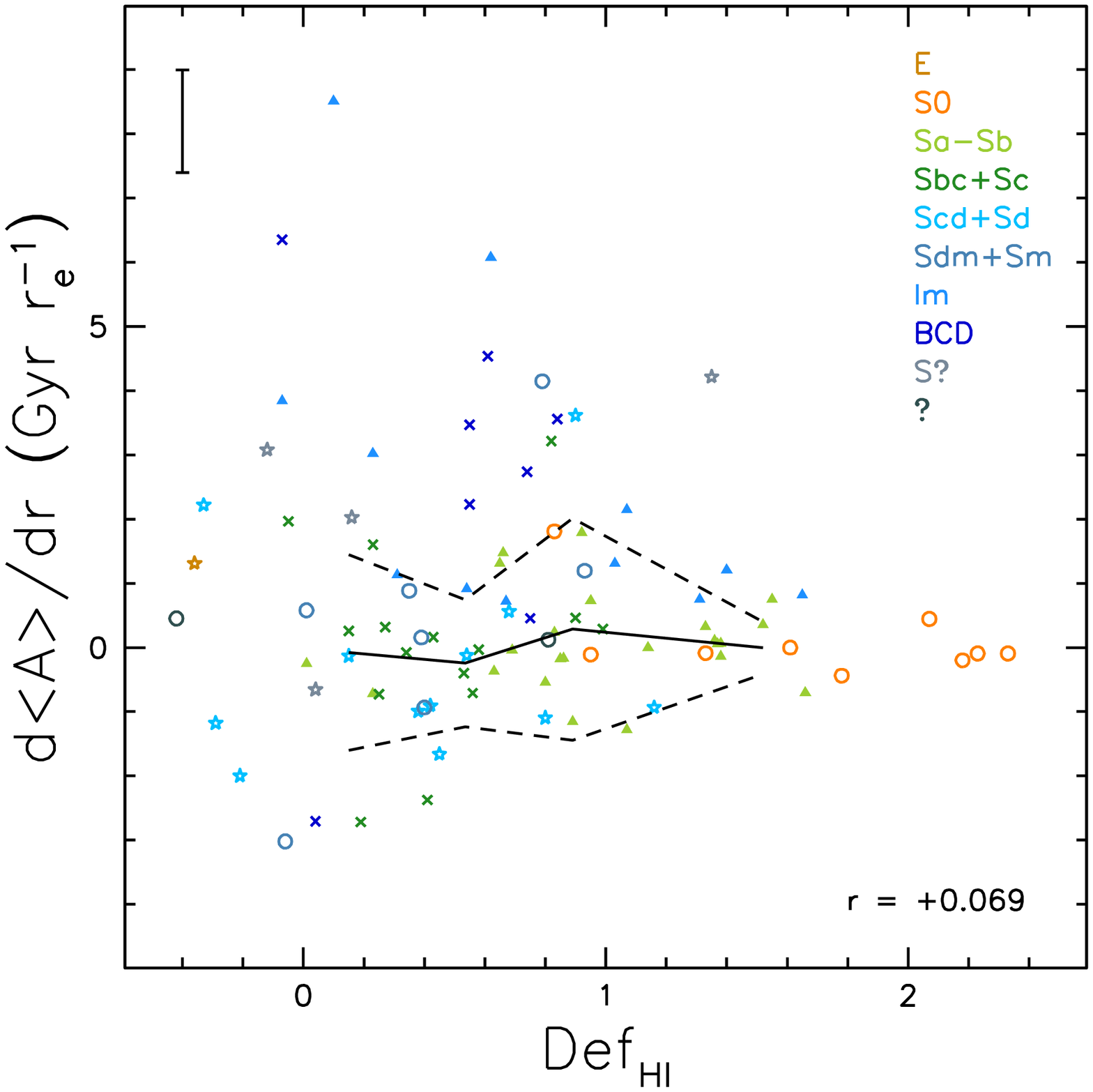} & 
   \includegraphics[width=0.45\textwidth]{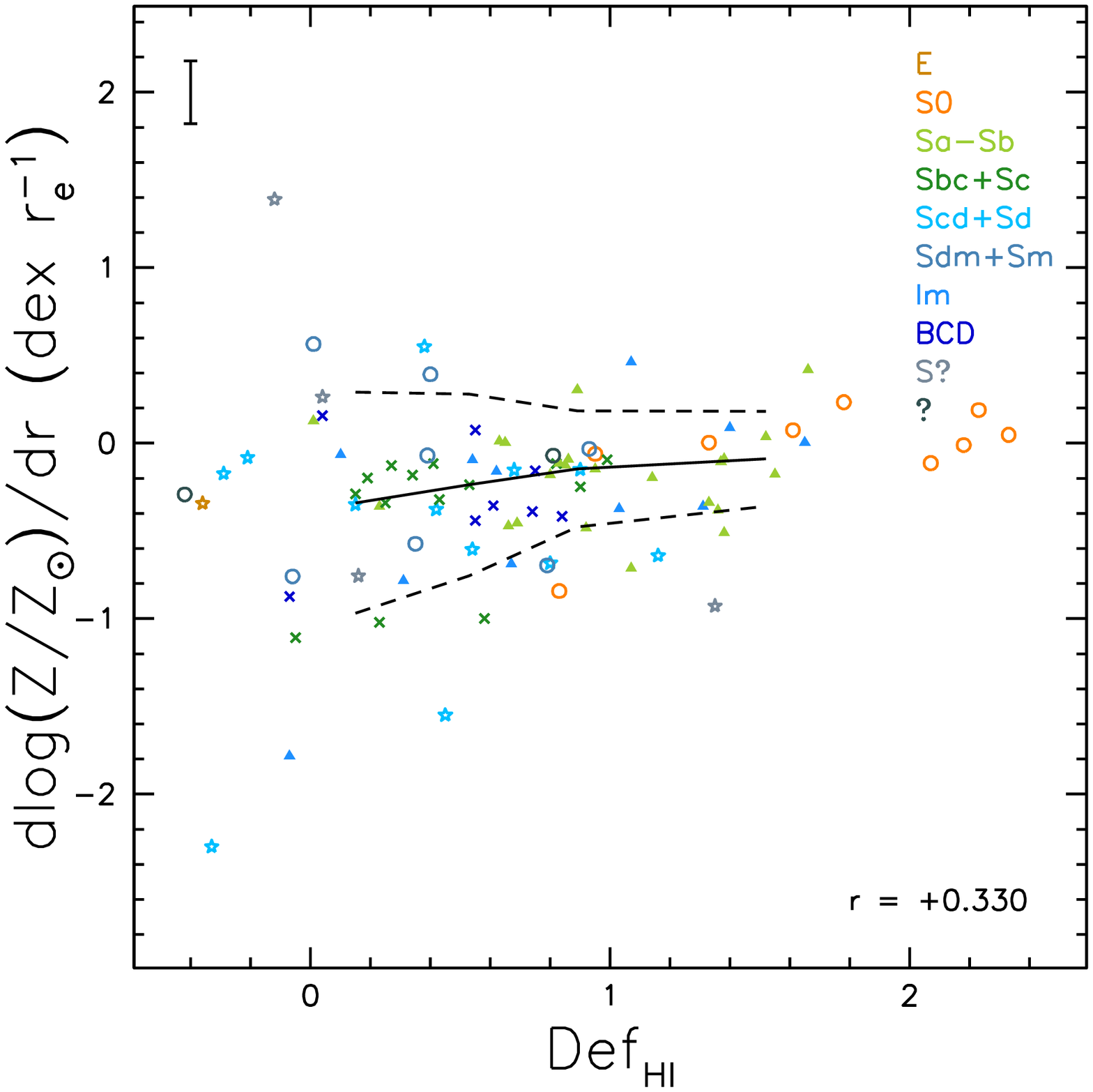}
  \end{tabular}
  \caption{As in \Fig{MAZ-C28-H} but versus cluster-centric distances 
(\textit{top}), galaxy surface densities (\textit{middle}), and \ion{H}{I} gas 
deficiencies (\textit{bottom}).}
  \label{fig:MAZ-Enviro}
 \end{center}
\end{figure*}

We can also use stellar population gradients of Virgo galaxies to investigate 
the possible role of galaxy structure and environment in setting radial 
variations in their SFHs and chemical evolution. As probes of structure, we 
have used the NIR parameters of concentration ($C_{28}$), apparent magnitude 
($m_H$), effective surface brightness ($\mu_e$) and effective radius ($r_e$). 
For correlations with environment, we have used (projected) cluster-centric 
distance ($D_{M87}$), local galaxy density ($\Sigma$), and \ion{H}{I} gas 
deficiency ($Def_{\hi}$). The definitions and sources of these parameters are 
provided in Paper I. The age and metallicity gradients of Virgo galaxies are 
plotted in Figures~\ref{fig:MAZ-C28-H}-\ref{fig:MAZ-Enviro} against the above 
parameters. In each figure, the galaxies are separated into basic groups 
corresponding to gas-poor (top-left), spiral (top-right), and irregular 
(lower-left) types, while the distributions of all three groups are 
superimposed at lower-right. For each group we show the median trend and its 
rms dispersion (solid and dashed lines, respectively), as well as the typical 
error per point and the Pearson correlation coefficient, $r$. These figures 
highlight that both structure and environment play little role in the stellar 
population gradients of Virgo galaxies.

\begin{figure*}
 \begin{center}
  \begin{tabular}{c c}
   \includegraphics[width=0.45\textwidth]{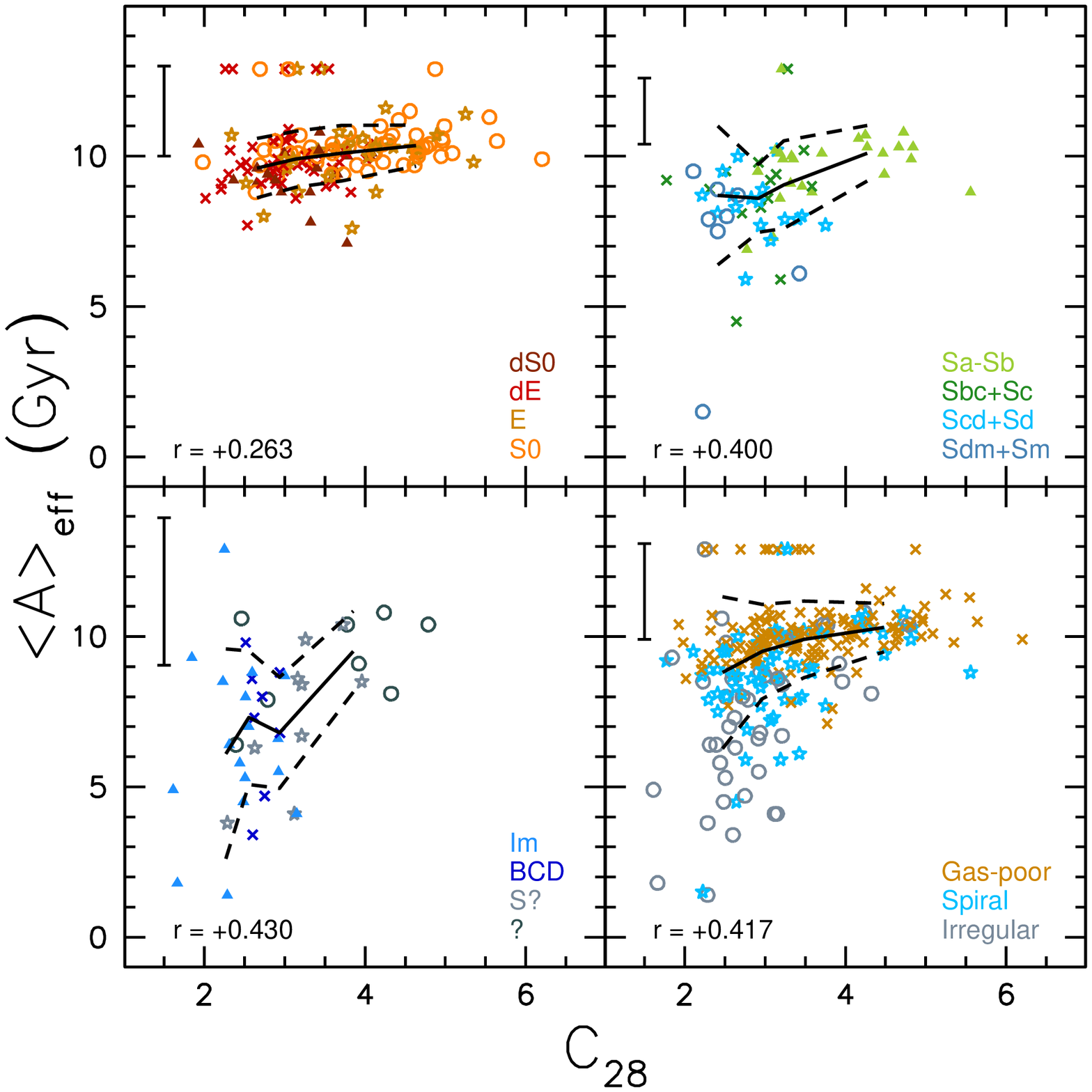} & 
   \includegraphics[width=0.45\textwidth]{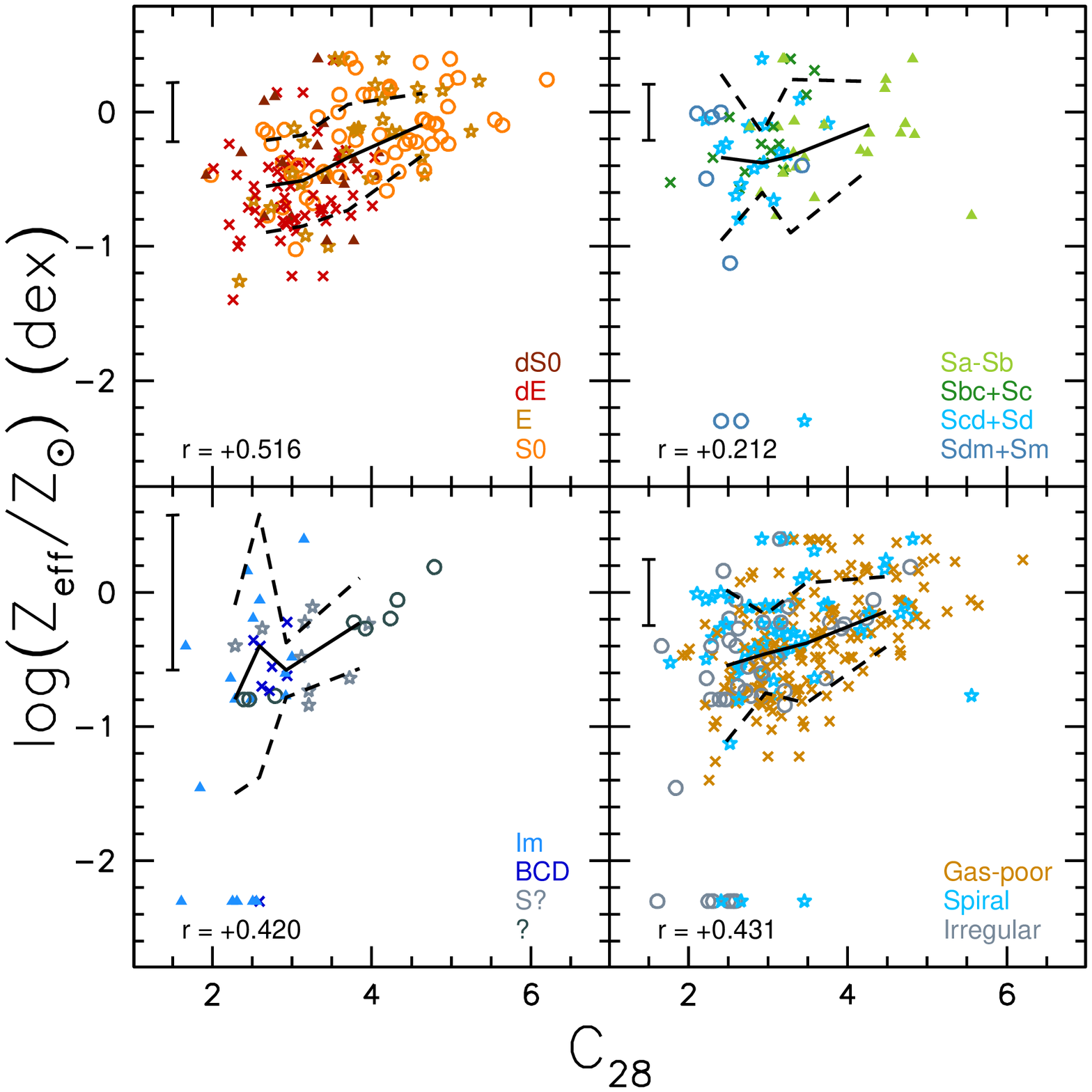} \\
   \includegraphics[width=0.45\textwidth]{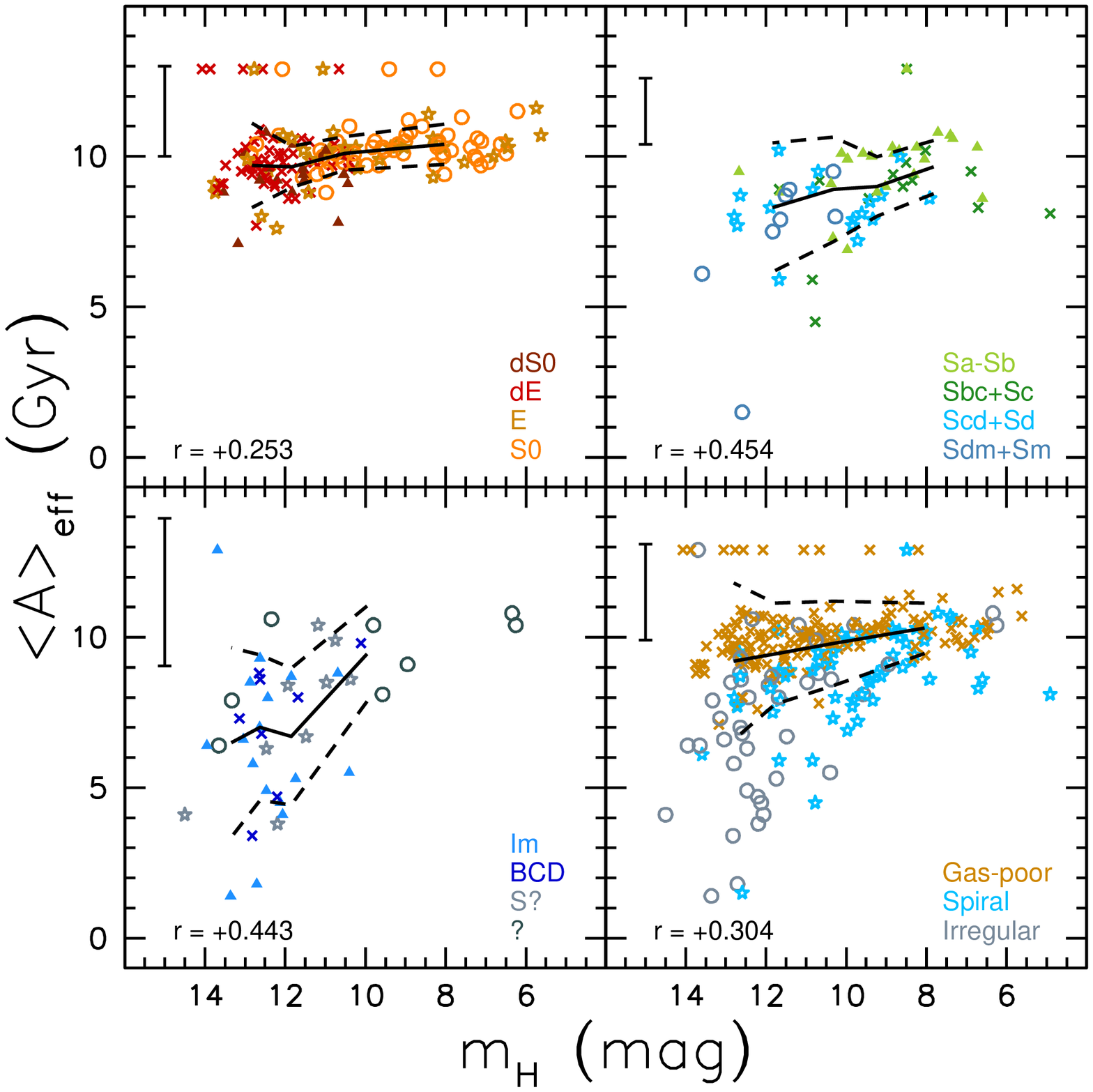} & 
   \includegraphics[width=0.45\textwidth]{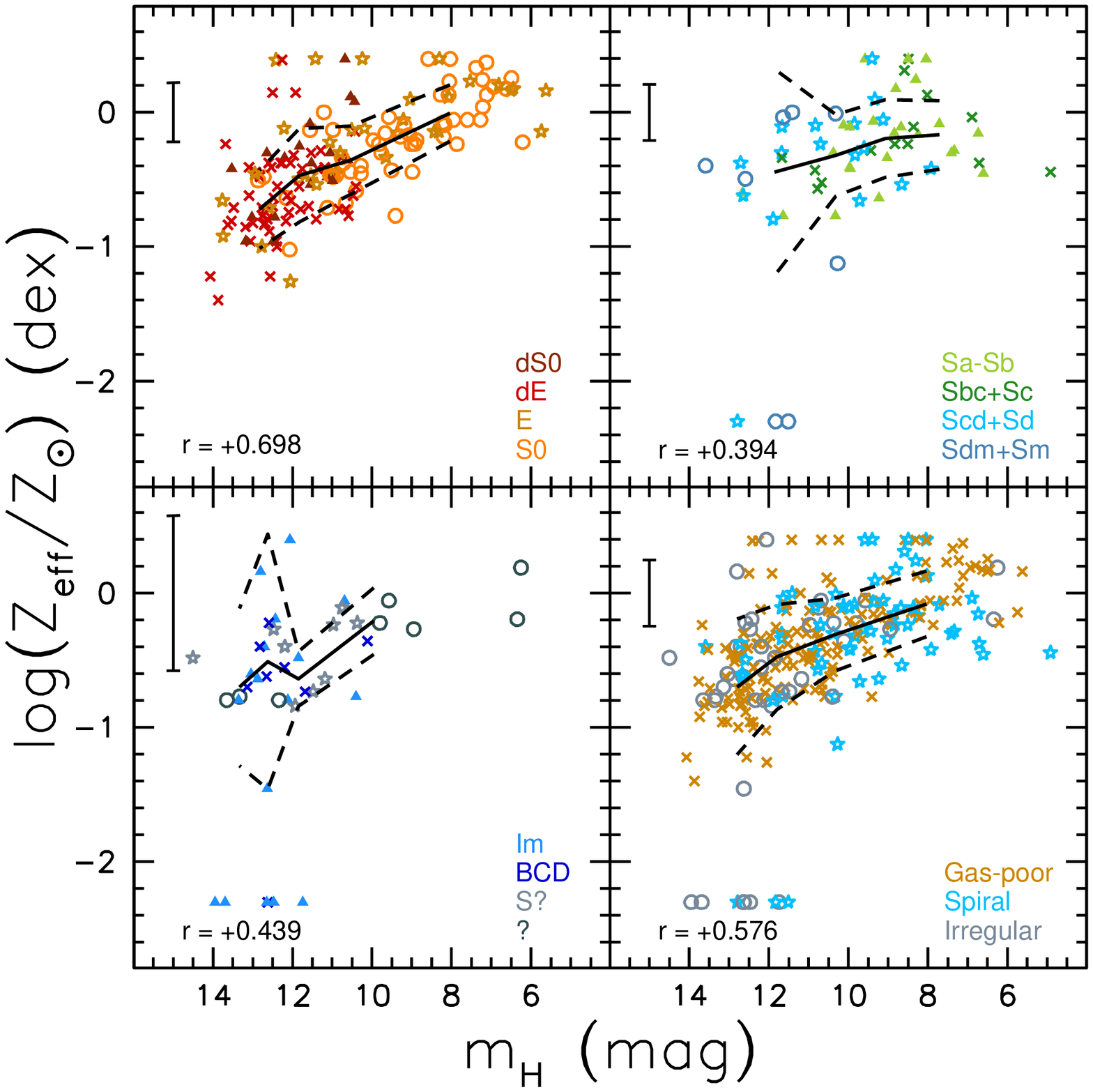}
  \end{tabular}
  \caption{As in \Fig{MAZ-C28-H} but for the (\textit{left}) 
mean ages and (\textit{right}) metallicities of Virgo galaxies
measured at $r_e$.}
  \label{fig:AZ-C28-H}
 \end{center}
\end{figure*}

\begin{figure*}
 \begin{center}
  \begin{tabular}{c c}
   \includegraphics[width=0.45\textwidth]{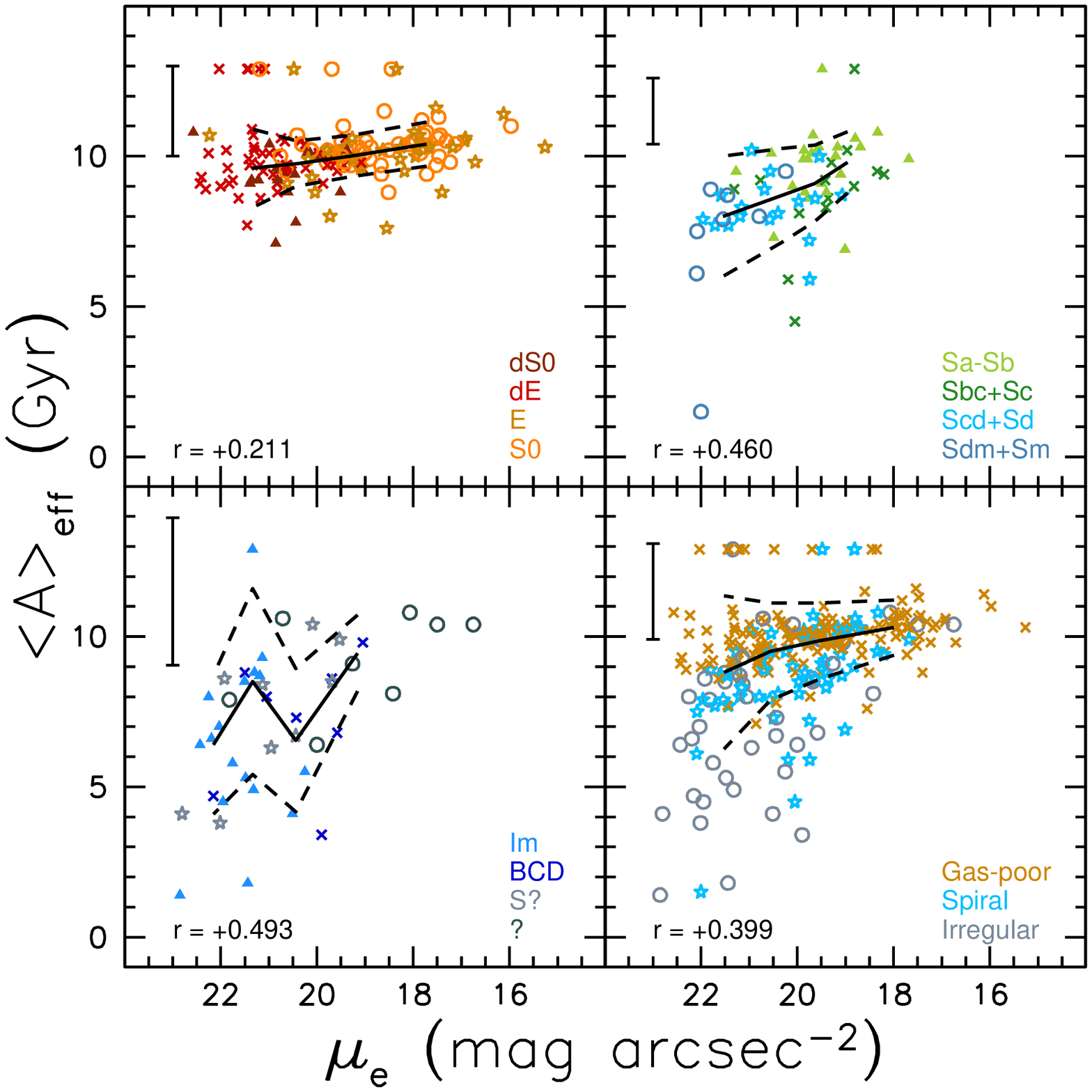} & 
   \includegraphics[width=0.45\textwidth]{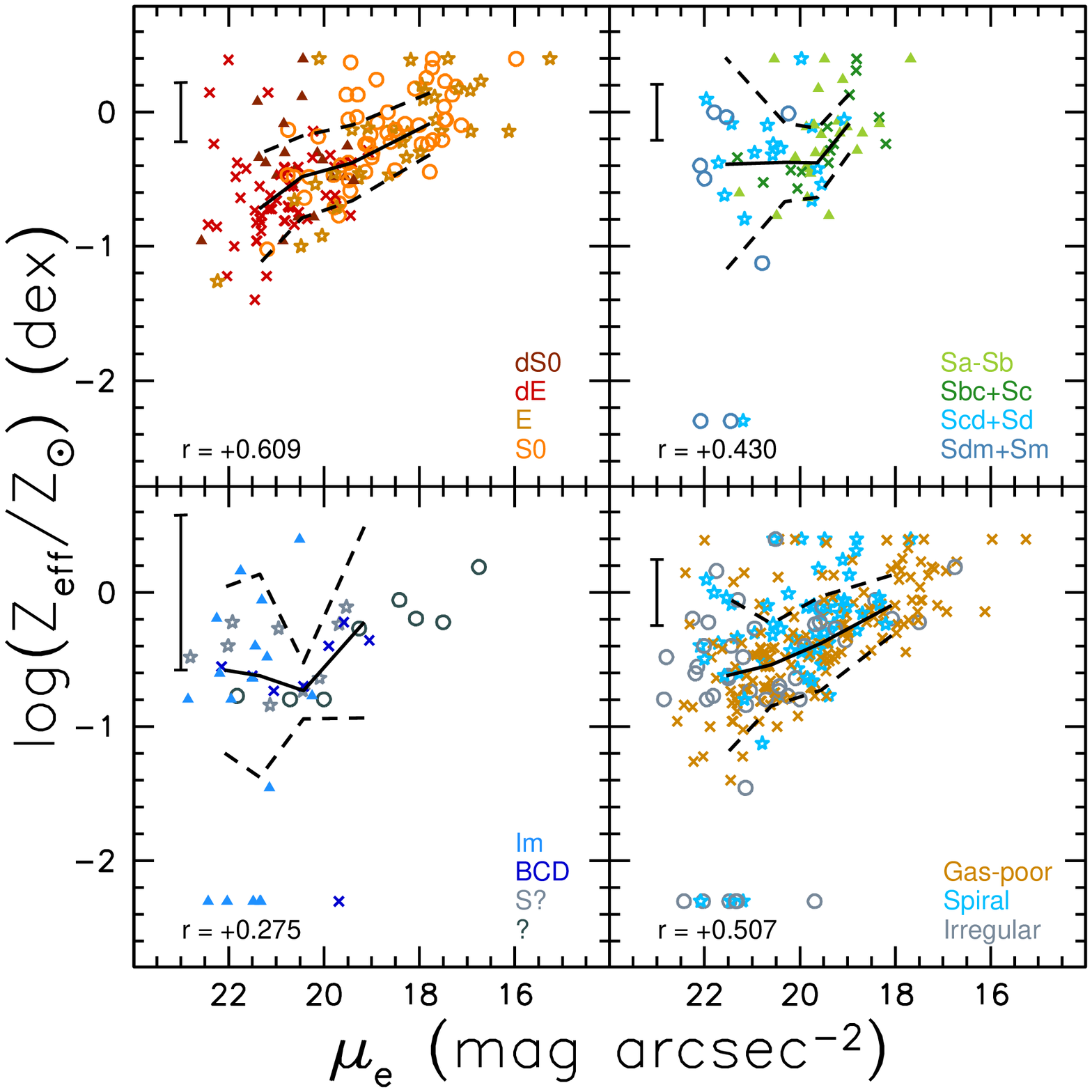} \\
   \includegraphics[width=0.45\textwidth]{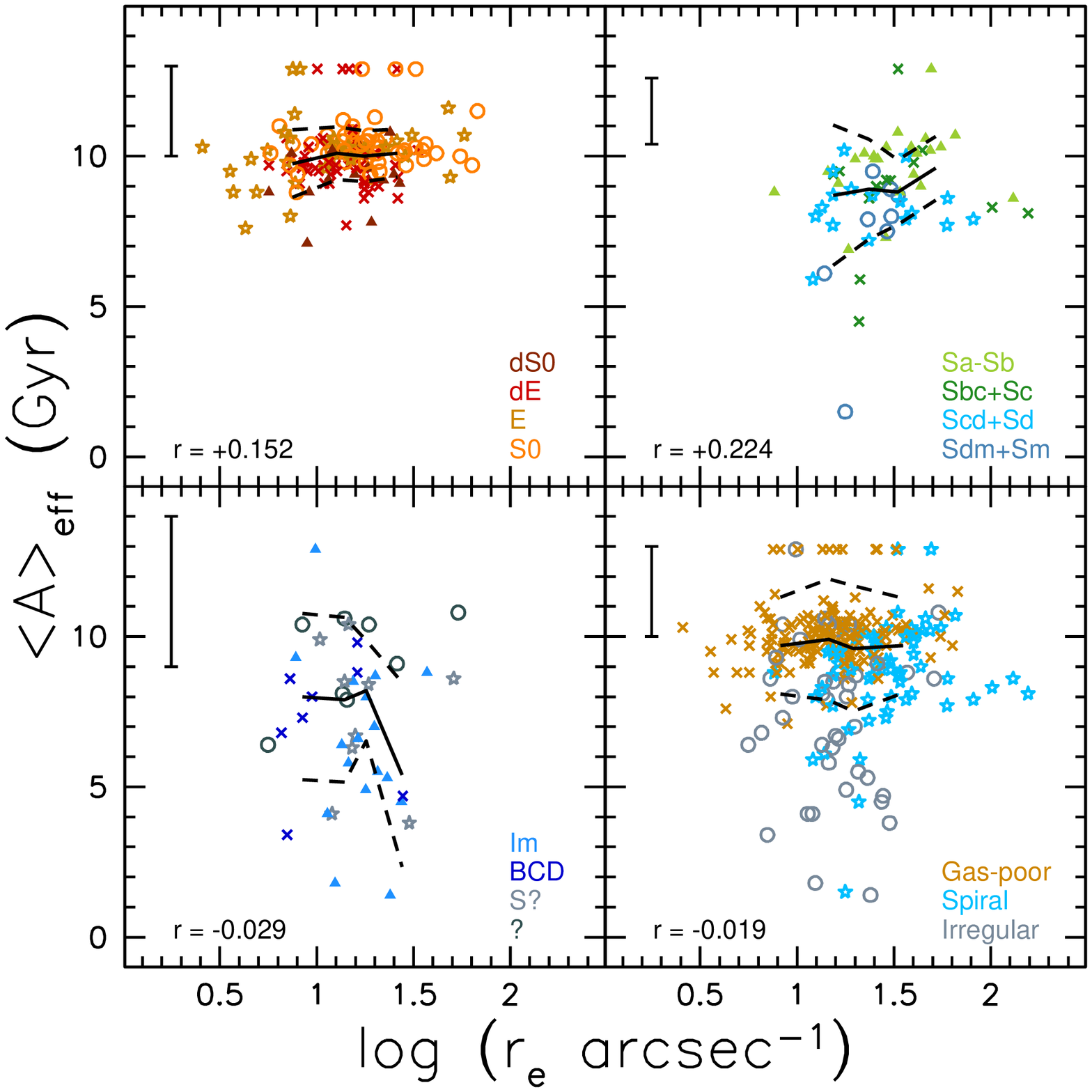} & 
   \includegraphics[width=0.45\textwidth]{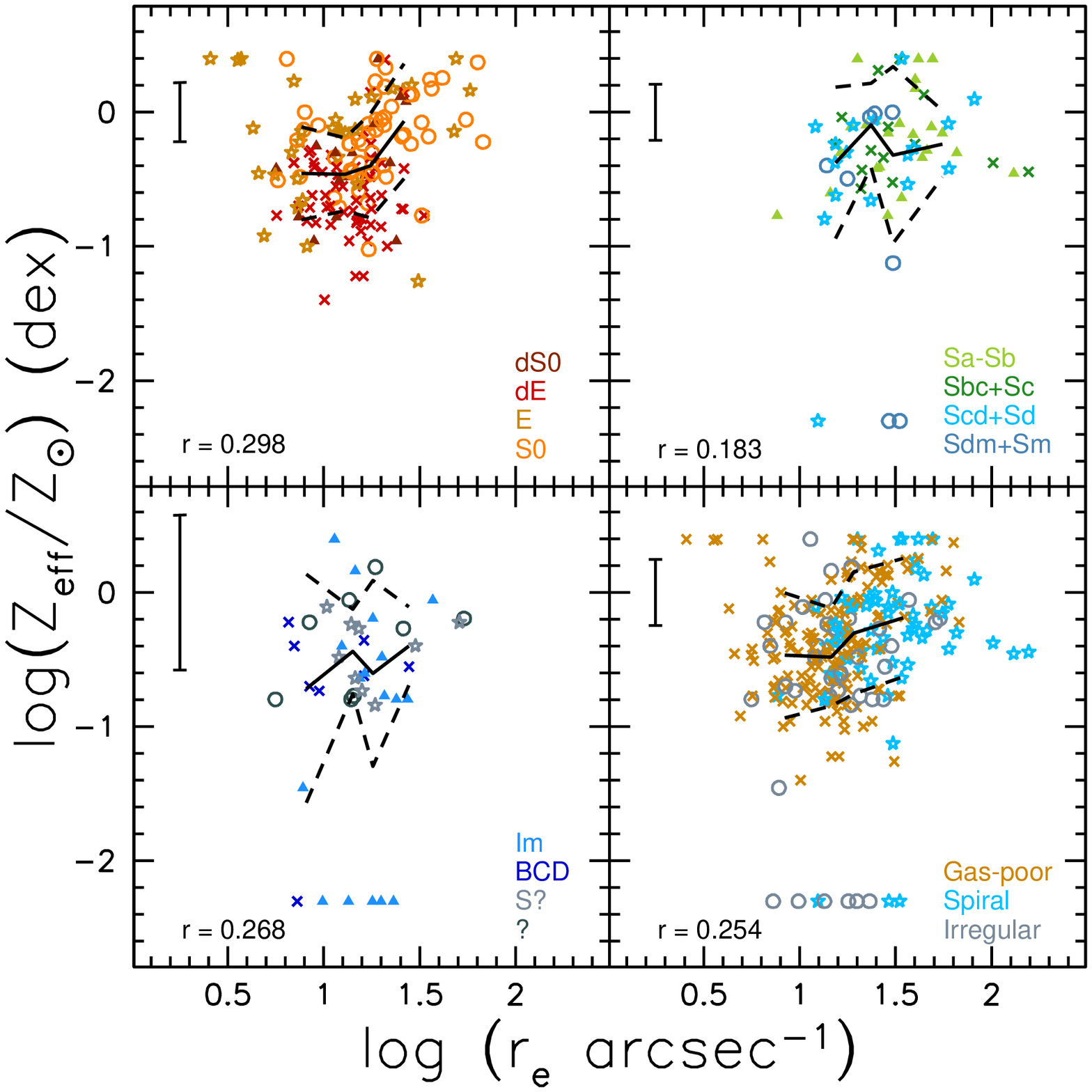}
  \end{tabular}
  \caption{As in \Fig{AZ-C28-H} but versus $H$-band effective surface 
brightnesses (\textit{top}) and effective radii (\textit{bottom}).}
  \label{fig:AZ-mue-re}
 \end{center}
\end{figure*}

\begin{figure*}
 \begin{center}
  \begin{tabular}{c c}
   \includegraphics[width=0.45\textwidth]{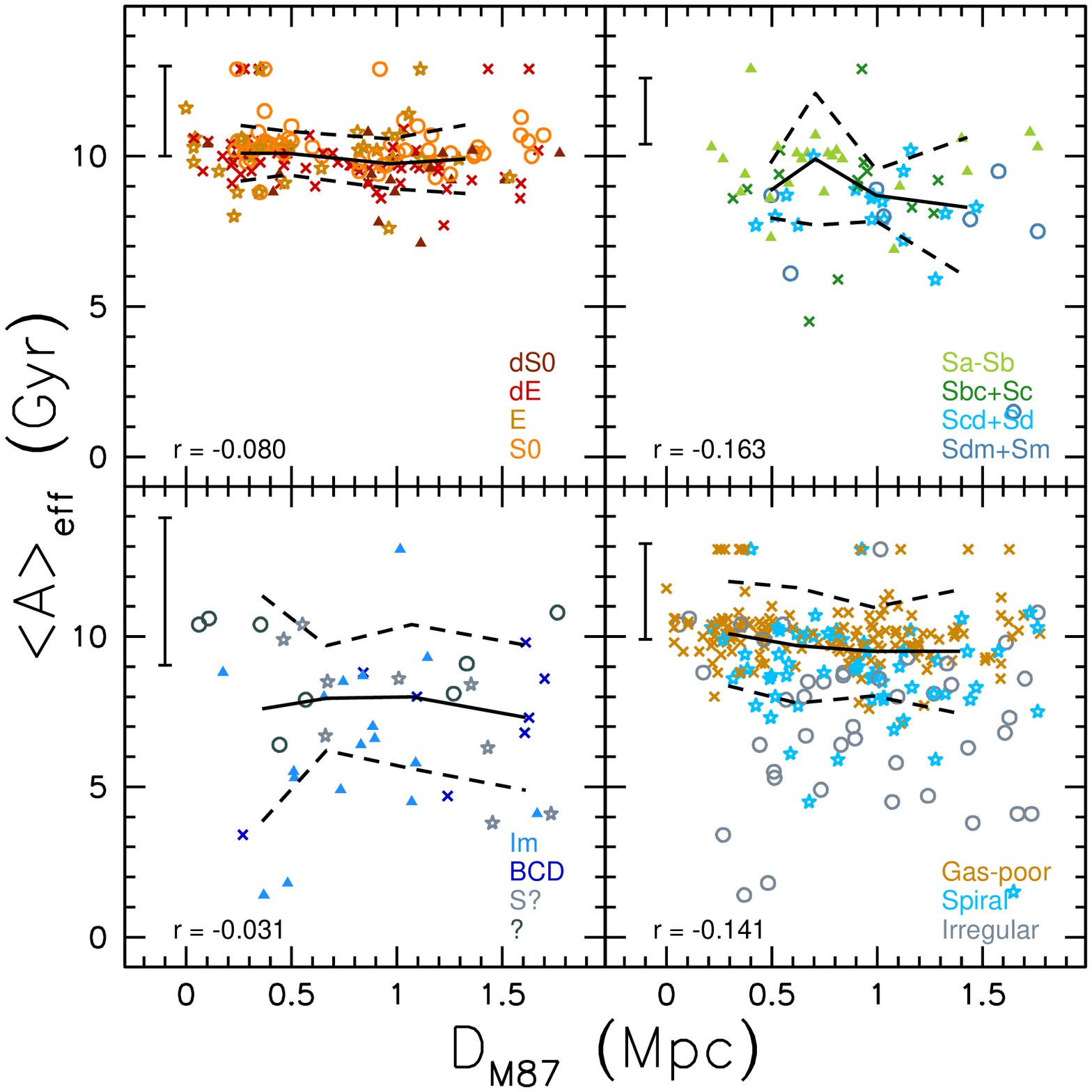} & 
   \includegraphics[width=0.45\textwidth]{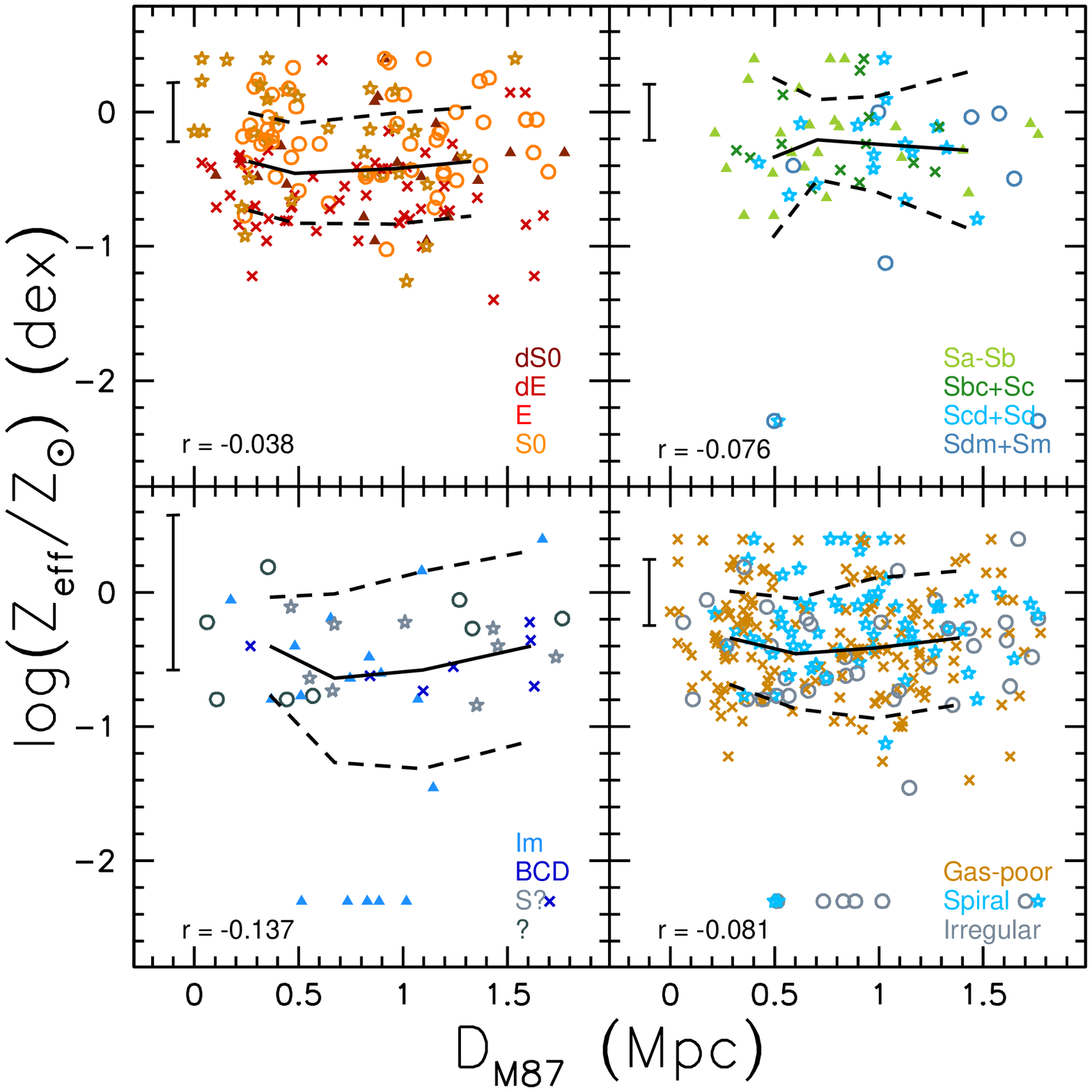} \\
   \includegraphics[width=0.45\textwidth]{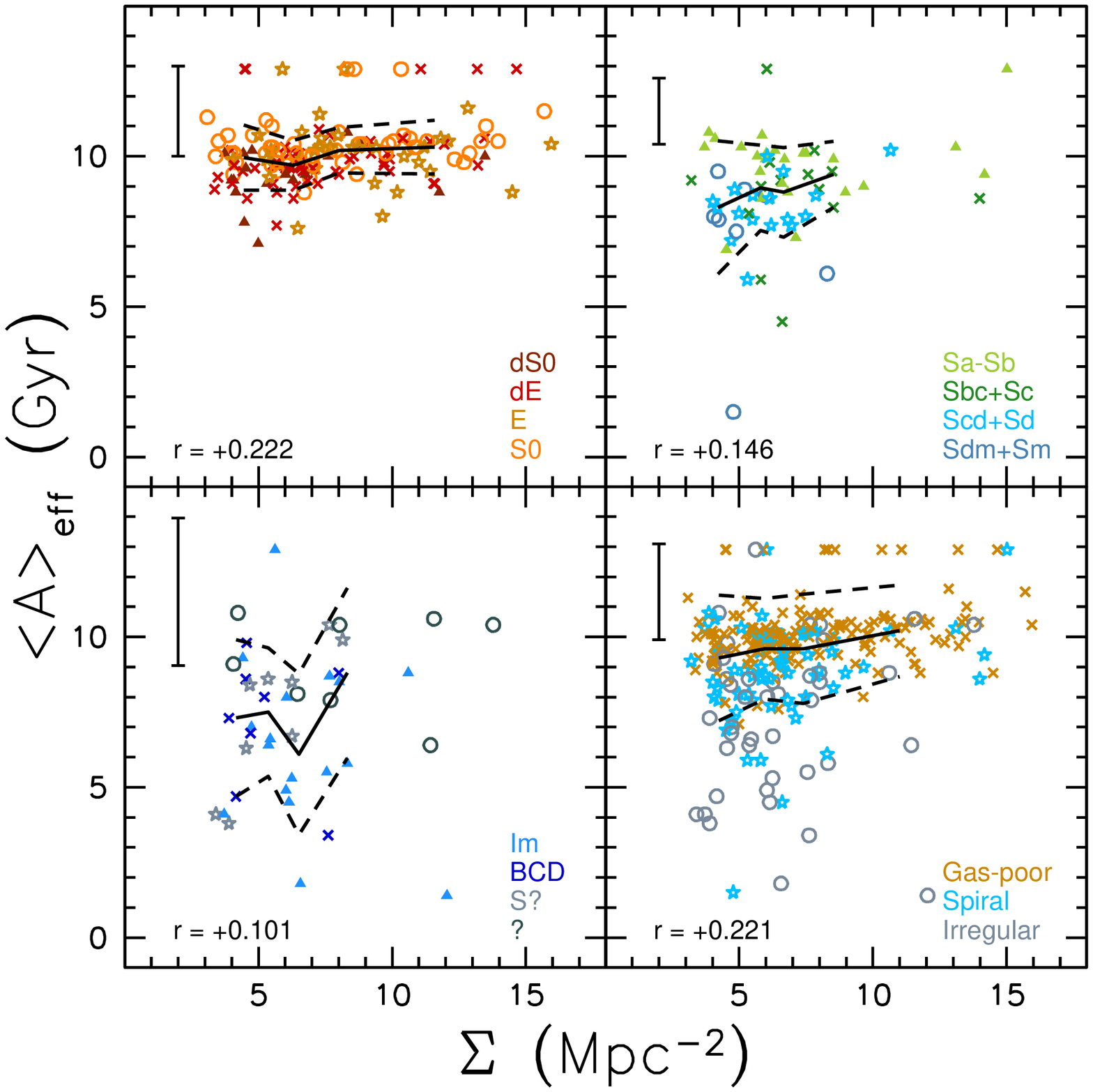} & 
   \includegraphics[width=0.45\textwidth]{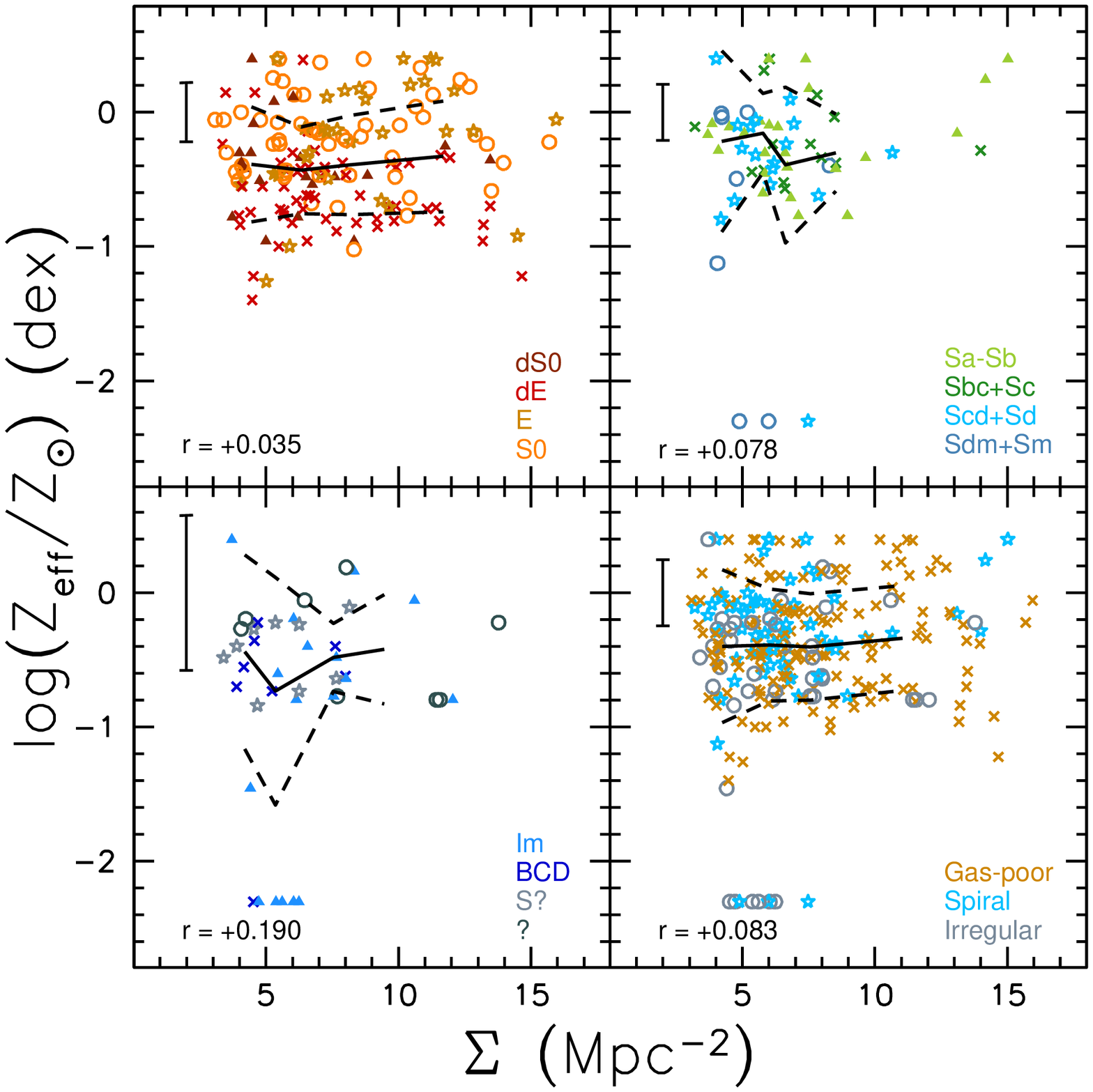} \\
   \includegraphics[width=0.45\textwidth]{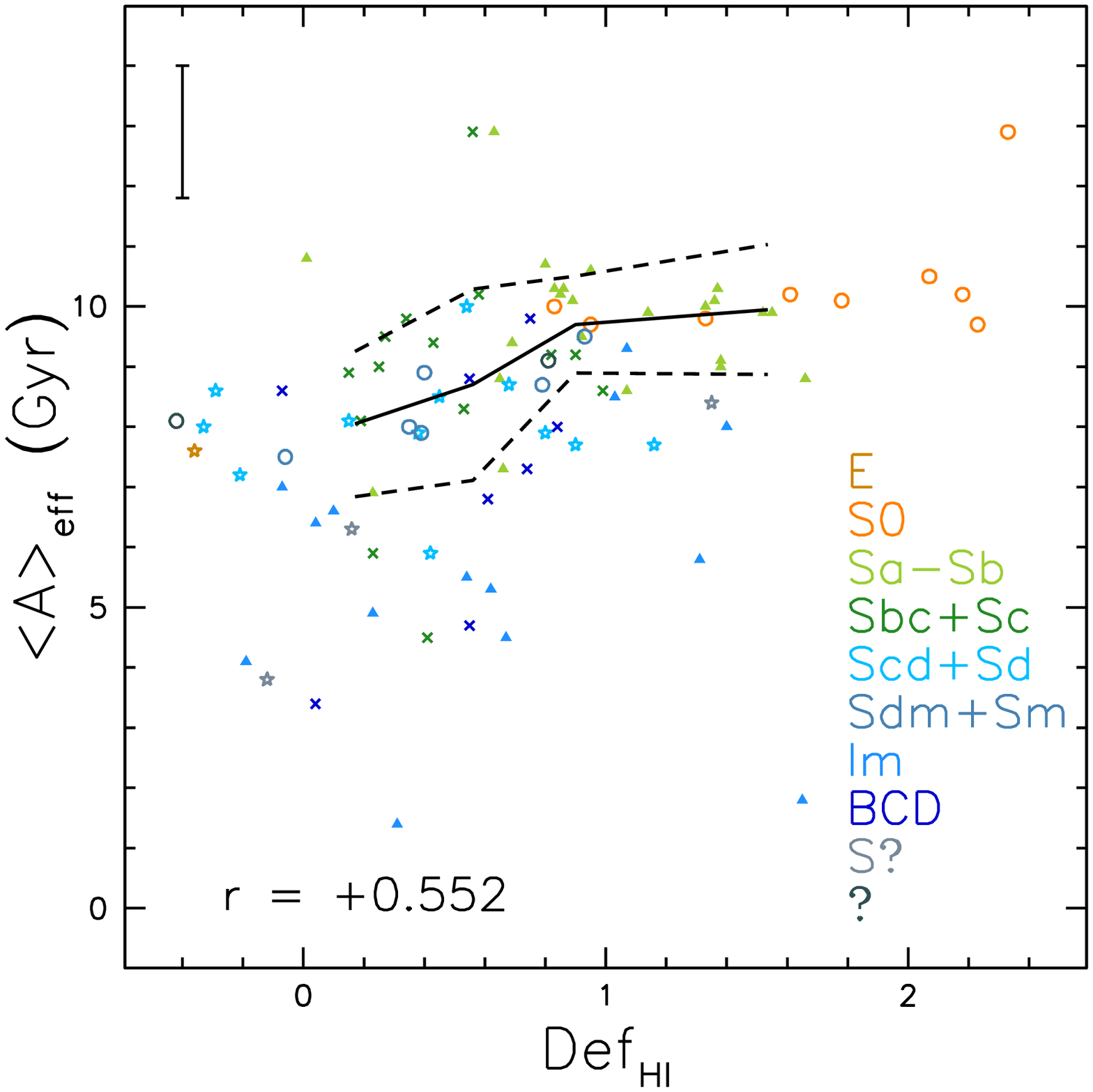} & 
   \includegraphics[width=0.45\textwidth]{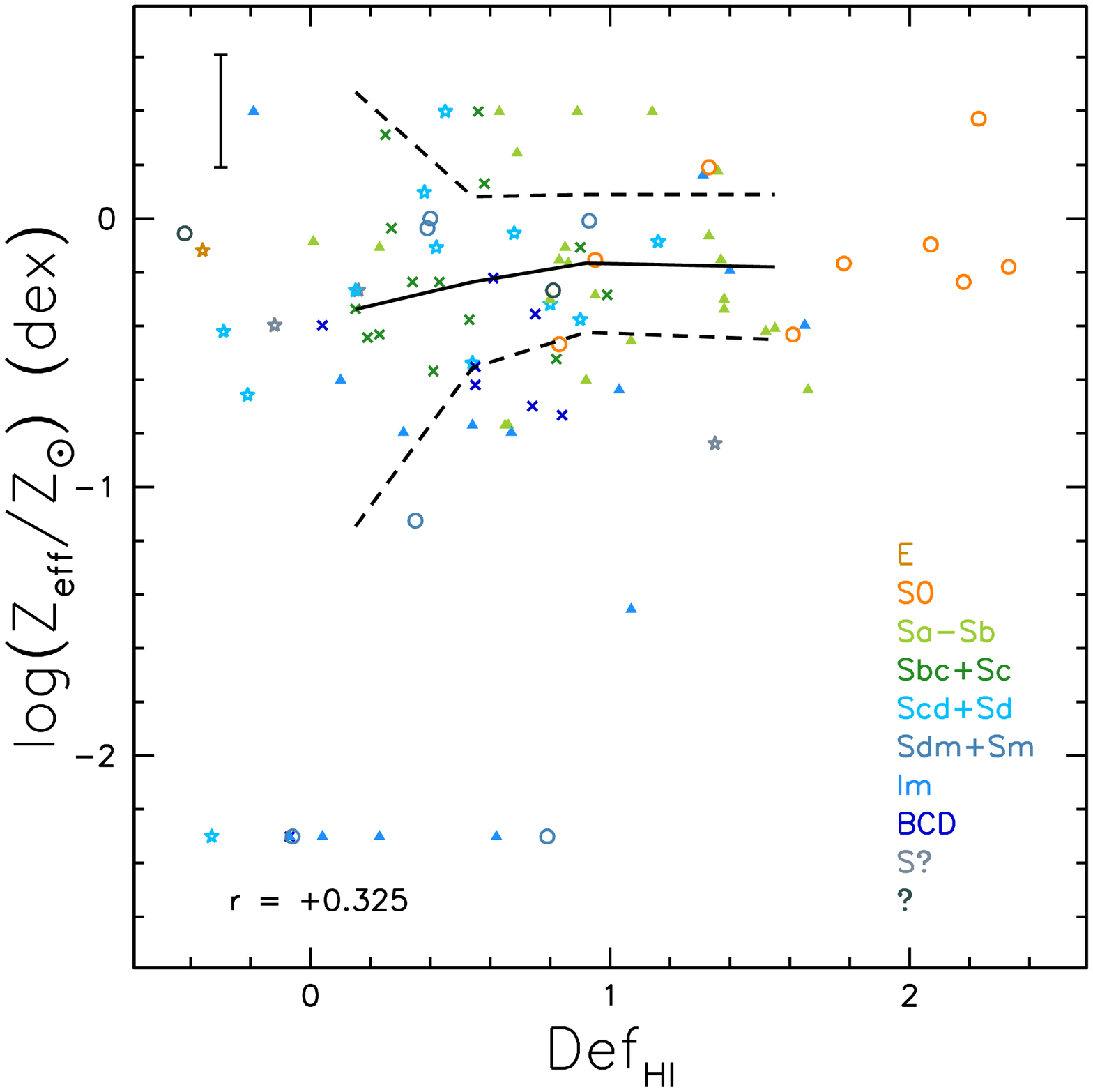}
  \end{tabular}
  \caption{As in \Fig{MAZ-C28-H} but versus cluster-centric distances 
(\textit{top}), galaxy surface densities (\textit{middle}), and \ion{H}{I} gas 
deficiencies (\textit{bottom}).}
  \label{fig:AZ-Enviro}
 \end{center}
\end{figure*}

A similar set of plots, but in terms of the effective ages and metallicities, 
of Virgo galaxies, is presented in 
Figures~\ref{fig:AZ-C28-H}-\ref{fig:AZ-Enviro}. As a tracer of local SFH and 
chemical evolution, we define the effective age, 
$\left\langle A\right\rangle_{\text{eff}}$, and metallicity, $Z_{\text{eff}}$, of a 
galaxy to be the value of these quantities at its effective radius. We now 
discuss the scaling relations of the above stellar population diagnostics for 
each galaxy group.

\subsubsection{Gas-poor galaxies}\label{Sec:SPSR-GP}
As mentioned above, we do not detect any structural or environmental trends in 
the stellar population gradients of Virgo gas-poor galaxies. Chief amongst 
these absentee trends is that the metallicity gradients of Virgo dE/dS0's are 
independent of luminosity, since \cite{Sp09} found a linear relation between 
these parameters for systems with masses $\lesssim$ 3.5$\times$10$^{10}$ 
M$_{\odot}$ \citep[$m_H \gtrsim$ 9.5 mag at the distance of Virgo;][]{Ga96}. 
However, the relatively small number of low-mass galaxies in those authors' 
sample (13) undermines the reliability of their result. Indeed, the existence 
of this correlation is far from settled in the literature 
\citep{dP05,Mi05,Og05,LB10,To10a}. For instance, with a slightly larger sample, 
\cite{Kol09} showed that gas-poor dwarfs are highly scattered in the 
metallicity gradient--velocity dispersion (--luminosity) plane, not unlike what 
we find for the Virgo dE/dS0's. The observed scatter in metallicity gradient 
for Virgo dE/dS0's with $m_H >$ 11.0 mag ($>$0.4 dex) cannot be explained by 
observational errors ($\lesssim$0.2 dex).

The ages of Virgo gas-poor galaxies also seem to be independent of their 
structure or environment (but see \se{SPSR-Sp}). The most intriguing of these 
null results are those involving the parameters $m_H$ and $\Sigma$. Many 
studies report that the central ages of E/S0's correlate with both their 
central velocity dispersion and local galaxy density 
\citep{Cal03,Ne05,Th05,Cl06,Sm07}. Further, a similar age trend exists for such 
galaxies against stellar mass indicators \citep{Gal05} (but see 
\citealt{Gr09a,Gr09b} and \citealt{Sm09}); likewise, the integrated ages of 
Virgo E's may increase with their $H$-band luminosity \citep{Ga02}. The 
absence of the above trends persists even when we consider individual 
morphologies of gas-poor galaxies. We return to this topic in \se{D-gGP-Down}. 

The metallicities of Virgo gas-poor galaxies exhibit positive correlations 
against their structural parameters of $C_{28}$, $m_H$, and $\mu_e$. The most 
statistically significant of these trends, that against $m_H$, reflects the 
known metallicity-luminosity relation of galaxies \citep{Za94} and corroborates 
a similar correlation in terms of these galaxies' integrated metallicities 
\citep{Ga02}. The steepness of our observed $Z_{\text{eff}}$-$m_H$ trend, and the 
contribution of all gas-poor types to it, suggest that the chemical enrichment 
of these galaxies depends principally on their stellar masses and supports the 
idea that the galaxian red sequence is due to metallicity variations 
\citep{Fa73,St01,Be04,Fa07}. Contrary to the above trends, the metallicities of 
Virgo gas-poor galaxies do not correlate with any of their environmental 
parameters. This result conflicts with the $Z$-$\Sigma$ trend \citep{Th05}; we 
also return to this issue in \se{D-gGP-Down}.

\subsubsection{Spirals}\label{Sec:SPSR-Sp}
None of the stellar population diagnostics for Virgo spiral galaxies exhibit 
significant correlations with their structural parameters. One worth noting is 
the null trend between age and surface brightness. However, \cite{BdJ00} and 
M04 both found a positive correlation between these quantities for field 
spirals, suggesting that the local potential regulates the SF in those 
galaxies. An obvious cause for the absence of this trend in our data is that 
our sample is drawn from a galaxy cluster, whereby environmental effects, such 
as gas stripping, could perturb the Virgo spiral SFHs (see below). Dust could 
also skew the age estimates for some of our galaxies.

The ages of Virgo spirals are most strongly affected by their environment, such 
that galaxies with a greater neutral gas deficiency tend to have older stellar 
populations. Although \cite{Ga02} showed that $Def_{\hi}$ also affects these 
galaxies' \textit{integrated} ages, this is secondary to the effect of $m_H$, 
contrary to our results. Also intriguing is the fact that we do not find a 
$Def_{\hi}$ trend amongst any of the other stellar population diagnostics for 
Virgo spirals, which would have been expected given the gas deficiency 
dependence of their local SFHs. For example, the traditional picture of gas 
stripping posits that the SF in disk outskirts should be truncated before that 
in the inner regions \citep{Vo01}, thereby changing that galaxy's age gradient. 
We return to this discrepancy in \se{Sp-GS}. It should finally be noted that 
while a significant $Def_{\hi}$-morphology correlation exists amongst Virgo 
spirals ($r$ = -0.68; Paper I), these galaxies alone do not drive the 
$Def_{\hi}$ trend observed in \Fig{AZ-Enviro}.

\subsubsection{Irregulars}\label{Sec:SPSR-Irr}
Akin to Virgo spirals, the ages of Virgo irregulars positively correlate with 
their gas deficiencies (\Fig{AZ-Enviro}), such that systems with larger 
$Def_{\hi}$ are typically older. Again, though, this trend is surprising given 
that the metallicities plus age and metallicity gradients for the Virgo 
irregulars bear no $Def_{\hi}$ imprint. Other than an age-gas deficiency trend, 
the Virgo irregulars collectively exhibit no other significant correlations in 
their stellar population diagnostics. The metallicities of Virgo peculiars 
(S?+?) though, increase with both their luminosities and surface brightnesses 
(Pearson $r$ = +0.66 and +0.48, respectively), similar to what was found for 
Virgo gas-poor galaxies.

\subsection{Integrated stellar populations}\label{Sec:ISPs}
\begin{figure*}
 \begin{center}
  \begin{tabular}{c c}
   \includegraphics[width=0.45\textwidth]{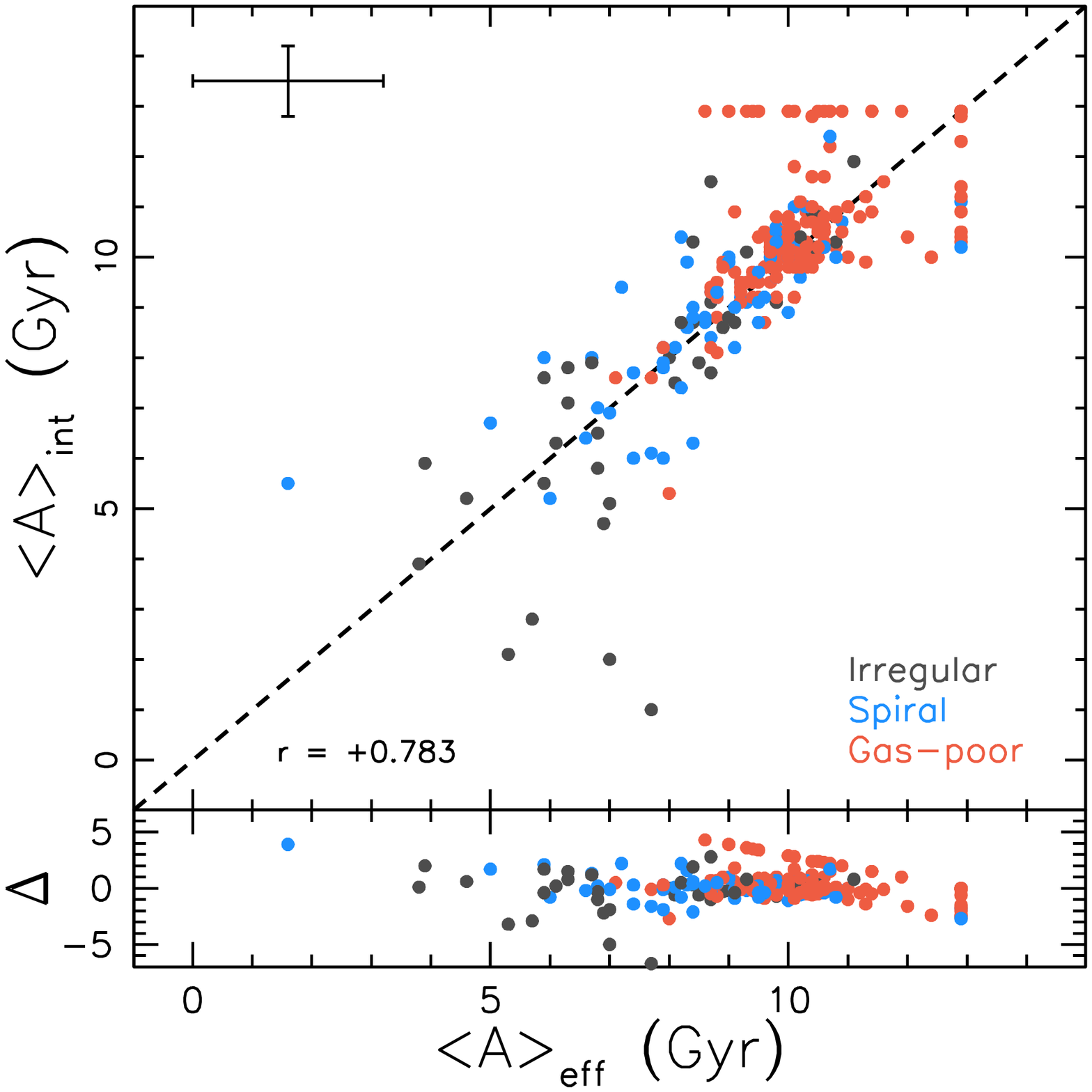} & 
   \includegraphics[width=0.45\textwidth]{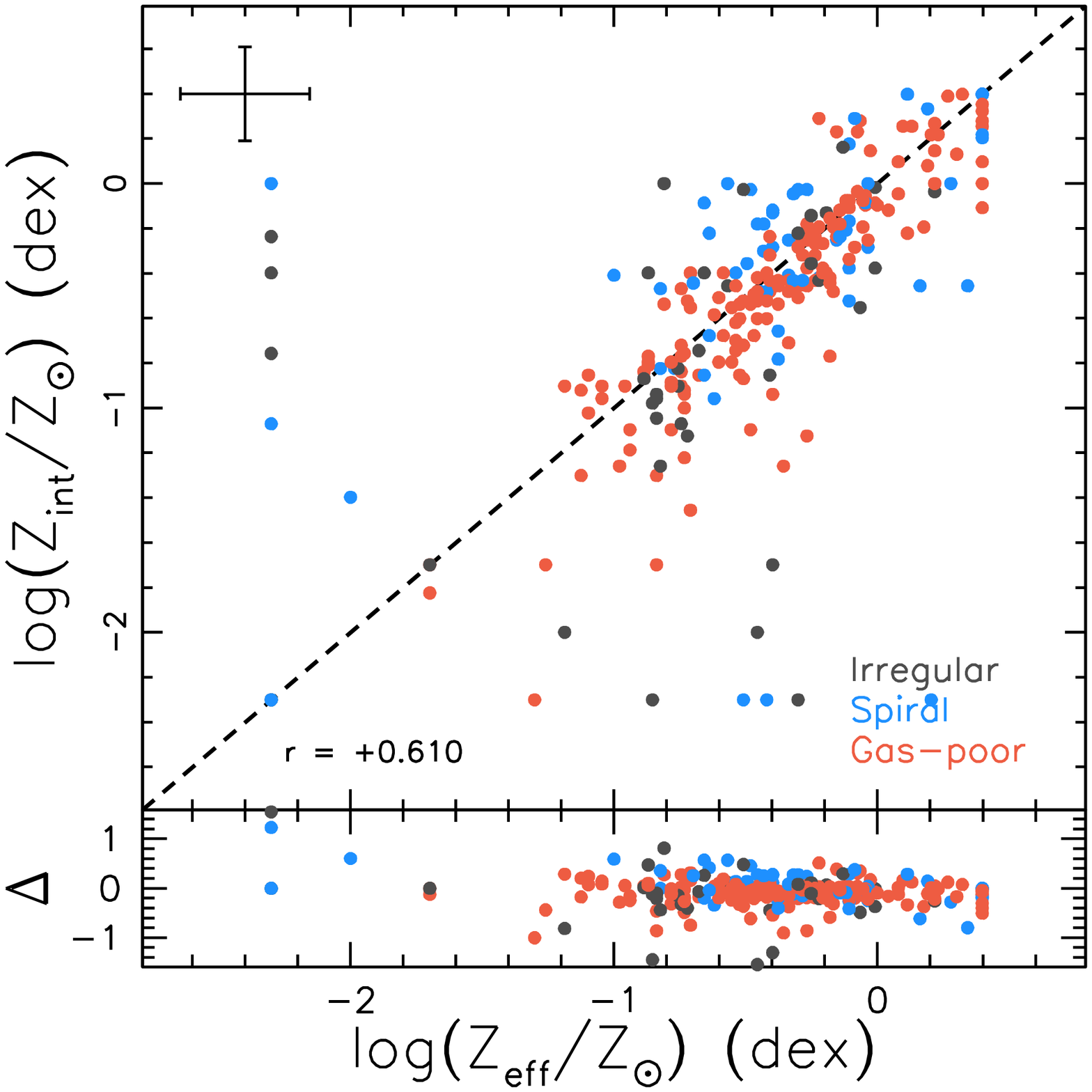}
  \end{tabular}
  \caption{(\textit{left}) Integrated mean ages versus mean ages
measured at $r_e$ for all Virgo galaxies. The data points for individual
galaxies have been coloured according to their specific morphology.
Direct equality (solid line), the Pearson correlation coefficient,
and the typical error per point are also shown in each plot. 
(\textit{right}) Same as (\textit{left}) but for metallicities.}
  \label{fig:AZ-int-eff}
 \end{center}
\end{figure*}

The stellar population scaling relations that we have discussed above apply to 
the SFHs and chemical evolution of Virgo galaxies on local scales. It is 
clearly of interest if these correlations also hold on galaxy-wide scales. To 
explore this, we must look at the integrated characterizations of our galaxies' 
stellar populations. \Fig{AZ-int-eff} shows that the integrated ages and 
metallicities of Virgo galaxies against their corresponding effective 
quantities are remarkably consistent with one another. On the other hand, 
substantial scatter exists between these galaxies' integrated and effective 
metallicities, which increases to low $Z_{eff}$ and is preferentially directed 
to lower $Z_{int}$. Despite this scatter, our galaxies are distributed in a 
rather monotonic fashion in the $Z_{int}$-$Z_{eff}$ plane. The integrated and 
effective stellar population estimates of Virgo galaxies thus agree with one 
another (at least in a relative sense), implying that their local scaling 
relations should apply globally as well.

%%%%%%%%%%%%%%%%%%%%%%%%%%%%%%%%%%%%%%%%%%%%%%%%%%%%%%%%%%%%%%%%%%%%%%%%%%%%%%%%
% SECTION 4: DISCUSSION
%%%%%%%%%%%%%%%%%%%%%%%%%%%%%%%%%%%%%%%%%%%%%%%%%%%%%%%%%%%%%%%%%%%%%%%%%%%%%%%%

\section{DISCUSSION}\label{Sec:Disc}

Below, we discuss the stellar population information of Virgo cluster galaxies 
in the context of popular galaxy formation and evolution scenarios \citep[i.e., monolithic collapse, hierarchical merging, and secular and environmental evolution;][]{La74,Co00,KK04,BG06} with the aim of establishing which one, or any 
combinations thereof, contributed most significantly to the inferred stellar 
content of each basic galaxy type. In this discussion, we focus only on those 
galaxies for which our sample is richest: gas-poor giants and dwarfs (E/S0 and 
dE/dS0, respectively), and spirals (Sa$-$Sd).

\subsection{Giant gas-poor galaxies}\label{Sec:D-gGP}
\subsubsection{Monolithic collapse versus hierarchical merging}
\label{Sec:D-gGP-MonoVSMerg}
As reviewed in \se{I-gGP}, spectroscopic studies of gas-poor giants have 
revealed downsizing trends in their central stellar populations which match 
monolithic collapse scenario predictions more readily than with hierarchical 
merging. While hierarchical models may be tweaked to reproduce said trends 
\citep[e.g.,][]{DeL06}, many studies have instead contrasted the above 
scenarios through the stellar population gradients that they predict in E/S0's 
\citep{KA99,TO03,Mi05,Og05,Wu05,SB07,Kol09,Sp09,To10a}, but there is no 
consensus. From the perspective of stellar populations, the the formation of 
gas-poor giants remains open to interpretation.

In order to explain the large central [$\alpha$/Fe] ratios found in local E's 
\citep{Th99}, a monolithic-like formation of these galaxies should have 
happened on a timescale of $\sim$1 Gyr. Such a timescale implies shallow, 
positive age gradients, and steep and negative metallicity gradients on account 
of the cumulative enrichment of the central regions \citep{Ko04,Pi08}. Although 
the recent monolithic collapse models of \cite{Pi10} successfully predict our 
observed metallicity gradients within Virgo E/S0's (-0.33 dex per decade 
radius, in the median), their ability to reproduce these galaxies' age 
gradients is suspect (A. Pipino, private communication). Also, simulations 
suggest that the metallicity gradients of collapse remnants should correlate 
with their stellar masses due to the effect of the potential well on both 
dissipation and chemo-dynamics during collapse \citep{KG03}; we recover no such 
trend. The stellar population information of Virgo E/S0's then suggests that 
these galaxies' origins were not monolithic-like in nature.

\subsubsection{Merger gas fractions}\label{Sec:D-gGP-GasFrac}
Given this tension between stellar population gradients of Virgo gas-poor 
giants and the monolithic collapse model, let us examine the possibility of a 
hierarchical origin. At face value, Virgo E/S0's having positive age and 
negative metallicity gradients are consistent with a scenario whereby their 
interiors was formed via compact starbursts \citep{Ho09b}, while their 
outskirts were accreted from (dwarf) satellites \citep{Ab06}. That is, these 
galaxies' stellar population gradients suggest the involvement of both gas-rich 
and gas-poor mergers in their formation. If this scenario is correct, then the 
continuous radial increase (decrease) in the age (metallicity) profiles of 
these Virgo E/S0's contrast with early merger simulations \citep{Wh80} which 
predicted that violent relaxation would flatten the stellar population 
gradients in the remnants' outskirts. More recent work has demonstrated, 
however, that non-negligible age and metallicity gradients, like those in Virgo 
gas-poor giants, may persist within merger remnants, but only if the 
progenitors' gradients were roughly twice as steep \citep{Di09,Ho09c}.

\begin{figure*}
 \begin{center}
  \includegraphics[width=0.9\textwidth]{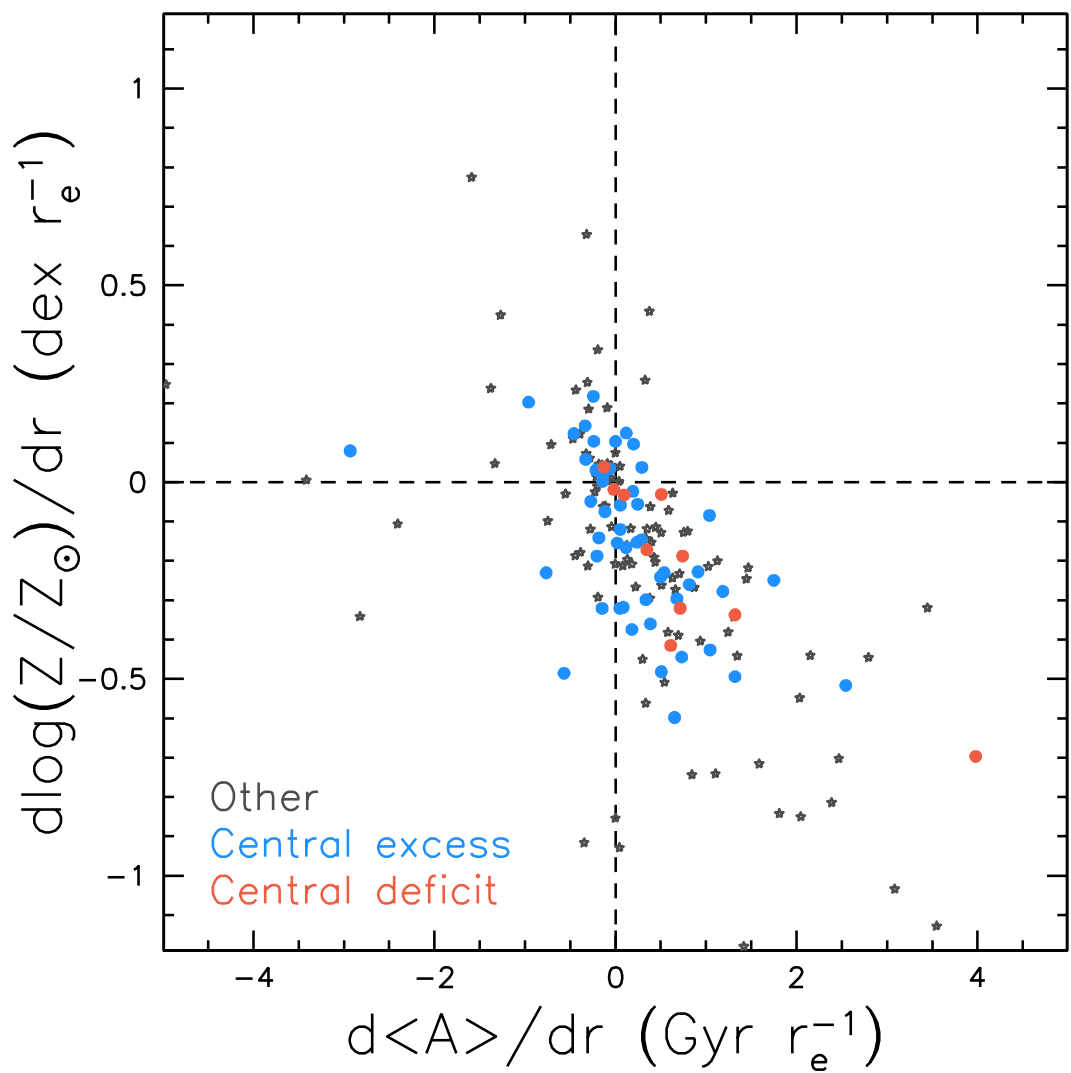}
  \caption{Metallicity gradients versus age gradients for Virgo gas-poor 
galaxies. The circular points represent those galaxies whose central light 
profiles have been analysed by \cite{Co06}; the colours of these data points 
denote a central light excess or deficit. The Virgo gas-poor galaxies whose 
central light behaviour has not yet been analysed are represented by dark grey 
stars.}
  \label{fig:MZ-MA-gETG}
 \end{center}
\end{figure*}

Recent photometric studies of Virgo E/S0's have suggested that the gas 
fractions of the mergers which produced them may be ascertained from the 
central behaviour (cusp/core) in their light profiles \citep{Co06,Kor09}. That 
is, gas-rich merger remnants should have cusped light profiles due to their 
centralized starbursts \citep{MH96}, while gas-poor merger remnants should have 
cored light profiles due to binary black hole scouring of the central regions 
\citep{Me06}. Given a cusp/core catalog for Virgo E/S0's \citep{Co06}, we can 
then explore the effect of gas fractions on merger remnants' stellar population 
gradients. The comparison of the age and metallicity gradients of cusped and 
cored Virgo E/S0's in \Fig{MZ-MA-gETG},suggests that neither of these two 
galaxy types distinguish themselves in this regard. This bolsters the above 
claim that progenitors' stellar population gradients may indeed survive merger 
events.

The stellar population scaling relations of Virgo E/S0's also offer insight 
into their merger gas fractions. We showed in \se{SPSR-GP} that the 
metallicities of Virgo gas-poor giants increase with both their concentration 
and luminosity. Although a gas-poor merger of low-$C_{28}$ galaxies could 
conceivably produce a more concentrated remnant, the absence of merger-induced 
SF implies that it still would not follow the established $Z$-$C_{28}$ relation 
(i.e., its metallicity would not exceed that of its progenitors). From this, we 
suggest that a significant fraction of the stellar mass in massive Virgo E/S0's 
may have formed via starbursts in gas-rich mergers. Coupled with the 
predominance of cusped light profiles in less luminous Virgo E/S0's 
\citep{Co06,Kor09}, this indicates that gas-rich mergers play a significant 
role in the formation of all the gas-poor giants in this cluster.

\subsubsection{Do Virgo E/S0's exhibit downsizing?}\label{Sec:D-gGP-Down}
The hallmark of recent spectroscopic studies of local E/S0's is that their 
central SSP ages and [$\alpha$/Fe] ratios increase with dynamical mass and 
local density \citep{Cal03,Ne05,Th05,Cl06,Sm07}. However, the authenticity of 
these downsizing trends remains contentious \citep{Tr00,Ke06,SB06a,Tr08}. For 
instance, \cite{Tr08} suggest that downsizing may actually represent a 
mass-dependent trend in the latest central SF episode within these galaxies, 
since SSP fits are biased by the system's youngest stars. As alluded to in 
\se{D-gGP-MonoVSMerg}, downsizing has major implications about giant gas-poor 
galaxy formation. 
% It is thus of interest whether the stellar populations of Virgo E/S0's also 
%exhibit downsizing.
As stated in \se{SPSR-GP}, we do not detect a statistically significant 
downsizing trend by stellar mass in the ages of Virgo E/S0's. This result 
agrees with other conflicting results from prior investigations into the 
stellar mass dependence of gas-poor giants' stellar populations 
\citep{Gal05,Gr09a,Gr09b,Sm09}. Recall also that our non-detection of 
downsizing refers to the effective radii of Virgo E/S0's, which suggests that 
these galaxies' integrated SFHs do not downsize with respect to their stellar 
masses. That is, this result casts doubt on downsizing by mass as a key 
characteristic of the formation of giant gas-poor galaxies.

We also demonstrated in \se{SPSR-GP} that the stellar populations of Virgo 
E/S0's do not exhibit downsizing trends with environment. While this is, in 
principle, contrary to the age and metallicity differences between field and 
cluster gas-poor giants advocated by \cite{Th05}, we speculate that our null 
result may be due to saturation of the Thomas et al. trend on the density 
scales of galaxy clusters. A plausible source of such saturation may involve 
the preprocessing of gas-poor galaxies in groups which arrive for the first 
time into the cluster environment \citep{Mi04}. Further exploration of the 
possible environmental downsizing of Virgo E/S0's will have to await the 
compilation of accurate distances ($<$10\% error) for these galaxies, to enable 
3D estimates of cluster-centric distances and local galaxy densities.

\subsubsection{Are S0's a unique galaxy population?}\label{Sec:D-gGP-S0}
While most statistical studies of the stellar populations of giant gas-poor 
galaxies group E's and S0's together \citep[e.g.,][]{Th05}, the structural 
differences between these galaxy types\footnote{By definition, S0's have 
dynamically fragile disks and, in Virgo, are on average more concentrated, 
brighter and larger than E's.} hint at a unique origin for the latter. The 
origins of S0's have thus been popularly ascribed to an evolution of spiral 
galaxy progenitors, driven by environment, which involves the removal of the 
latter's gas (via hydrodynamic and/or tidal processes) and the subsequent 
fading of their stellar disks \citep{BG06}. The increasing fraction of blue 
cluster galaxies to high redshifts \citep{BO78} is perhaps the strongest 
evidence in favour of such a scenario. Since a hydrodynamic or tidal 
transformation of a spiral galaxy should leave its remnant with a central gas 
reservoir (see below), one would expect to uncover environmental signatures in 
S0's stellar populations (e.g., positive age gradients) if this scenario was 
true. Studies along these lines, however, have so far failed to reveal 
conclusive results \citep[and references therein]{Ch11}.

The positive age and negative metallicity gradients found in some Virgo S0's 
conform with their possible evolutionary origin. Younger and more enriched 
central regions in galaxies that have experienced either ram pressure stripping 
or harassment are a generic expectation from simulations of these effects 
\citep{Mo96,Vo01}. These simulations show that remnants of these processes are 
left with central gas reservoirs due to either the stronger gravitational 
restoring force in those regions (re: ram pressure stripping), or 
tidally-induced central gas flows (re: harassment). The participation of Virgo 
S0's in the age-gas deficiency trend for the gas-rich systems in this cluster 
(\Fig{AZ-Enviro}) also supports an evolutionary origin for these galaxies. This 
trend and the lack of $D_{M87}$ or $\Sigma$ signatures in the stellar population 
diagnostics of Virgo S0's, strongly suggest that ram pressure stripping would 
be the preeminent cause of their evolution from a gas-rich spiral progenitor.

Alternatively, Virgo S0's may have formed, like Virgo E's, through hierarchical 
merging. Both the prominence of S0 bulges \citep{Dr80} and the evidence for 
(small) disks in the light profiles and kinematic maps of classified E's 
\citep{Kra08,Mc09} suggest that these two galaxy types represent a continuum in 
gas-poor giant structure. Furthermore, the stellar population diagnostics for 
Virgo E's and S0's follow similar scaling relations \citep[see also][]{Ja10}. 
If Virgo E's and S0's share a common origin, as these commonalities suggest, 
then the positive age gradients we find in many of the latter imply that their 
disks existed prior to their bulges. While stellar disks can survive a minor 
merger \citep{HM95}, it is unlikely that such events could build the massive 
bulges of S0's. Conversely, major mergers may explain the formation of Virgo 
S0's given the simulations \citep{Ho09a} and observations \citep{Ka09,Ca10} 
that both support the idea of disk survival (or even regrowth) after such an 
event, provided that it is of a high gas fraction. Gas-rich merging may be 
problematic though as the gas disk must be promptly removed after the merger to 
keep the disk's stellar population older than that of the bulge; this problem, 
of course, does not exist for Virgo S0's having negative age gradients. Mergers 
can consume gas through centripetal gas flows and subsequent starbursts 
\citep{MH96}, but it is unlikely that this mechanism will deplete the remnant's 
entire gas disk. Thus, the merger scenario of S0 formation may still require an 
external agent (e.g., ram pressure stripping) to remove gas from these 
galaxies' progenitors; that is, no one mechanism may explain these galaxies' 
origins \citep{Be08}.

\subsection{Dwarf gas-poor galaxies}\label{Sec:D-dGP}
The gradual realization that gas-rich and gas-poor dwarf galaxies overlap in 
terms of their structure, dynamics, gas content, substructure and SFHs 
\citep{LF83,Pe02,Co03a,Li06a,Tols09} has lent considerable weight to the idea 
of an evolutionary link between these two galaxy types. External mechanisms 
that could drive the evolution of a gas-rich dwarf into a gas-poor dwarf 
include ram pressure stripping or thermal evaporation of its ISM, and galaxy 
harassment \citep{GG72,CS77,Mo96}. An internal mechanism for such a depletion 
is unlikely due to dwarf galaxies' (implied) inefficiencies in both SF 
\citep{Gre03} and gas expulsion \citep{MLF99,We08,Le09,Tols09}. Despite the 
extensive overlap of their physical properties, important differences remain 
between gas-rich and gas-poor dwarfs in terms of their surface brightnesses, 
metallicities and globular cluster statistics \citep{Bo86,Gre03,St06}. A 
primordial origin, like that for gas-poor giants (i.e., monolithic collapse or 
hierarchical merging), thus remains viable for gas-poor dwarfs, and is supported 
by the very old ages and low metallicities of the nucleated dwarfs in both the 
Coma and Fornax galaxy clusters \citep{RS04}.

\subsubsection{A primordial origin?}\label{Sec:dGP-Prim}
We have found that a majority of Virgo gas-poor dwarfs in our sample have 
positive age and negative metallicity gradients (admittedly though, 
\Table{Agrd} indicates that the former are statistically consistent with being 
flat). Given such a stellar population gradient combination, it seems unlikely 
that these galaxies could be primordial in origin. For instance, the 
metallicity gradients in Virgo dE/dS0's are collectively steeper than those of 
Virgo E/S0's, contrary to the predicted flattening of this diagnostic towards 
low masses in monolithic collapse models \citep{KG03}. The absence of such a 
trend in our data is supported by the Pearson test ($r=$ +0.34) shown in the 
lower-right panel of \Fig{MAZ-C28-H} for Virgo gas-poor galaxies. The 
primordial origin hypothesis for Virgo dE/dS0's is also discredited on account 
that these galaxies do not exhibit comparable stellar population scaling 
relations with those of Virgo E/S0's. The stellar population data Virgo 
dE/dS0's thus argue against either a hierarchical or monolithic-like origin for 
them.

\subsubsection{The evolutionary scenario}\label{Sec:D-dGP-Evol}
Much of the stellar population data for the Virgo gas-poor dwarfs is readily 
interpreted in the context of an evolutionary origin. While our data cannot 
confirm which environmental mechanism is most likely to have removed these 
galaxies' gas, the absence of trends in their stellar population diagnostics 
with both galaxy density and cluster-centric distance disfavours it being 
harassment. On the other hand, the fact that gas stripping of (more massive) 
Virgo spiral galaxies is observed \citep[and references therein]{Chu09} and 
should increase in efficiency with lower masses makes this mechanism appealing. 
Given this, negative age gradients in Virgo dE/dS0's may arise from the 
strangulation of gas-rich progenitors' haloes, while positive age gradients may 
result from ram pressure stripping of progenitors' gas disks \citep{Bo08a}.

Not all the stellar population information for Virgo dE/dS0's agrees with their 
possible evolutionary origin. While we find broad consistency in the central 
metallicities of all Virgo dwarfs (\Table{Zgrd}), the young ages of our 
Im+BCD's suggests that our (light-biased) estimates sample only their most 
recently formed stars, such that their fitted metallicities are better thought 
of as being upper limits. This, coupled with the more negative metallicity 
gradients within the Im+BCD's, then implies that the true metallicities of 
Virgo gas-rich and gas-poor dwarfs are likely discrepant, as found amongst 
Local Group dwarfs as well \citep{Gre03}. The fact that we find mild agreement 
between the central metallicities of all Virgo dwarfs, however, implies that 
Im+BCD's are at least capable of enriching themselves to the same degree as 
dE/dS0's. This suggests a solution to discrepant metallicity problem which the 
evolutionary scenario faces, whereby the progenitors of Virgo dE/dS0's 
sufficiently enrich themselves before completely losing their gas supplies 
(e.g., through triggered SF). This can happen once a dwarf galaxy has been 
stripped of its gas, as only evolved stars can replenish its ISM at that point 
\citep{Bo08a}. Conversely, the extremely metal-poor outskirts in Virgo Im's may 
be understood as a recent accretion of pristine gas onto those regions (i.e., 
those estimates are lower limits). Considering our poor statistics in the 
stellar population diagnostics of Virgo Im+BCD's, we refrain from pursuing this 
comparison further.

In a manner similar to the gas-poor giants, the Virgo dS0's exhibit several
structural differences relative to the Virgo dE's, suggesting a different
origin for these two gas-poor dwarf types. This is supported by KS tests
with low probabilities ($P <$ 0.5) between their respective stellar
population diagnostics. 
\citet[and references therein]{Li09}, \citet{Tolo09}, and \cite{Pa10a} have 
also explored the idea of distinct classes of Virgo gas-poor dwarfs and argued 
that the younger, flatter, less centrally clustered population is the prototype 
for the evolutionary remnant of a low-mass spiral or gas-rich 
dwarf\footnote{\footnotesize This population also has the highest frequency of 
substructure (e.g., bars) and blue centers.}. These authors also suggest that 
the older, rounder, more centrally clustered gas-poor 
dwarfs\footnote{\footnotesize This population also has the highest frequency of 
nucleation.} may have a primordial origin. The many structural and stellar 
population differences that we find between Virgo dS0's and dE's thus bolster
the main conclusion of the above studies; namely, that the Virgo gas-poor dwarfs
likely formed through at least two different channels.

\subsection{Spiral galaxies}\label{Sec:D-Sp}
As discussed in \se{I-Sp}, the formation of spiral galaxy bulges remains
unsettled, particularly with respect to the relative contributions of
hierarchical merging \citep{Be92} and secular evolution \citep{KK04} as a 
function of Hubble type. Making matters worse, to distinguish these 
processes is non-trivial as satellite accretion is likely to excite disk 
instabilities \citep{Gau06}.  Furthermore, while the formation of spiral
galaxy disks is well understood in principle as the self-gravitating collapse
of a rotating gas cloud \citep{FE80}, the local environment may play a significant
role in the evolution of this component in cluster spirals.  Below, we try
to understand the formation of the bulges and disks of Virgo spirals via 
comparisons of their relative ages and metallicities and assuming that 
bulge light dominates these galaxies' interiors ($\lesssim$1 $r_e$).  
Decompositions of Virgo spirals' light profiles indeed show that the 
bulge-to-total light ratio is typically above 50\% within their
effective radii \citep{Mc09}.

The negative metallicity gradients found in most Virgo spirals indicates
that their bulges are more enriched than their disks.  Conversely, the age
gradients in Virgo spirals range from being typically positive for
early-types (Sa-Sbc) to negative for late-types (Scd+Sd). Each spiral 
(sub-)class, however, contains a mixture of positive and negative age
gradients (\Table{Agrd}), such that no one scenario can describe the
origin of any one galaxy type. We therefore address the origins of Virgo
spirals based on age gradient sign rather than morphological type. The
definite existence of positive and negative age gradients in these systems
is confirmed by their measured error envelopes which do not typically
overlap with zero.

\subsubsection{Positive age gradients}\label{Sec:D-Sp-PAG}
The combination of a young, metal-rich bulge and an old, metal-poor disk (found 
in $\sim$35\% of Virgo spirals, most commonly Sa$-$Sc's) suggests a formation 
scenario for the bulge which involves the accumulation of a central gas 
reservoir. Both secular evolution and a recent gas-rich minor 
merger\footnote{\footnotesize We stipulate that such a merger must be of a low 
mass ratio if the (old) disk is to survive such an event.} could create such a 
reservoir \citep{PN90,HM95}.  However, the former scenario would likely fail to
produce a positive age gradient for two reasons. First, the gas reservoir that
a bar creates via a centripetal flow from the disk would be spatially confined
to the galaxy's nucleus and not spread over the entire bulge. Second, the
scattering of stars into the pseudo-bulge by a bar should actually flatten a 
spiral galaxy's age gradient \citep{KK04}. Secular evolution is thus unlikely 
the source of positive age gradients in Virgo (early-type) spirals.

While secular evolution may fails to explain the positive age gradients
in Virgo spirals, the success of the gas-rich minor merger scenario is
contingent upon at least two constraints.  First, some mechanism is needed
to restrict the funneling of gas into the nucleus as this will not likely
lead to the creation of a positive gradient over a large radial extent
within the remnant. This funneling may derive from the weak torques acting
on the gas during a minor merger \citep{Ho09a} or from tidal disruption
of the secondary before coalescence, such that the merger turns into
an accretion event. Second, the merging of a gas-rich dwarf with a spiral
galaxy should create a positive metallicity gradient, due to the deposition
of metal-poor gas and stars into the bulge. This challenge for the merger
scenario may be overcome if the accreted gas has either been pre-enriched
or is consumed over a long timescale.

Although gas-rich minor mergers may explain the origin of bulges in Virgo 
spirals with positive age gradients, it offers no insight as to why stellar 
ages and metallicities continue to increase and decrease, respectively, outward 
in these galaxies' disks. The most compelling scenarios for the creation
of such disks include outside-in formation and disk fading, wherein the latter 
occurs after depletion of the gas disk by external agents. While the formation 
of outside-in disks has been seen in simulations \citep{SL03}, current 
observational evidence suggests that this is a rare mechanism for the growth of 
spiral disks \citep{Mu07}.  Therefore, the existence of positive age gradients
in most Virgo bulges and disks in our sample may be explained by a combination
of gas-rich minor merging and disk fading.  We shall revisit the topic of gas
disk depletion in \se{Sp-GS}.

\subsubsection{Negative age gradients}\label{Sec:Sp-NAG}
The alternate combination of an old, metal-rich bulge with a young, metal-poor 
disk (found in $\sim$35\% of Virgo spirals, most commonly Sc$-$Sd's) implies 
that the gas supply of the disk accumulated later and/or was converted to 
stars over a longer timescale than that of the bulge (i.e., the bulge formed 
before the disk). Bulges may deplete their gas supplies most efficiently at 
early epochs, thereby making them relatively old, by undergoing major mergers. 
This suggestion contradicts the popular wisdom that late-type spirals possess 
pseudo-bulges \cite{Ba03,Ma03,KK04}, but recent spectroscopic stellar 
population analyses have shown that the stellar masses of late-type bulges are 
indeed predominantly old \citep[M09;][]{Sa10}, as expected in a classical sense. 
Also, the existence of a negative age gradient in a spiral galaxy that has
experienced the stellar mixing and radial gas flows associated with secular 
evolution seems counter-intuitive. The bulges of such galaxies are then most 
consistent with having originated through mergers.

A negative age gradient throughout a spiral galaxy also places constraints on 
the formation mechanism and epoch of its disk. Such radial behaviour in stellar 
age, coupled with a negative metallicity gradient, suggests that the outskirts 
of these galaxies were formed through a more extended SFH, from a more pristine 
ISM and suffered metal-rich outflows, while their central regions would have 
resulted from a more intense but short-lived SFH which involved substantial gas 
recycling. Such an inside-out origin of Virgo late-type disks agrees well with 
results from $\Lambda$CDM simulations of disk formation \citep{Mo98,SL03}
% Samland & Gerhard (2003)? Tissera+ (2001)?
and supports the idea that the gravo-thermal collapse of gas clouds within them 
(and thus their star formation rates) depends on the local surface mass density 
\citep{Ke89}. The fact that the disks in Virgo late-type spirals are younger 
than their bulges also suggests that these gaseous disks have not yet 
(or only recently) been stripped by the Virgo ICM. If true, then these galaxies 
are likely recent additions to this cluster.

\subsubsection{Gas stripping}\label{Sec:Sp-GS}
The existence of positive age gradients and a statistically significant 
age-$Def_{\hi}$ scaling relation in Virgo spirals both argue that gas removal 
plays a fundamental role in the evolution of these galaxies. We ascribe the 
agent of gas removal to ram pressure since several observational evidences 
favour its influence on gas-rich cluster galaxies (for a summary, see 
\citealt{Chu09}, and references therein), its efficiency at quenching the SF in 
spiral disks is high \citep{Po99,RH05,Kro08}, and the ages of Virgo spirals are 
evidently independent of both their cluster-centric distances and local galaxy 
densities\footnote{\footnotesize However, projection effects in the diagnostics 
$D_{M87}$ and $\Sigma$ and the possibility that ram pressure stripping can 
occur at large cluster-centric radii in Virgo due to a dynamic or ``lumpy'' ICM 
\citep{CK08} may also be playing a role in generating these null results.}. 
Although cluster galaxies are expected to quickly lose their gas haloes to 
hydrodynamical interactions at large cluster-centric radii 
\citep{Ba00,Be02,Mu09}, such a strangulation scenario \citep{La80} should not 
produce positive age gradients within them if the local star formation rate is 
density-dependent (M04).

The removal of a cluster spiral's gas supply should not only affect its SFH, 
but other factors related to its stellar populations as well. Ram pressure 
stripping gradually removes a spiral galaxy's gas disk from the outside-in 
\citep{Vo01}, but the deeper potential well towards its center prevents the 
complete loss of its \ion{H}{I} supply, ultimately leaving it with a large 
anemic annulus about its center. Given such a range in the gas retention 
properties within cluster spirals, the age gradient of a highly deficient 
system might be expected to be more positive than that of a less deficient 
system, since SF will undoubtedly be quenched in the former's outer disk but 
not in its central regions. Moreover, if SF is quenched in a galaxy's 
outskirts, so too should the enrichment of the local ISM, such that highly 
deficient spirals should be more metal-poor and have more negative metallicity 
gradients than less deficient spirals.  Finding no $Def_{\hi}$ effect 
whatsoever on either the metallicities or the stellar population gradients
of Virgo spirals is thus disconcerting.  Many factors may contribute to
wash out these trends, however, such as a rapid stripping timescale, a recent 
stripping epoch for the highly deficient spirals, and/or triggered SF or gas 
fallback accompanying the stripping event 
\citep[as seen in simulations;][]{Vo01}. The absence of $Def_{\hi}$ trends 
expected in the stellar population diagnostics of cluster spirals may be 
evidence that gas stripping does not simply entail a gradual outside-in 
removal of their gas and subsequent quenching of their SF.

\begin{figure*}
 \begin{center}
  \includegraphics[width=0.9\textwidth]{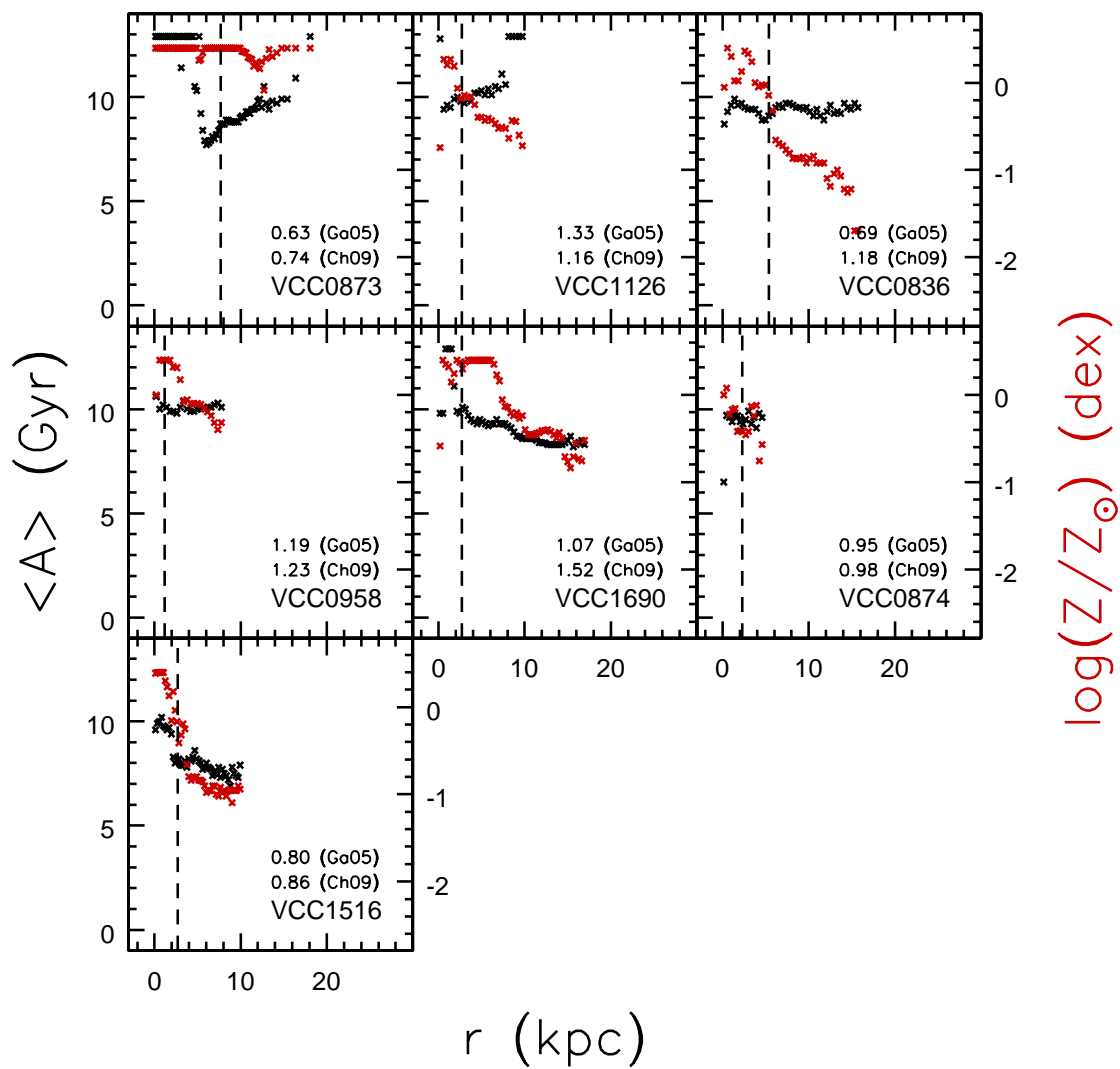}
  \caption{Age (black) and metallicity (red) profiles for Virgo spiral galaxies 
which overlap with the Crowl \& Kenney (2008) sample. For each galaxy, we 
provide in the bottom right corner of each window the Virgo Cluster Catalog 
identification number and the $Def_{\hi}$ values from both \citet{Ga05} and 
\citet[Ch09]{Chu09}.}
  \label{fig:CK08}
 \end{center}
\end{figure*}

To bolster our claim that the effects of gas stripping on cluster spirals
is more complex than the simple (intuitive) picture described above, we show
in \Fig{CK08} the stellar population profiles for Virgo spirals that we 
have in common with \cite[; hereafter CK08]{CK08}. CK08 analysed the stellar 
populations located just beyond the truncated H$\alpha$ disks in several Virgo 
spirals to determine when SF was quenched in those regions. They concluded that 
the highly \ion{H}{I}-deficient galaxies must have lost their gas disks recently
($\le$500 Myr) without any triggering of SF. Seven of CK08's spirals overlap with
our sample, thereby enabling a qualitative comparison of our age profiles with 
them.  We find that the age profiles of only two of these galaxies 
(VCC0873 and VCC1126) notably increase beyond their H$\alpha$ disks, implying 
that their disks were stripped gradually, not recently. Interestingly, the age 
profile of VCC1690 ($Def_{\hi}$ = 1.07) actually decreases just beyond its 
H$\alpha$ disk, suggesting that ram pressure has actually triggered SF during 
the stripping process. The age profiles of the other four galaxies remain flat 
beyond their H$\alpha$ disks, implying that these galaxies were instead 
stripped rapidly. The age profiles for Virgo spirals and our comparison with 
CK08 suggest that the gas stripping timescales for gas-rich galaxies falling
into the Virgo cluster may not be uniform \citep{Po99,Vo01,RH05,Kro08} and that
ram pressure may indeed promote SF in galaxies before ultimately quenching it.

%%%%%%%%%%%%%%%%%%%%%%%%%%%%%%%%%%%%%%%%%%%%%%%%%%%%%%%%%%%%%%%%%%%%%%%%%%%%%%%%
% SECTION 5: CONCLUSIONS
%%%%%%%%%%%%%%%%%%%%%%%%%%%%%%%%%%%%%%%%%%%%%%%%%%%%%%%%%%%%%%%%%%%%%%%%%%%%%%%%

\section{Conclusions}\label{Sec:Concs}

We have used high-quality optical and near-infrared ($griH$) imaging
of $\sim$300 Virgo cluster galaxies to determine radial stellar population
variations for a representative sample spanning the full range of galaxian
parameter space.  We have also compared stellar population diagnostics
(ages, metallicities, and their gradients) with structural and environmental
parameters to investigate their contribution to galaxies' SFHs and chemical
evolution. All these data considered, we have developed plausible formation
and evolution scenarios for each of the major galaxy types (gas-poor giants
and dwarfs, and spirals) within a cluster setting.  Here is a summary of our
main results:
\begin{itemize}
 \item Most galaxies' colour gradients are due to variations in metallicity; 
   ellipticals, Sbc--Sd spirals, gas-rich dwarfs, and peculiars also show 
   non-negligible age variations;

 \item Although reddening can skew stellar population estimates from colours, 
   realistic dust modelling cannot be achieved through colour information alone;

 \item Of the SFHs considered, only an exponential SFH can simultaneously 
   account for the wide colour range exhibited by all galaxy types, especially
   the gas-poor giant galaxies;

 \item The metallicity gradients of Virgo galaxies become increasingly negative 
   towards later types along the Hubble sequence; 

 \item Significant correlations in the stellar population diagnostics
   of Virgo galaxies indicate that:
 \begin{description}
  \item[-] the chemical evolution of gas-poor galaxies is determined by
    the interplay between stellar masses (nucleosynthesis sites)
    and surface densities (potential well depth),
  \item[-] the higher metallicities of E/S0's that experienced more
    dissipation during their formation suggests that their stellar
    masses were built up \textit{in situ},
  \item[-] the SFHs of gas-rich cluster galaxies are strongly affected
    by complicated gas removal processes that likely trigger SF as well
    as inhibit it.
 \end{description}

 \item The stellar population data for Virgo galaxies suggest the 
   following formation and evolution scenarios for the major galaxy types
   in this cluster:
 \begin{description}
  \item[-] Giant gas-poor (E/S0) galaxies grew mostly via gas-rich
    mergers but some gas-poor mergers, as well as gas removal processes
    in S0's, also likely play a role;
  \item[-] Dwarf gas-poor (dE+dS0) galaxies do not result from a single 
    formation mechanism but the dominant production channel likely
    involves an evolution from a gas-rich progenitor (e.g., Im+BCD)
    that is driven by environment;
  \item[-] Spiral (Sa$-$Sd) galaxies having a bulge that is older
    and more enriched than the disk likely formed their bulge through
    merging and grew their disk inside-out afterward, free from
    environmental effects. Conversely, spirals with a disk older
    and less enriched than their bulge, whilst still having
    a merger-built bulge, likely underwent stripping of their gas disk.
  \end{description}
\end{itemize}

\bigskip
We are grateful to L. P. Chamberlain, C. Conroy, E. Emsellem, L. Ferrarese,
and C. Maraston for insightful discussions. This research has made use of
the NASA/IPAC extragalactic and GOLDMine databases. J. R. and S.C. acknowledge
financial support from the National Science and Engineering Council of Canada.

%%%%%%%%%%%%%%%%%%%%%%%%%%%%%%%%%%%%%%%%%%%%%%%%%%%%%%%%%%%%%%%%%%%%%%%%%%%%%%%%
% APPENDIX
%%%%%%%%%%%%%%%%%%%%%%%%%%%%%%%%%%%%%%%%%%%%%%%%%%%%%%%%%%%%%%%%%%%%%%%%%%%%%%%%

\section{APPENDIX: Dust}\label{App:Dust}

The presence of dust in galaxies affects broadband studies of their stellar 
populations since a change in reddening can mimic variations in both age
and metallicity \citep{By94,dJ96}. The weak SF characteristics of gas-poor 
galaxies \citep{ST07} imply that reddening of their colours by dust should
be minimal since their dust reservoirs should not have been recently replenished 
by stellar feedback \citep{Em08}.  The same cannot be said for gas-rich 
galaxies, as evidenced by the extremely red optical-NIR colours that we
find in some Virgo spirals (Paper I); however, \cite{Ma10} show that such
red colours could also result from the presence of a significant TP-AGB
population.  Rather than correct our photometry using idealized dust models 
\citep[e.g.,][]{Ga02}, we instead address the impact that dust would have
on galaxy colours.

To gauge the effect of reddening, we examine three popular dust models: (1) 
foreground screen, (2) face-on triplex \citep{Di89}, and (3) clumpy medium 
\citep{CF00}. The screen model follows a simple prescription, $A_{\lambda}$ = 
1.08$\tau_{\lambda}$, where $A_{\lambda}$ and $\tau_{\lambda}$ are the 
extinction and optical depth in band $\lambda$, respectively. The triplex 
model assumes exponentially declining distributions of stars and dust in both 
the radial and vertical directions. We calculate triplex extinctions according 
to Eq.~(14) of M04, which invokes a simple treatment of scattering \citep{dJ96}. 
M04 argued that a large reddening gradient is only achieved with the triplex 
model when the stellar and dust scale lengths are equal, contrary to 
observations \citep{Xi99}. For both the screen and triplex models, we use the 
Milky Way extinction curve and albedos from \cite{Go97}. The clumpy medium 
model mimics the preferential but short-lived extinction of young star 
clusters \citep{Wi92,Go97} due to the higher dust concentrations in molecular 
clouds than in the diffuse ISM. The power-law dependence of the extinction 
curve in this model has been verified in a multi-band study of SDSS galaxies 
\citep{Jo07}.  However, as Charlot \& Fall contend, the neglected scattering
in their model make its applicability to galaxian light profiles questionable. 
Indeed, the absence of a detailed scattering treatment in most analytic dust
models is another valid reason to forego reddening corrections to our data.

\begin{figure*}
 \begin{center}
  \includegraphics[width=0.9\textwidth]{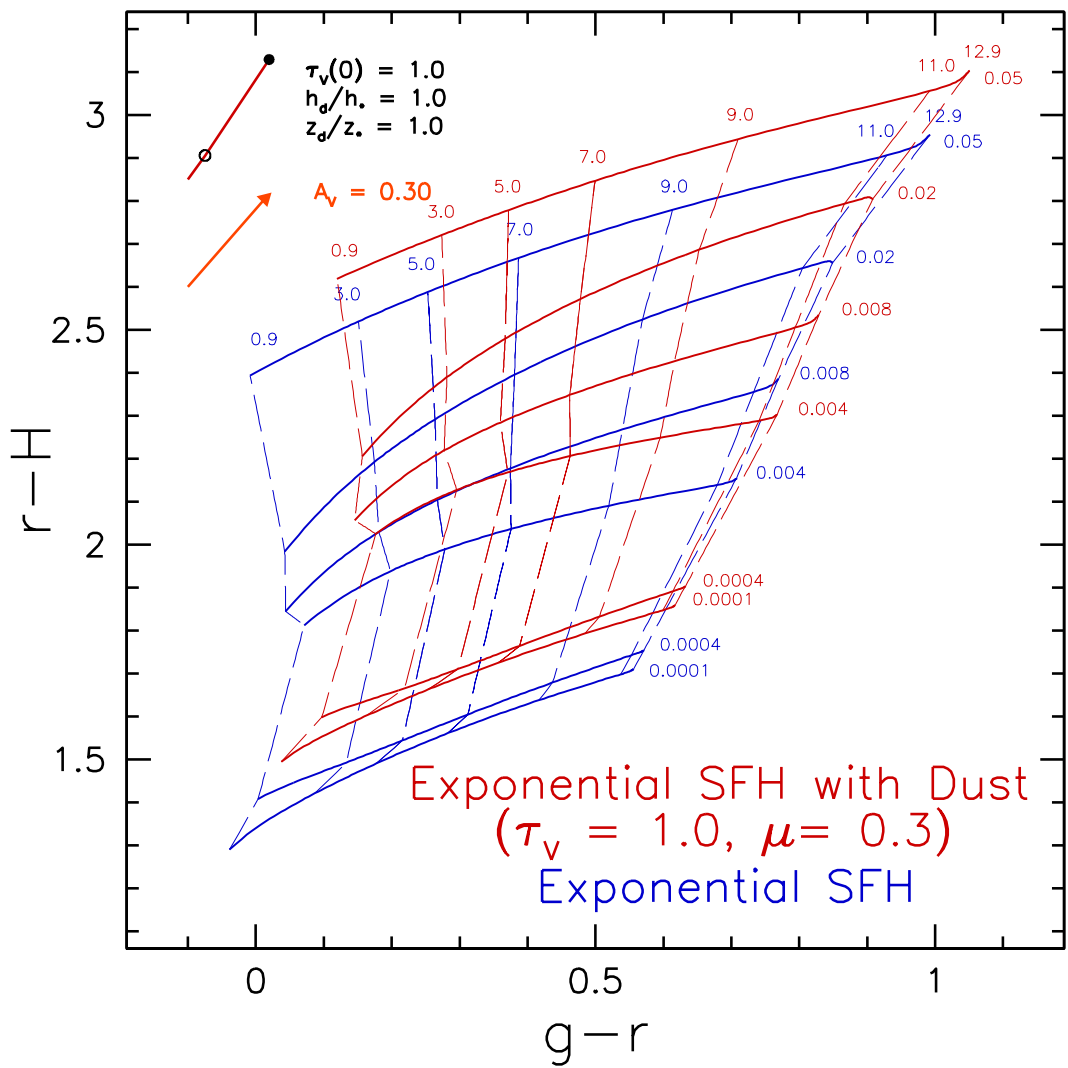}
  \caption{As in \Fig{SFH-Comp-1} but comparing an exponential 
SFH model with (red) and without (blue) reddening, whereby a clumpy 
medium dust model was used for the former.  The reddening vectors for 
the foreground screen and face-on triplex models (orange-red and dark-red, 
respectively) are also shown. Parameter values for the clumpy model are
indicated at lower-right in each panel ($\tau_V$ = $V$-band optical depth
of molecular cloud; $\mu$ = relative optical depth of diffuse ISM), while
those for the other models are adjacent to the vectors themselves.
The position of the galaxy's center and half-light radius in the
triplex vector are indicated by the solid and open circles, respectively.}
  \label{fig:Dust-Comp}
 \end{center}
\end{figure*}

We compare in \Fig{Dust-Comp} the predictions for each of the above 
dust models against our preferred exponential SFH model. The predicted 
reddening mostly affects optical-NIR colours and thus mostly skews metallicity 
estimates \citep{BdJ00}. Considering our typical measurement errors however 
(Paper I), the models are degenerate with respect to colours and even 
themselves. Although the dust extinction vectors lie nearly parallel to the 
median colour gradients of our Virgo spirals (Paper I), unfeasibly large 
attenuations would be required to fully explain the magnitude of these 
galaxies' gradients (M04). With exception to a few edge-on Virgo spirals 
that lie completely redward of the exponential model grid (e.g., VCC0873),
we can thus rule out dust as a dominant contributor to the colours of Virgo
gas-rich galaxies. This point is supported by our finding in Paper I that 
these galaxies' central colours are independent of inclination.
Constraints on the dust properties of Virgo cluster galaxies should
be significantly improved with the highly anticipated completion of
several ongoing dedicated X-ray, UV, optical, IR-FIR, and sub-mm
surveys of this cluster.

%%%%%%%%%%%%%%%%%%%%%%%%%%%%%%%%%%%%%%%%%%%%%%%%%%%%%%%%%%%%%%%%%%%%%%%%%%%%%%%%
% REFERENCES
%%%%%%%%%%%%%%%%%%%%%%%%%%%%%%%%%%%%%%%%%%%%%%%%%%%%%%%%%%%%%%%%%%%%%%%%%%%%%%%%

\clearpage

%%%%%%%%%%%%%%%%%%%%%%%%%%%%%%%%%%%%%%%%%%%%%%%%%%%%%%%%%%%%%%%%%%%%%%%%%%%%%%%%
% TABLES
%%%%%%%%%%%%%%%%%%%%%%%%%%%%%%%%%%%%%%%%%%%%%%%%%%%%%%%%%%%%%%%%%%%%%%%%%%%%%%%%

% TABLE 1
\begin{deluxetable}{ccccc}
% \tablecolumns{10}
 \tabletypesize{\scriptsize}  
 \tablewidth{0pc}  
 \tablecaption{Explored star formation histories for stellar population modeling}
% \break (...)} 
 \tablehead{
  \colhead{SFH} &
  \colhead{$\Psi(t)$} &
  \colhead{$\left\langle A \right\rangle$ ($\tau$, $A$)} &
  \colhead{$\tau$ range} &
  \colhead{$\left\langle A \right\rangle$ range} \cr
  \colhead{} &
  \colhead{(M$_{\odot}$ Gyr$^{-1}$)} &
  \colhead{(Gyr)} &
  \colhead{(Gyr)} &
  \colhead{(Gyr)} \\ [2pt]
  \colhead{(1)} &
  \colhead{(2)} &
  \colhead{(3)} &
  \colhead{(4)} &
  \colhead{(5)}
 }
 \startdata
  Constant & constant = $\dfrac{1}{\tau}$ & $A$ - $\dfrac{\tau}{2}$ & [0.2, 13] & 6.5 - 12.9 \cr
  Delayed & $\dfrac{t}{\tau^2}$e$^{(-t/\tau)}$ & $A$ - 2$\dfrac{\tau - e^{-A/\tau}\left(\tau + A + \dfrac{A^2}{2\tau}\right)}{1 + \dfrac{A}{\tau}(1 - e^{-A/\tau})}$ & [-$\infty$, $\infty$] & 0.9 - 12.9 \cr
  Exponential & $\dfrac{1}{\tau}$e$^{-t/\tau}$ & $A$ - $\tau$ $\dfrac{1 - e^{-A/\tau}\left(1 + \dfrac{A}{\tau}\right)}{1 - e^{-A/\tau}}$ & [-$\infty$, $\infty$] & 0.9 - 12.9 \cr
  Linear & $\dfrac{2}{\tau}\left[ 1 - \left( \dfrac{t}{\tau}\right)\right]$, for $t < \tau$ & $A$ - $\dfrac{\tau}{3}$ & [0.3, 36.3] & 0.9 - 12.9 \cr
         &                                                  \hspace*{15mm}0, for $t \ge \tau$ & & & \cr
  Sandage & $\dfrac{t}{\tau^2}$e$^{(-t^2/2\tau^2)}$ & $A$ - $\dfrac{(1/2)\tau\sqrt{2\pi}\text{ erf}(A/\sqrt{2}\tau) - Ae^{-A^2/2\tau^2}}{1 - e^{-A^2/2\tau^2}}$ & [0.08, $\infty$] & 4.4 - 12.9 \cr
  Single-burst & $\delta$($t$) & $A$ - $t$ & [0.0, 12.1] & 0.9 - 13.0 \\
 \enddata
 \label{tbl:SFHs}
\end{deluxetable}

\bigskip

% TABLE 2
\begin{deluxetable}{cccccc}
% \tablecolumns{10}
 \tabletypesize{\scriptsize}  
 \tablewidth{0pc}  
 \tablecaption{Median age gradients of Virgo galaxies}
% \break (...)} 
 \tablehead{
  \colhead{Morphology} &
  \colhead{} &
  \colhead{$\dfrac{\text{d}\left\langle A \right\rangle}{\text{d}r}$} &
  \colhead{} &
  \colhead{$\left\langle A \right\rangle_0$} &
  \colhead{$N$} \\ \\ [-5pt]
  \cline{2-4} \\ [-5pt]
  \colhead{} &
  \colhead{$r/r_e$ (Gyr $r_e^{-1}$)} &
  \colhead{$r$ (Gyr kpc$^{-1}$)} &
  \colhead{log $r$ (Gyr dex$^{-1}$)} &
  \colhead{(Gyr)} &
  \colhead{} \\ [2pt]
  \colhead{(1)} &
  \colhead{(2)} &
  \colhead{(3)} &
  \colhead{(4)} &
  \colhead{(5)} &
  \colhead{(6)}
 }
 \startdata
  dS0     &  0.05 $\pm$ 1.13 &  0.03 $\pm$ 0.89 &  0.12 $\pm$ 1.90 &  9.65 $\pm$ 0.99 & 19 \cr
  dE      &  0.22 $\pm$ 1.45 &  0.30 $\pm$ 1.26 &  0.34 $\pm$ 2.50 &  9.68 $\pm$ 1.14 & 49 \cr
  E       &  0.30 $\pm$ 0.91 &  0.27 $\pm$ 1.96 &  0.84 $\pm$ 3.05 & 10.15 $\pm$ 0.82 & 31 \cr
  S0      &  0.00 $\pm$ 0.62 &  0.00 $\pm$ 0.58 & -0.07 $\pm$ 1.50 & 10.20 $\pm$ 0.70 & 53 \cr
  Sa$-$Sb &  0.02 $\pm$ 0.73 &  0.01 $\pm$ 0.55 &  0.02 $\pm$ 1.59 & 10.00 $\pm$ 1.27 & 24 \cr
  Sbc+Sc  &  0.26 $\pm$ 1.49 &  0.13 $\pm$ 0.60 &  0.41 $\pm$ 2.25 &  9.20 $\pm$ 1.44 & 15 \cr
  Scd+Sd  & -0.13 $\pm$ 1.19 & -0.10 $\pm$ 0.84 & -0.53 $\pm$ 2.40 &  8.95 $\pm$ 1.70 & 17 \cr
  Sdm+Sm  &  0.58 $\pm$ 1.98 &  0.28 $\pm$ 0.93 &  0.71 $\pm$ 2.38 &  7.20 $\pm$ 2.37 &  9 \cr
  Im      &  1.31 $\pm$ 3.08 &  1.03 $\pm$ 2.86 &  2.05 $\pm$ 4.55 &  5.15 $\pm$ 2.98 & 13 \cr
  BCD     &  2.07 $\pm$ 2.68 &  1.92 $\pm$ 4.85 &  4.25 $\pm$ 5.64 &  5.60 $\pm$ 2.08 &  8 \cr
  S?      &  1.16 $\pm$ 1.55 &  1.01 $\pm$ 1.03 &  2.80 $\pm$ 2.60 &  6.50 $\pm$ 3.19 &  9 \cr
  ?       &  0.58 $\pm$ 0.99 &  0.18 $\pm$ 2.15 &  1.26 $\pm$ 2.44 &  9.15 $\pm$ 1.68 &  8 \\
 \enddata
 \label{tbl:Agrd}
\end{deluxetable}

\bigskip

% TABLE 3
\begin{deluxetable}{cccccc}
% \tablecolumns{10}
 \tabletypesize{\scriptsize}  
 \tablewidth{0pc}  
 \tablecaption{Median metallicity gradients of Virgo galaxies}
 \tablehead{
  \colhead{Morphology} &
  \colhead{} &
  \colhead{$\dfrac{\text{dlog}(Z/Z_{\odot})}{\text{d}r}$} &
  \colhead{} &
  \colhead{log($Z_0/Z_{\odot}$)} &
  \colhead{$N$} \\ \\ [-5pt]
  \cline{2-4} \\ [-5pt]
  \colhead{} &
  \colhead{$r/r_e$ (dex $r_e^{-1}$)} &
  \colhead{$r$ (dex kpc$^{-1}$)} &
  \colhead{log $r$} &
  \colhead{(dex)} &
  \colhead{} \\ [2pt]
  \colhead{(1)} &
  \colhead{(2)} &
  \colhead{(3)} &
  \colhead{(4)} &
  \colhead{(5)} &
  \colhead{(6)}
 }
 \startdata
  dS0     & -0.10 $\pm$ 0.30 & -0.05 $\pm$ 0.27 & -0.16 $\pm$ 0.54 & -0.21 $\pm$ 0.29 & 19 \cr
  dE      & -0.23 $\pm$ 0.43 & -0.18 $\pm$ 0.43 & -0.46 $\pm$ 0.84 & -0.49 $\pm$ 0.22 & 49 \cr
  E       & -0.17 $\pm$ 0.25 & -0.14 $\pm$ 0.60 & -0.60 $\pm$ 0.89 & +0.05 $\pm$ 0.36 & 31 \cr
  S0      & -0.06 $\pm$ 0.24 & -0.04 $\pm$ 0.21 & -0.27 $\pm$ 0.47 & +0.01 $\pm$ 0.25 & 53 \cr
  Sa$-$Sb & -0.13 $\pm$ 0.26 & -0.05 $\pm$ 0.10 & -0.34 $\pm$ 0.53 & -0.07 $\pm$ 0.27 & 24 \cr
  Sbc+Sc  & -0.25 $\pm$ 0.43 & -0.09 $\pm$ 0.18 & -0.48 $\pm$ 0.62 & -0.08 $\pm$ 0.32 & 15 \cr
  Scd+Sd  & -0.25 $\pm$ 0.70 & -0.11 $\pm$ 0.57 & -0.38 $\pm$ 2.03 & -0.09 $\pm$ 0.65 & 17 \cr
  Sdm+Sm  & -0.15 $\pm$ 0.93 & -0.08 $\pm$ 0.69 & -0.28 $\pm$ 1.44 & -0.25 $\pm$ 0.34 &  9 \cr
  Im      & -0.68 $\pm$ 1.02 & -0.30 $\pm$ 1.06 & -0.86 $\pm$ 1.78 & -0.40 $\pm$ 0.41 & 13 \cr
  BCD     & -0.32 $\pm$ 0.36 & -0.36 $\pm$ 0.56 & -0.63 $\pm$ 0.67 & -0.40 $\pm$ 0.19 &  8 \cr
  S?      & -0.31 $\pm$ 0.55 & -0.26 $\pm$ 0.34 & -0.42 $\pm$ 0.74 & -0.47 $\pm$ 0.38 &  9 \cr
  ?       & -0.20 $\pm$ 0.45 & -0.08 $\pm$ 0.97 & -0.54 $\pm$ 1.13 & +0.01 $\pm$ 0.57 &  8 \\
 \enddata
 \label{tbl:Zgrd}
\end{deluxetable}

%%%%%%%%%%%%%%%%%%%%%%%%%%%%%%%%%%%%%%%%%%%%%%%%%%%%%%%%%%%%%%%%%%%%%%%%%%%%%%%%
% FIGURES
%%%%%%%%%%%%%%%%%%%%%%%%%%%%%%%%%%%%%%%%%%%%%%%%%%%%%%%%%%%%%%%%%%%%%%%%%%%%%%%%

% FIGURE 1
\clearpage

% FIGURE 2
\clearpage

% FIGURE 3
\clearpage

% FIGURE 4
\clearpage

% FIGURE 5
\clearpage

% FIGURE 6
\clearpage

% FIGURE 7
\clearpage

% FIGURE 8
\clearpage

% FIGURE 9
\clearpage

% FIGURE 10
\clearpage

% FIGURE 11
\clearpage

% FIGURE 12
\clearpage

% FIGURE 13
\clearpage

% FIGURE 14
\clearpage

% FIGURE 15
\clearpage

% FIGURE 16
\clearpage

\end{document}